\documentclass[10pt, a4paper]{article}
\usepackage{amsmath} 
\usepackage{mathtools} 
\usepackage{amsfonts}
\usepackage{amssymb}
\usepackage[main = english, ngerman]{babel}
\usepackage{graphicx}
\usepackage{microtype} 
\usepackage{footnote}
\usepackage{blindtext}
\usepackage{booktabs,caption}
\usepackage[flushleft]{threeparttable}
\usepackage{etoolbox}
\usepackage{lettrine}
\usepackage[pdfborderstyle={/S/U/W 1}, colorlinks, linkcolor = black, citecolor = black, filecolor = black, urlcolor = black]{hyperref} 
\usepackage{csquotes}
\usepackage{acro}
\usepackage{hyperref} 
\usepackage{setspace} 

\usepackage{titlesec}

\usepackage{geometry}

\usepackage[authoryear]{natbib}
\setcitestyle{authoryear,open={(},close={)}} 

\usepackage{rotating}
\usepackage{tabularx}
\usepackage{adjustbox}
\usepackage{blindtext}
\usepackage{epsfig}
\usepackage{longtable}
\usepackage{dcolumn}
\usepackage{here}
\usepackage{floatflt}
\usepackage{wrapfig}
\usepackage{dsfont}
\usepackage{graphics}
\usepackage{tikz}
\usepackage{epstopdf}
\usepackage{subcaption}
\usepackage{caption}
\usepackage{ragged2e}
\usepackage{xcolor}
\usepackage{soul} 
\usepackage{listings}
\usepackage[hang,flushmargin]{footmisc}
\usepackage{lipsum}
\usepackage{bbm}
\usepackage{eurosym} 
\usepackage{fancyhdr}

\graphicspath{{images/}}

\newgeometry{left=3cm,right=2cm,bottom=2cm,top=2cm}

\AtBeginDocument{\addtocontents{toc}{\protect\setlength{\parskip}{0pt}}}
\AtBeginDocument{\addtocontents{lof}{\protect\setlength{\parskip}{0pt}}}
\AtBeginDocument{\addtocontents{lot}{\protect\setlength{\parskip}{0pt}}}

\acsetup{first-style=short}

\newcommand{\thankyounote}[1]{%
    \begingroup%
    \let\thefootnote\relax%
    \footnotetext{#1}%
    \endgroup%
}

\newcounter{pageCount}
\begin{document}

\begingroup
\fontsize{12pt}{13.5pt}\selectfont 

\begin{titlepage}
	\newfont{\smc}{cmcsc10 at 12pt}
	\newgeometry{left=3cm,right=3cm,bottom=2cm,top=2cm}
	\linespread{1.5} 

\begin{center}
$~~$\\
$~~$\\
{\Large 
Characterizing Measurement Error in the German Socio-Economic \\[-0.5em] 
Panel Using Linked Survey and Administrative Data \\[3cm]
} 
\begin{singlespacing}
 \large  \textbf{Nico Thurow}\footnote{E-Mail: nthurow@uni-bonn.de} \\
 \normalsize Bonn Graduate School of Economics \\ 
 University of Bonn \\[3cm]
\end{singlespacing}
 Working Paper \\
 This version: \today \\[3cm]
\end{center}

\noindent \textbf{Abstract}

\begin{singlespacing}

\begin{justify}

\begin{changemargin}{0.9cm}{0.9cm} 

This paper exploits the linkage of German administrative social security data (German: \textit{Integrierte Erwerbsbiografien}) and survey data from the socio-economic panel (\textit{Sozio-ökonomisches Panel, SOEP}) for the characterization of measurement error in metrics quantifying individual-specific labor earnings in Germany. We find that survey participants' decision whether to consent to linkage is non-random based on observables. In that sense, the studied sample does not constitute a random sample of SOEP. Further, measurement error is not classical and differential: We observe underreporting of income on average, autocorrelation, and non-zero correlation with the true signal and other observable characteristics. In levels, calculated reliability ratios above 0.94 hint at a relatively small attenuation bias in simple linear univariate regressions with earnings as the explanatory variable. For one-period changes in income the bias from measurement error is exacerbated.

\thankyounote{I thank Christoph Breunig, Joachim Freyberger, Ingo Isphording, Amelie Schiprowski, Jakob Wegmann and participants of the Bonn Econometrics \& Statistics Brown Bag Seminar and the Applied Micro Coffee seminar at the University of Bonn for comments and helpful discussions. I am very grateful to have been able to use the infrastructure at GESIS Köln, which facilitated most of the data analysis. All errors are my own.}

\end{changemargin} 

\vspace{5mm}




\end{justify} 
\end{singlespacing}

\restoregeometry
\end{titlepage}


\newpage
\pagenumbering{roman} 

\tableofcontents

\newpage
\listoffigures
\listoftables

\newpage
\setcounter{pageCount}{\value{page}}
\pagenumbering{arabic}
\setcounter{page}{1}

\section{Introduction}\label{sec:introduction}

This research project exploits the linkage of administrative social security data (\textit{Integrierte Erwerbsbiografien}, IEB) and survey data from the \textit{Sozio-ökonomisches Panel} (SOEP) -- jointly called SOEP-CMI-ADIAB -- for the characterization of measurement error in metrics quantifying individual-specific labor earnings in Germany. This is crucial as measurement error can affect the accuracy of statistical analyses by biasing estimates. Evidence-based policy relies on empirical evidence establishing correlational associations as well as cause-effect relationships. Thus, if the underlying data informing policy makers is subject to measurement error, this can result in suboptimal policy decisions.

The linked dataset provides two measures of an individual's income: self-reported gross labor earnings of the month preceding the interview (SOEP) and average gross daily income of the current employment spell as reported by employers for the purpose of determining social security contributions (IEB). In principle, neither measure is necessarily free from measurement error. However, as commonly done in the validation study literature, we will assume that the register data represents the true underlying signal. Labor earnings are a key determinant for social security contributions and reporting by employers to the respective agencies are mandatory by law. In the case of deliberate or careless incorrect earnings reporting by employers, imposed fines can amount to \euro$50'000$ per report (c.f.\ German Social Act (\textit{Sozialgesetzbuch}) IV, §111). Arguably, this provides an incentive to report incomes correctly. Following a harmonization of the two variables, we define measurement error as the log difference between the two measures. In essence, this implies that the thereby constructed variable corresponds to reporting errors by SOEP respondents. The resulting panel enables us to explore the distribution of measurement error in SOEP as well as its dynamics over time.

Our contribution is threefold: Firstly, we are the first to provide a thorough characterization of measurement error in SOEP by being able to exploit the recent linkage to administrative data. Secondly, our findings complement the validation study literature, which has been dominated by evidence from the US, thereby allowing for a more comprehensive understanding of reporting errors in survey data. In addition, information on the decision to consent to linkage allows us to assess whether the resulting subsample of matched individuals can be considered a random subsample of SOEP. This may be of interest in settings which require a researcher to extrapolate the measurement error distribution to samples other than the validation dataset.

An emphasis is placed on data preparation by explaining the steps taken in detail. Our analysis is concerned with reporting errors in survey data and, thus, should not be polluted by noise stemming from poor data management. It is crucial that the two measurements of income reflect one and the same concept: gross monthly earnings from the main job in the month prior to the interview excluding any other monetary benefits. This is particularly intricate when trying to assess the extent of measurement errors among individuals holding multiple jobs at the same time or for occupations with untypical pay structures, which presumably are more common in the lower parts of the income distribution' support. In either case, it is not clear what responses made by SOEP participants in relation to income refer to exactly. We claim that our chosen sample restrictions imposed on SOEP-CMI-ADIAB suffice to guarantee a proper comparison, with the result of boiling things down to the pure reporting error.

A brief analysis of the consent decision reveals that linkage consent is not random. This has implications for estimation techniques based on extrapolating the measurement error distribution of an auxiliary sample (e.g., SOEP-CMI-ADIAB) to a larger primary dataset (e.g., SOEP).\footnote{C.f., \cite{chenetal2005}.} Further, we find that measurement error in SOEP is non-classical, exhibiting negative correlation with the true signal. This is in line with previous studies. For the sample constructed, we observe underreporting of labor income across the income distribution on average. This cannot be reconciled unambiguously with the phenomenon of mean-reversion found in previous studies. However, the proportion of individuals in a certain income quantile, who overreport, is declining in the income level. In addition, we use reliability ratios as the key determinant of the degree of attenuation bias due to measurement error. The results further stress the non-classical nature of measurement error in SOEP. However, the implied reliability when using survey-based income data is at the upper bound of the range of reliability ratios established in previous studies. Thus, according to these statistics, earnings data in SOEP appears to be reliable, and we expect the bias from measurement error to be small in univariate regressions.

The paper proceeds as follows: In Section \ref{sec:motivation}, we continue to present our main motivations for this paper and briefly discuss the underlying measurement error model as well as possible pitfalls associated with measurement error in the linear regression model. Section \ref{sec:literature} provides an overview of related literature. A detailed description of the two underlying data sources, the data linkage procedures and consenting behavior, and of sample processing steps are the focus of Section \ref{sec:data}. Section \ref{sec:analysis} presents the main results of this paper in two steps. First, we characterize the measurement error distribution and provide preliminary evidence for the severity of measurement error using Mincer-type regressions. Secondly, we provide estimated reliability ratios, which are subject to robustness tests. A discussion of obtained results and scope for extension follows in Section \ref{sec:discussion}. Section \ref{sec:conclusion} concludes. Additional figures and tables alluded to in the main text are provided in the \hyperref[sec:appendix]{Appendix}.

\section{Motivation}\label{sec:motivation}

In the Economics literature, a large part of well-received empirical studies concerning the German labor market relies on administrative data sources. Facilitated by the database of the platform \textit{Web of Science}, a first investigation resting upon the three keywords "\textit{German*}", "\textit{Labo*}", and "\textit{market}" aims to provide an overview of the different data sources used in publications from Top-5 economics journals for the years 2000-2023.\footnote{For the purpose of the analysis, we considered the \textit{American Economic Review}, \textit{The Quarterly Journal of Economics}, the \textit{Journal of Political Economy}, \textit{The Review of Economic Studies}, as well as \textit{Econometrica}. The asterisk \textit{*} used in the keywords allows the search engine to include words which start with the respective stem, e.g.\ \textit{Germany}, \textit{Labor}, and \textit{Labour}.} The analysis reveals that no study makes use of SOEP survey data: Out of 14 hits by the search engine not a single paper made use of German survey data but that most (7) relied on German administrative data.\footnote{Of the remaining studies, 3 made use of aggregate or firm-level data, 2 used non-German data sources, another self-gathered historical data, and one study was not empirical in nature. Note that this exercise is not restricted to income as the variable of interest.}

Administrative data are typically of large sample sizes, which arguably allows for a more precise estimation of focal parameters in finite samples. In many cases, a main drawback, however, is the lack of additional information about the cross-sectional units. The reverse appears to hold for the average survey dataset, which is smaller but allows for a plethora of observables to be included in analyses. However, the underlying reporting behavior is assumed to be noisy (if available at all due to item nonresponse), which implies that the resulting variable might not capture the true concept of interest. 

There exist several conjectures for the occurrence of mismeasurements of income: Firstly, suppose that the respondent understands the underlying concept, but simply does not have any or at most limited information about it. This may result in rough estimates, e.g.\ rounding or entirely random guesses. Secondly, suppose the respondent does not understand the underlying concept and states the true value of a different characteristic. For a regular employee, this could arise when the sum of gross income and the subjective value of any non-pecuniary benefit is stated instead of pure gross monthly labor earnings or when income from the previous position is reported. Another purported explanation is social desirability. Certain societal norms may seem valuable for an individual to comply with. In the context of income, this can refer to the attributes of being (financially) poor or rich, which corresponds to labor earnings at the lower and upper end of the income distribution, respectively. Empirically, it has been shown that relatively poor individuals tend to overstate their earnings whereas richer ones tend to understate their earnings in survey reports (c.f. Section \ref{sec:literature}). Hence, this channel is particularly appealing for explaining the phenomenon of mean-reversion in reported income data.

In general, administrative data can also be subject to noise and, therefore, deviate from the true signal. If income reports sent to public authorities by employers are not generated by automated software but entered manually, the submitted forms may not report the true amount of earnings from a given employment spell. This could be either due to transposed numbers or because of erroneously reporting the income level of a coworker of the individual in question. Further, the process of linking a survey respondent to her respective administrative data may result in linkage error, whereby an unsuitable observation of the administrative dataset is attached to that individual. However, as indicated above, we abstract from errors of this kind for now and consider register data to correspond to the true signal.

\subsection{Measurement error ($u$) in the linear model}\label{sec:ME_linear_model}

Formally, we can define a mismeasured variable $X$ as a function of the true signal $X^*$ and some noise $u$:
\begin{equation*}
X = g(X^*, u).
\end{equation*}
\noindent In general, we can think of $g(\cdot)$ to be a function of arbitrary form. As commonly practiced in the measurement error literature, we impose an additive error structure, i.e.\
\begin{equation}\label{eq:additive_measurement_error}
X = g(X^*, u) = X^* + u,
\end{equation}
\noindent which is also standard in the validation study literature. It is in this setting that we speak of $u$ as measurement error, which is therefore given by the difference between the mismeasured variable and the true signal. 

\subsubsection*{Classical measurement error}

Conventionally, the field distinguishes among classical and non-classical measurement error.\footnote{The following considerations largely follow \cite{pischke2007}.} The first type assumes independence of the measurement error with the true signal and a mean of zero. For the linear model, it has been shown that a consistent estimation of a parameter of interest depends on the role of the mismeasured variable in that model. If the observed (mismeasured) variable poses as the dependent variable, the parameters in that model can be estimated consistently provided the assumptions usually made in the context of ordinary least squares (OLS) estimation hold. In particular, we require that the measurement error is unrelated to the residual term. However, the presence of classical measurement error involves a larger variance to be explained by the model, which can loosely be interpreted as a loss in "efficiency". If the mismeasured variable is among the regressors, the parameters associated with this variable cannot be estimated consistently but suffer from attenuation bias: They are shrunk towards zero. If other independent variables are correlated with the mismeasured regressor, the bias carries over to the respective parameter estimates. Transformations of the mismeasured variables -- e.g., including $X^2$ among the regressors -- can exacerbate the attenuation bias.

As soon as the measurement error digresses from classical measurement error characteristics, we speak of non-classical measurement error. Exemplarily, this is the case if the reporting behavior exhibits mean-reversion, in which case $cov(X^*, u)<0$. As we will see, this is a common theme in the context of reported income in survey data. It turns out that non-classical measurement error in the dependent variable -- as opposed to solely in the covariates -- introduces bias in the linear model.

In order to assess the presence of a certain type of measurement error, the ideal yet infeasible setting would provide a researcher with the mismeasured variable as well as the measurement error. In particular, the marginal distribution of the error, i.e. the deviation between the mismeasured variable and the true signal, as well as joint distributions across different observable dimensions are then identified and can be analyzed. Moreover, knowledge about the variance of measurement error and about the variance of the mismeasured variable, respectively, allows for a correction of the estimator in the linear model, as elaborated below. Thus, if one believes in the case of classical measurement error, validation data allows for consistent estimation of both variances and, hence, for consistent estimation of the parameter of interest.

To depict possible pitfalls from measurement error, we review the linear regression model. We will see that the results of this simple exercise are fundamental to reliability statistics frequently used in validation studies. Consider the univariate linear model $Z_i^* = \alpha + \beta X_i^* + \epsilon_i$, where all parameters and random variables are scalar. Suppose we are given an independently and identically distributed sample of size $n$, i.e.\ $(Z_i^*,X_i)_{i=1}^n$, drawn from some joint distribution $f_{Z^*,X}$ with existing moments of first- and second-order. In particular, we do not observe the \textit{true} independent variable $X^*$ but a noisy measure $X$, where $X_i = X^*_i + u_i$. It is well known that the OLS estimator of $\beta$ in the univariate context using the mismeasured variable $X$ is given by
\begin{align*}
\hat{\beta}_{OLS} = \frac{\frac{1}{n}\sum_{i=1}^n(X_i - \bar{X})(Z_i^* - \bar{Z^*})}{\frac{1}{n}\sum_{i=1}^n(X_i - \bar{X})^2} \overset{p}{\longrightarrow} \frac{cov(Z_i^*, X_i)}{var(X_i)} \eqcolon \frac{\sigma_{Z^*,X}}{\sigma^2_{X}},
\end{align*}
\noindent where $\overset{p}{\longrightarrow}$ denotes convergence in probability. Assuming independence of the residual term $\epsilon$ and $X$ (e.g., $\epsilon_i \perp \left(X_i^*, u_i \right)$) and plugging in the structural equation for $Z_i$ yields $\sigma_{Z^*,X} = \sigma^2_{X} + \sigma_{X^*,u}$ in the nominator and $\sigma^2_X = \sigma^2_{X^*} + \sigma^2_u + 2\sigma_{X^*,u}$ in the denominator. In the presence of classical measurement error ($\mathbb{E}[u] = 0$ and $cov(u, X^*) = 0$), this results in the so-called \textit{attenuation bias} of the OLS estimator:
\begin{equation}\label{eq:bias_class_ME}
\hat{\beta}_{OLS} - \beta \overset{p}{\longrightarrow} \frac{-\sigma^2_u}{\sigma^2_{X^*} + \sigma^2_u} \beta \eqcolon \left(\lambda - 1\right)\beta.
\end{equation}
That is, the estimate will be inconsistent and biased towards $0$ since $\frac{\sigma^2_u}{\sigma^2_{X^*} + \sigma^2_u} > 0$ in general. Trivially, the OLS estimator is consistent in the absence of measurement error ($\sigma^2_u = 0$) or if there is no relation between the covariate of interest and the dependent variable ($\beta = 0$). In all other cases, the bias intensity depends on the variance of the measurement error relative to the variance of the true signal.\footnote{Note that we will also be able to identify $\beta$ if $\sigma^2_{X^*} \longrightarrow \infty$ while $\sigma^2_u$ remains finite. However, we require the existence of second-order moments for inference. Hence, it is not clear how useful this special case is.} In the context of panel data, the commonly applied estimators based on first differences or demeaning do not solve this problem either but potentially exacerbate it (see below). Further, this issue extends to the multivariate case as well, where the bias may spread to other estimated parameters in a linear model depending on the correlation between the mismeasured variable and the remaining (correctly measured) covariates (c.f.\ \citeauthor{boundetal1994}, \citeyear{boundetal1994}).

Note that if the ratio $\lambda = \frac{\sigma^2_{X^*}}{\sigma^2_{X^*} + \sigma^2_{u}}$ is known or of estimable nature, one can adjust the OLS estimator by pre-multiplying with the inverse of $\lambda$ (or a consistent estimate thereof) to obtain a consistent estimate of $\beta$. It is partly for this reason that the \textit{reliability ratio} $\lambda$ has played a central role in previous validation studies with attempts to estimate this parameter using information on the measurement error and the error-prone variable $X$. Another likely reason for its popularity is the great weight attached to linear models and OLS estimation in economic training and applied empirical studies. Loosely speaking, the reliability ratio can be interpreted as the share of the variation in the mismeasured variable stemming from the true signal. Knowledge about this statistic provides us with an indicator of the severity of the bias in the linear model.

\subsubsection*{Non-classical measurement error}

In general, we cannot be certain that the assumption of classical measurement error holds. Let us consider the same setting as above with measurement error in the independent variable only, now allowing for $cov(X^*,u) \neq 0$. Under this assumption, the probability limit of the OLS estimator of $\beta$ is given by
\begin{align*}
plim \textrm{ } \hat{\beta}_{OLS} = \frac{\sigma^2_{X^*} + \sigma_{X^*,u}}{\sigma^2_{X^*} + \sigma^2_{u} + 2\sigma_{X^*, u}} \beta \eqcolon \left(1-b_{Xu}\right) \beta.
\end{align*}
In this case, the sign of the bias $-b_{Xu} \beta$ cannot be pinned down and may as much as reverse signs of the estimate relative to the true parameter of interest. To see this, suppose $\beta>0$.\footnote{The same condition to maintain sign invariance applies if $\beta < 0$. For $\beta = 0$, measurement error has no attenuating effect in the first place.} In order for the probability limit of $\hat{\beta}_{OLS}$ to be of equal sign, one can show that this is equivalent to requiring $r_{X^*,u} > -\sqrt{\frac{\sigma^2_{X^*}}{\sigma^2_u}}$, where $r_{A,B}$ denotes the correlation coefficient of the two random variables $A$ and $B$. In general, violations of this condition cannot be ruled out. For a sufficiently large variation in the measurement error relative to $\sigma^2_{X^*}$ and large negative correlation of the measurement error with the true signal, the condition will no longer be satisfied. Previous research has shown that in the context of labor earnings, measurement error is negatively correlated with the true signal. If mismeasured income variables were used in the setting described above in addition to large variation of the measurement error, we would expect the sign of the estimate to differ from the sign of the parameter of interest.

Moreover, the presence of non-classical measurement error in addition to a negative correlation of the measurement error and the true signal may lead to estimates with a positive bias, making the terminology \textit{attenuation bias} no longer suitable in this context. The condition, for which a positive bias can show up, is given by $r_{X^*,u} < -\sqrt{\frac{\sigma^2_u}{\sigma^2_{X^*}}}$. This condition is me for a large negative correlation and a small relative variation of the measurement error.  Also note that in the special case with $\sigma_{X^*,u} = 0$ we are back to the classical case discussed above. 

Very similar to the classical case, estimation of $b_{XX^*} \coloneq (1-b_{Xu})$ appears useful to adjust for the bias from measurement error arising in the linear model when using OLS estimation. Note that in this model, this amounts to 
\begin{equation}\label{eq:bias_ME_nonclass}
b_{XX^*} = (1-b_{Xu}) = \frac{\sigma^2_{X^*} + \sigma_{X^*,u}}{\sigma^2_{X^*} + \sigma^2_{u} + 2\sigma_{X^*, u}} = \frac{cov(X^* + u,X^*)}{var(X^* + u)} = \frac{cov(X, X^*)}{var(X)},
\end{equation}
which, purely based on the mechanics associated with OLS, can be estimated by regressing the true signal $X^*$ on $X$ and a constant. That is, we impose $X^*_i = \gamma + b_{XX^*} X_i + \eta_i$ with orthogonal residuals and estimate this equation using OLS.\footnote{This relationship is apparent in the notation of $b_{XX^*}$ and $b_{Xu}$ as well, since the projection coefficients correspond to a linear projection of $X$ on $X^*$ and $u$, respectively.} We will make use of this idea when computing reliability ratios in the context of non-classical measurement error in Section \ref{sec:analysis}.

Very frequently, income related variables appear as dependent variables of interest in empirical analyses. In the non-classical case, measurement error results in biased OLS estimates. Take the univariate model $Z_i^* = \alpha + \beta X_i^* + \epsilon$ as above, but now suppose that we observe the random sample $(Z_i,X_i^*)_{i=1}^n$ with $Z_i = Z_i^* + \eta$ and the following structural measurement error model: $\eta_i = \delta Z_i^* + \nu$ with $\nu \perp (Z_i^*, X_i^*)$. In this case, the OLS estimator based on the regression of $Z_i$ on $X_i^*$ converges in probability to $(1+\delta)\beta$. Again, due to the empirical observation that the measurement error in income variables tends to be negatively correlated with the true signal, we expect this bias to be negative yielding an attenuated estimate of $\beta$.

\subsubsection*{Combinations of mismeasured variables and measurement error}

A combination of mismeasured variables potentially exacerbates the effect of measurement error. This also holds when the operation linking two mismeasured variables is linear. Consider a panel setting, in which a researcher is interested in questions related to changes of the mismeasured variable, i.e.\ $\Delta X_{it} = X_{it} - X_{it-1}$. This arises frequently in estimation of the fixed-effects model, which is of the following form: $Z_{it} = \mu_{i} + \gamma X_{it}^* + \nu_{it}$, where $\gamma$ shall be the parameter of interest. The model allows for individual-specific intercepts, which capture time-invariant levels of the outcome variable across cross-sectional units. Unlike the covariates $X_{it}$, these fixed effects are assumed to be correlated with the residuals $\nu_{it}$. One way to achieve consistent estimation of $\gamma$ is by computing first differences of the model, i.e., $\Delta Z_{it} = \gamma \Delta X_{it}^* + \Delta \nu_{it}$, and to estimate the parameter of interest using conventional parametric estimation methods. 

Now, suppose that we only have access to a mismeasured version of $X_{it} = X_{it}^* + u_{it}$ with an additive measurement error structure as in Equation \ref{eq:additive_measurement_error}. If we regress the differenced outcome variable on the differenced mismeasured variable, this amounts to estimating the following model: $\Delta Z_{it} = \gamma \Delta X_{it} + (\Delta \nu_{it} - \gamma \Delta u_{it})$. For simplicity, let us assume that the residuals $\nu_{it}$ are independent of the measurement error and that the available variables in first differences have been normalized so that they are equal to zero in expectation. For the case of classical measurement error, it can then be shown that the OLS estimator of $\gamma$ suffers from attenuation bias:
\begin{equation*}
\hat{\gamma}_{OLS} \overset{p}{\longrightarrow} \gamma \frac{\sigma^2_{\Delta X_{t}^*}}{\sigma^2_{\Delta X_{t}^*} + \sigma^2_{\Delta u_{t}}}
\end{equation*}
Let us impose some more structure on the temporal relationship of the individual variables and assume that measurement error is not autocorrelated whereas the true signal is, namely, $X_{it}^* = \rho X_{it-1}^* + \xi_{it}$ with $\xi_{it}$ white noise and $\rho \neq 0$. Further, we simply apply our static measurement error model to the dynamic case. That is the measurement error of $\Delta X_{it}$ is equal to first differences in $u_{it}$, which is another assumption on the error model. Moreover, the variances are assumed to be homoskedastic across time, i.e.\ $\sigma^2_{X_{t}^*} = \sigma^2_{X^*}$ and similarly for $u_{it}$. This yields
\begin{equation*}
\hat{\gamma}_{OLS} \overset{p}{\longrightarrow} \gamma \frac{(1-\rho)\sigma^2_{X^*}}{(1-\rho)\sigma^2_{X^*} + \sigma^2_{u}}.
\end{equation*}
Note that for $\rho = 0$ we are back to the previously determined attenuation bias $\lambda$, as defined from Equation \ref{eq:bias_class_ME}. If the autocorrelation of the independent variable is large, this results in a large bias. Thus, by first differencing we incur the cost of introducing measurement error twice while reducing the variance of the differenced independent variable subject to mismeasurement. This increases the share of the measurement error variance in the total variance of the mismeasured variable. In the context of log-income as our variable of interest, we observe a large degree of persistence in administrative income across time while measurement error is less correlated. Thus, this special case appears to be particularly worrisome. In the general case, allowing for serial correlation among measurement errors as well, differencing can be desirable if the autocorrelation of $u_{it}$ exceeds that of $X_{it}$ (\citeauthor{pischke2007}, \citeyear{pischke2007}).

\section{Literature review}\label{sec:literature}

This paper falls within the strand of literature, which aims to characterize measurement error in income data using linked survey and administrative data. We make a distinction between two common approaches: a parametric and a mostly nonparametric approach. In the following, we discuss contributions in each respective realm and place our paper in the latter.

The parametric approach relies on the modelling assumption that different types of measurement error are jointly normally distributed.\footnote{In addition to the pure reporting error, reference period error on the survey participant's side, linkage error due to erroneous matching of survey and administrative data during the linkage process, or a mistaken declaration of income by the employing party are considered (c.f.\ \citeauthor{jenkinsriosavila2023}, \citeyear{jenkinsriosavila2023}).} This implies a mixture model with latent classes to be estimated from a validation dataset using maximum likelihood. In contrast to most nonparametric validation studies, the analyses in this group do not rely on the assumption that the administrative data represents the true signal but allow for measurement error in register data as well. Moreover, the measurement error distribution is allowed to vary with observable covariates.

Using validation data from Sweden, \cite{kapteynypma2007} find large errors in administrative data due to linkage errors and argue that survey data should be considered as more reliable. \cite{meijeretal2012} formalize the previous analyses by mapping them to a mixture factor model with the question of factor score prediction at its core. The authors also attribute a larger reliability to Swedish survey data and make out a lower mean squared error compared to register data. More recently, \cite{jenkinsriosavila2023} extend the framework suggested by \citeauthor{kapteynypma2007} by explicitly allowing for survey participant's reference period errors resulting from instable employment conditions and apply it to UK data. Again, the authors find that UK survey data is more reliable and attribute only a small share of the discrepancy to the newly introduced reference period error. Moreover, the task of predicting income is said to benefit from the inclusion of both variables in the analysis. Together, these findings question the nonparametric approach, which often relies on the assumption that the true signal is given by the register data. Further, the three studies do not find evidence for the existence of mean reversion, which arguably presented the main rationale for engaging in a more thorough analysis of the measurement error's supposedly non-classical structure. However, results by \cite{jenkinsriosavila2020} suggest that some of these results are sensitive to the initial share of observations assumed to be error-free.

The nonparametric approach does not rely on distributional assumptions regarding the diffusive error processes. However, most analyses of this type have relied on the assumption that in addition to the mismeasured variable the true signal is available for juxtaposition. Typically, this is assumed to be provided by register data. The first contributions were based on US data and relied heavily on male survey participants. In their seminal paper, \cite{boundkrueger1991} coin the notion of the \textit{reliability ratio}, which is defined as the ratio of the signal's variance to the mismeasured variable's variance. Their study of the \textit{Current Population Survey} (CPS) for the years 1977 and 1978 finds that 82\% (92\%) of the mismeasured variable's variance can be attributed to the true underlying signal. Moreover, the authors document a positive correlation between the reporting error and the true signal (defined as \textit{mean-reversion}) as well as serial correlation of reporting error observations up to two periods apart. In addition, the effect of the error accumulates when considering first-differences, finding its expression in lower reliability ratios. Using a validation study of the \textit{Panel Study of Income Dynamics} (PSID) for the years 1982 and 1986, \cite{boundetal1994} find similar evidence for non-classical measurement error in reported income variables. To rationalize these findings, \cite{pischke1995} devises an autoregressive-moving-average (ARMA) model, which relies on the decomposition of income into a transitory and a permanent part, respectively. Based on the same PSID validation study, the author finds decreasing reliability ratios over time and argues that this can be attributed to business cycle fluctuations. Further, the author hypothesizes that this results from the underreporting of non-salient transitory earnings changes.

Relying on the same data used by \cite{boundkrueger1991}, the study by \cite{bollinger1998} estimates the conditional expectation function of mismeasured earnings conditional on the true signal nonparametrically. The hypothesized mean-reversion of reported income variables is largely confirmed and turns out to be more pronounced among men. Zooming in further, the results from estimating conditional quantiles reveal that a small share of survey participants at the lower end of the income distribution highly overstate their earnings relative to the register data, making these individuals largely responsible for the observed mean-reversion of earnings reports. Although the focus is on exploring correlations of measurement error with observables, \cite{kimtamborini2014} give evidence for mean-reversion in the US's \textit{Survey of Income and Program Participation} (SIPP). In contrast to those findings, \cite{bingleymartinello2014} use data gathered by the \textit{Survey of Health, Ageing and Retirement in Europe} (SHARE), which was linked to Danish register data for the years 2003 and 2004, to give evidence of classical measurement error in the survey responses with regard to income. However, they find lower reliability ratios than observed in US surveys ($55\%$-$68\%$). 

To this date, three studies have used validation datasets based on German survey and administrative data. \cite{valetetal2019} investigate the effect of interviewer presence on non-reporting and misreporting biases in earnings variables in the \textit{Legitimation of Inequality Over the Life-Span} (LINOS) survey, which was linked to administrative data processed by the German Federal Institute for Employment Research (IAB). The main finding is that interviewer presence is associated with higher rates of non-reporting but does not have a significant effect on misreporting behavior. More importantly, they find that measurement error is negatively correlated with income, which is once more evidence for mean-reversion. Whereas the aforementioned study is of descriptive nature, \cite{schmillenetal2024} use data from the surveys \textit{Panel Arbeitsmarkt und soziale Sicherung} (PASS, engl.: \textit{Panel Study Labor Market and Social Security}) and \textit{Weiterbildung und Lebenslanges Lernen} (WeLL, engl.: \textit{Continuing Education and Lifelong Learning}) linked to IEB data, respectively, to replicate more fundamental classification analyses of measurement error based on \cite{boundkrueger1991} and \cite{pischke1995}. In addition, they are able to extend the dynamic measurement error by exploiting a richer panel structure ($T=4$). The authors find evidence for mean-reversion and attribute as much as $16\%$ of the total variance of the reported nominal income to measurement error. 

\cite{caliendoetal2024} are the first to examine the linked SOEP-CMI-ADIAB dataset in the context of measurement error. In their working paper, the authors provide evidence for mean-reverting measurement error in SOEP. However, their study is not based on econometric models or conventional measurement error theoretical considerations but mostly relies on the estimation of pooled covariances between measurement error and selected observable characteristics. They proceed to estimate "commonly studied economic relationships" (p.\ 18) in labor economics using either income data source and find discrepancies in the estimates of returns to education, i.e.\ income as dependent variable, and the association of income with indicators of life satisfaction, i.e.\ income as independent variable.

The analysis in this paper diverges from \citet{caliendoetal2024} in several ways. First, we extend the characterization of measurement error in SOEP by building upon the considerations by \cite{schmillenetal2024} with own exercises interspersed. In particular, the richer panel structure in the SOEP data allows for an analysis of the dynamics of measurement error for four time periods, which is not exploited by \cite{caliendoetal2024}.\footnote{The authors, however, provide summary statistics of the measurement error for each available survey year without explicitly referring to repeated observations of a cross-sectional unit over time.} Secondly, the steps taken to prepare the data and to obtain the core sample differ. Moreover, their analyses make use of data up until 2019, whereas we are able to draw on information up until 2021. This has the effect that our main sample contains almost twice as many cross-sectional units compared to \cite{caliendoetal2024}.

\section{Data}\label{sec:data}

The empirical analysis in this paper is based on the socio-economic panel (SOEP) survey dataset hosted by the German Institute for Economic Research (DIW) linked to administrative data processed by the research data center (FDZ) at the Institute for Employment Research (IAB). The final data product used is SOEP-CMI-ADIAB (c.f.\ \citeauthor{SOEPCMIADIAB}, \citeyear{SOEPCMIADIAB}). In the following, we describe both data sources in detail, shed light on the linkage procedure, and guide the reader through the sample processing steps.

\subsection{Survey data -- The Sozio-ökonomisches Panel (SOEP)}\label{sec:surveydata}

The \textit{Sozio-ökonomisches Panel} (SOEP, engl.: \textit{socio-economic panel}) relies on a representative survey of the German population and covers topics related to health, employment, education, and personal attributes, among many others.\footnote{For more information on SOEP and its modules, i.e.\ SOEP-Core and SOEP-IS, please refer to \cite{SOEP_CORE} and \cite{SOEP_IS}.} Over the period of 1984 to 2021, a total of $116'278$ individuals from $63'996$ households have participated in the survey. In order to track dynamics in the constitution of the population, to account for attrition, and, thus, to maintain representativeness, the SOEP frequently draws “refresher” samples, lastly taken place in 2017. It is for this reason that SOEP is an unbalanced panel dataset. Moreover, an emphasis is put on specific subgroups, which are then oversampled in some waves.\footnote{Among others, these include members of the LGBTQIA+ community, individuals with migration background, refugees, and inhabitants of urban areas.} As we are interested in features of the measurement error's distribution among the employed part of the general population -- as e.g.\ the expected value, the variance, or other statistics characterizing distributions -- sampling probabilities can be used to adjust for this non-random sampling scheme.

A short note on how the survey is implemented: Although there has been a transition from “Pen-and-Paper Interviews” (PAPI) to “Computer-Assisted Personal Interviews” (CAPI), most interviewing modes involve a direct exchange between the interviewer and the respondent.\footnote{The SOEP allows unsupervised interview completion only in rare cases, e.g., by mail or remote access. Exemplarily, in 2019, these two modes were deployed in approx. $10\%$ of all interviews.} Hence, we argue that there is little variation in interviewing modes and, unlike e.g.\ \cite{valetetal2019}, we will abstract from this factor in the analysis below. Moreover, the questions always refer to the person interviewed, i.e.\ SOEP does not allow for responses by a proxy. This rules out misreporting arising from the lack of knowledge about the characteristics of others.

One caveat should be noted in the weighting context: In essence, the linkage occurred for two mutually exclusive (in terms of cross-sectional units) modules within the SOEP, namely the SOEP-Core (active since its inception in 1984) and the SOEP Innovation Sample (SOEP-IS, active since 1998, not conducted in 2021). The latter addresses important research questions that the former cannot. In that sense, SOEP-IS is more flexible than SOEP-Core and allows for adaptation to specific topics over time. However, some key questions are part of both modules, including queries on employment status and income-related information. Therefore, participants from both data sources are valuable to our analysis. The weighting issue arises when estimating population parameters by means of weighted averages based on survey weights. To our knowledge, it is advertised against pooling the samples with the corresponding weights and weighting the estimates using the combined data without harmonizing across modules. Further, adequate adjustments appear to be non-trivial. For that reason, we provide either unweighted statistics or apply the weights to estimates based on data of each module, respectively. We leave further considerations in this regard to future research.

Using the provided sampling weights, Figure \ref{fig:SOEP_sampling} highlights the slightly different sampling structures across modules by sketching the weighted and unweighted empirical density function of gross monthly incomes in 2020. Top-earners were omitted to improve the presentation. The rightward shift of the weighted income distribution in SOEP-Core hints at an oversampling of low-income individuals. Neglecting sampling weights for now, the SOEP-IS sample appears to cover the German working population more broadly, which mirrors its more representative sampling nature. This is also corroborated by the fact that the weighted estimates of the variance of the income distribution in SOEP-Core is much smaller than its unweighted counterpart whereas the decrease is much more attenuated in SOEP-IS. In both modules, there is a lot of mass in the interval of \euro$0$ to \euro$500$ in reported gross monthly earnings. This is arguably due to the popularity of marginal employment contracts, which have the benefit of being exempt from social insurance contributions. Since the empirical density functions are smooth at the upper assessment limit for social security contributions of \euro$6900$ (in gray), this threshold does not appear to affect the income level.

\begin{figure}[H]
    \centering
    \caption{Representativeness of SOEP}
    \includegraphics[scale=0.55]{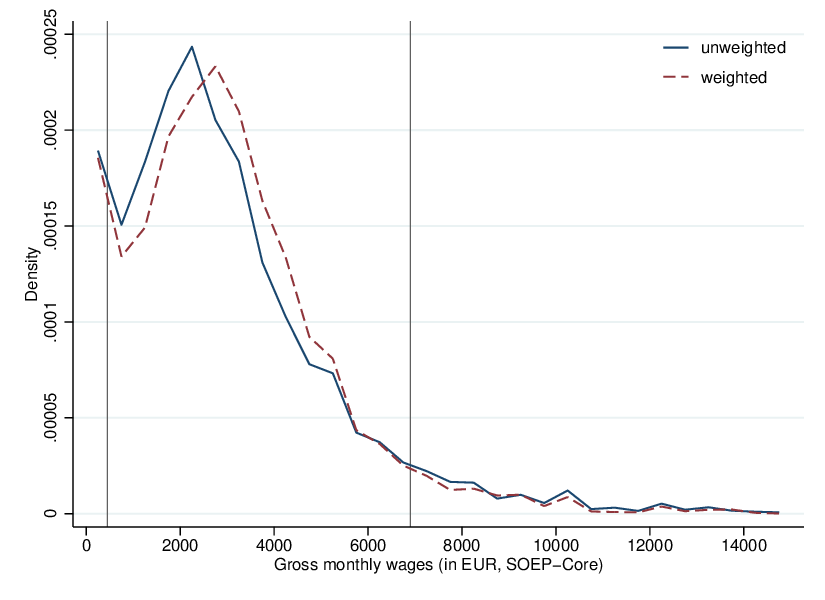}
    \includegraphics[scale=0.55]{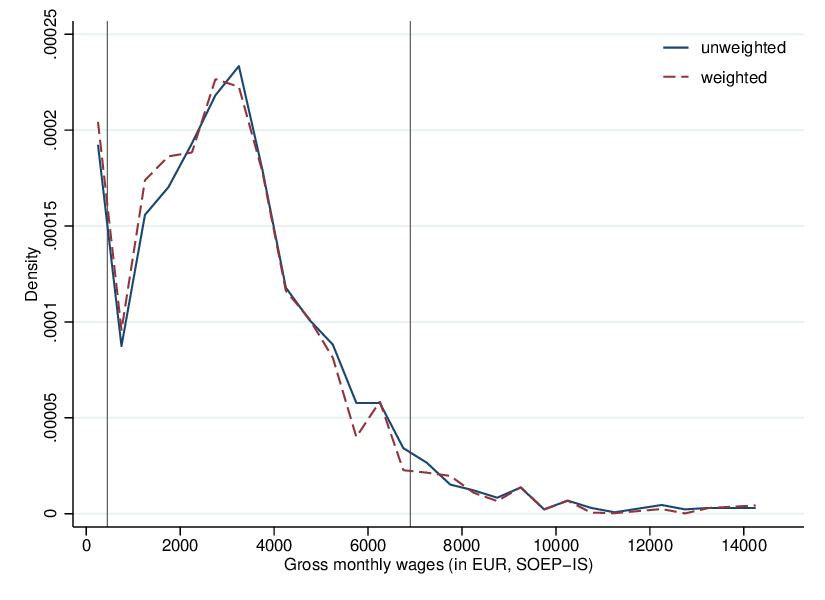}
    \label{fig:SOEP_sampling}
    \begin{minipage}{\textwidth}
        \begin{singlespace}
            \footnotesize{Figure \ref{fig:SOEP_sampling} shows the unweighted and weighted empirical density functions of gross monthly earnings in the modules SOEP-Core (left panel) and SOEP-IS, respectively, using a histogram with bins of size \euro$500$ for working individuals surveyed in 2020. Approx.\ $1\%$ of the observations (in the right tail) were omitted from the graphs. The gray vertical bars denote the lower (\euro$450$) and upper (\euro$6'900$) social security assessment limit for the year 2020. Mean: \euro$3131.21$ (\euro$3'058.37$), \euro$3'228.14$ (\euro$3'088.46$); standard deviation: \euro$3'437.65$ (\euro$2'612.58$), \euro$2'435.21$ (\euro$2'345.81$) (left to right, weighted statistics in brackets).}
        \end{singlespace}
    \end{minipage}
\end{figure}

The final sample used to characterize measurement error in the SOEP will be restricted to income and employment types subject to social security contributions, which is mostly due to sample criteria of the administrative data source (see below). For comparison, Figure \ref{fig:SOEP_sampling_main} presents the weighted and unweighted empirical density functions restricted to observations from our main sample.\footnote{The plots are based on pooled observations across available survey years. According to SOEP, the weights within each module refer to the underlying population in each survey year. If the population changes over time, this approach may not be adequate any longer. We will leave a more thorough discussion of this issue for future considerations not covered in this paper.} Although less pronounced, the results from Figure \ref{fig:SOEP_sampling} carry over to the main sample. The less representative nature of unweighted SOEP-Core sample is now also corroborated by the fact that the weighted estimates of the mean of the income distribution increases in both modules relative to the unweighted density. However, the graph on the right suggests that low-to-middle and middle-to-high earners may be oversampled in the restricted \textit{Innovation Sample}, which is arguably the result of the small sample size. Importantly, as opposed to \cite{schmillenetal2024}, SOEP resembles the population more than PASS, since the latter targets low-income households. This speaks for a more encompassing perspective provided by our results if the reporting behavior of the general working population is of interest.

\subsection{Administrative data -- Integrierte Erwerbsbiografien (IEB)}\label{sec:admindata}

As the research division of the German Federal Employment Agency (BMAS), the Institute of Employment Research (IAB) preprocesses and analyzes information on part of the economically active German population, i.e., the German labor force. In particular, data covering both unemployed but available for work, and employed individuals, who are subject to social security contributions, are used in aggregate statistics and facilitate public research. Arguably the most commonly known data product is the \textit{Survey of Integrated Labor Market Biographies} (SIAB), which constitutes a $2\%$ random sample of employees subject to social security contributions -- the \textit{Integrierte Erwerbsbiografien} (IEB, engl.: Integrated Employment Biographies). The data includes information related to the calculation of social security contributions for a given employment spell, e.g., personal information (nationality, average gross daily earnings, full-time vs.\ part-time employment, place of residency, etc.) and employer-related characteristics (industry class, firm age, establishment size, location of the employing establishment, etc.).

In light of our endeavors here, it is important to note that self-employed workers are not covered by IAB data products. In that sense, our analyses may not generalize to the entire German workforce nor to the economically inactive part of the population as a whole, as e.g.\ children, pensioners, or those unable to work at all are also not covered. Further, as we require the availability of the two income variables in the first place, our analysis is conditional on being in an active employment relationship. This implies that we cannot consider jobseekers in our study either. More euphemistically speaking, we shed light on the distribution of measurement error in survey income data for regularly employed individuals living in Germany.

In Germany, social security contributions are incurred as a proportion of individual earnings. However, this proportional fee is not levied uniformly across income groups but is capped from above and below. Moreover, the limits are adjusted annually to inflation, and vary for East and West Germany, respectively. In practice, this implies that individuals with earnings larger than the assessment limit incur a lower share of social security contributions relative to their income. Therefore, earnings larger than the assessment limit are not relevant for the calculation of social security contributions in a given year and were top coded, i.e.\ set to be equal to the assessment limit. According to \cite{drechsleretal2023}, this affects $10$ to $13\%$ of full-time employment spells.\footnote{The lower bound corresponds to cases of marginal employment (“Minijob”), which are characterized by a low level of income while working few hours. Earnings from such engagements are not subject to social security contributions. However, earnings data for this type of employment are available in the administrative data source since 1999. In 2020, earnings up to \euro$450$ per month were allowed in order to still be considered marginally employed.} These restrictions of the register data limit our analysis in that we cannot compare reported and employer-based data for units in the upper tail of the income distribution. In 2020, the assessment limit amounted to approximately \euro$6900$ of gross monthly labor income.

\subsection{Consent decision and data linkage}\label{sec:consent}

The procedure of linking cross-sectional units from SOEP to administrative data was designed as follows: In a first step, $45'345$ SOEP respondents participating at least once from 2019 to 2021 were asked to give their consent to linkage of their responses to register data from current as well as previous years.\footnote{This question was part of SOEP-IS only in 2019.} The answer of $33'733$ ($74\%$) individuals was affirmative. In a second step, the administrative data was matched to the group of consenters using information on surname, first name, date of birth, gender, and postal address. Among the consenters, $90\%$ could be successfully linked either through an exact match ($26'686$ individuals), or a probabilistic or manual match ($3'667$). This yields the SOEP-CMI-ADIAB sample consisting of $30'353$ survey participants.

Table \ref{tab:consent_SOEPtotal} presents summary statistics for the SOEP in column $(1)$ and the respective subsamples based on the linkage consent procedure (columns $(2)$ to $(9)$). All variables considered stem from the survey. The calculated statistics are based on the cross-section and do not incorporate the panel structure available in SOEP. Although the variables observed are marked by large standard deviations, we argue that differences across groups, to which the question of linkage consent was presented (\textit{Asked}) or not (\textit{Not asked}), reflects the temporal dimension of the query and the different composition of refresher waves within SOEP. For example, the focus on individuals with a migration background has intensified in recent years, which finds its expression in the observation that almost half of those asked for linkage consent have a history of migration. Further, this is reflected in the share of persons with German nationality, which is 24 percentage points (pp.) lower compared to the group of survey participants from the years 2018 and earlier not having been asked. Moreover, this group exhibits higher average monthly incomes -- likely due to inflation and increasing income levels over time -- and a later average year of first participation in the survey. Thus, we observe an oversampling of late-joiners due to the rather recent addition of the linkage consent query to the questionnaire. The relatively high variation in gross earnings among individuals not asked for their consent can be ascribed to the much larger sampling period of $2018-1984 + 1 = 35$ years, thus more likely reflecting income dynamics over time.

\begin{sidewaystable}[!htbp] 
    \begin{adjustbox}{scale=0.95,center}
    \begin{threeparttable}
        \caption{Consent decision -- Unweighted summary statistics}
        \label{tab:consent_SOEPtotal}
        \begin{tabular}{lccccccccc}
            \toprule 
            \toprule
            & \multicolumn{9}{c}{\small{Subsamples of the SOEP}} \\
            & \small{Total} & \small{Not asked} & \small{Asked} & \small{Not consented} & \small{Consented} & \small{Not matched} & \small{Matched} & \small{Exact match} & \small{Other match}\\
            & \footnotesize{(1)} & \footnotesize{(2)} & \footnotesize{(3)} & \footnotesize{(4)} & \footnotesize{(5)} & \footnotesize{(6)} & \footnotesize{(7)} & \footnotesize{(8)} & \footnotesize{(9)} \\[0.3ex]
            \hline \\[-1.8ex]
Female    &     0.50&     0.51&     0.50&     0.51&     0.49&     0.49&     0.49&     0.49&     0.47\\
          &\footnotesize(0.50)&\footnotesize(0.50)&\footnotesize(0.50)&\footnotesize(0.50)&\footnotesize(0.50)&\footnotesize(0.50)&\footnotesize(0.50)&\footnotesize(0.50)&\footnotesize(0.50)\\
Single    &     0.30&     0.31&     0.29&     0.30&     0.29&     0.29&     0.29&     0.29&     0.30\\
          &\footnotesize(0.46)&\footnotesize(0.46)&\footnotesize(0.45)&\footnotesize(0.46)&\footnotesize(0.45)&\footnotesize(0.45)&\footnotesize(0.45)&\footnotesize(0.45)&\footnotesize(0.46)\\
German    &     0.78&     0.87&     0.63&     0.64&     0.63&     0.50&     0.64&     0.68&     0.40\\
          &\footnotesize(0.42)&\footnotesize(0.33)&\footnotesize(0.48)&\footnotesize(0.48)&\footnotesize(0.48)&\footnotesize(0.50)&\footnotesize(0.48)&\footnotesize(0.47)&\footnotesize(0.49)\\
Migration background&     0.34&     0.25&     0.49&     0.52&     0.49&     0.61&     0.47&     0.44&     0.71\\
          &\footnotesize(0.48)&\footnotesize(0.43)&\footnotesize(0.50)&\footnotesize(0.50)&\footnotesize(0.50)&\footnotesize(0.49)&\footnotesize(0.50)&\footnotesize(0.50)&\footnotesize(0.45)\\
East      &     0.18&     0.19&     0.17&     0.15&     0.18&     0.15&     0.19&     0.19&     0.15\\
          &\footnotesize(0.39)&\footnotesize(0.39)&\footnotesize(0.38)&\footnotesize(0.36)&\footnotesize(0.39)&\footnotesize(0.35)&\footnotesize(0.39)&\footnotesize(0.39)&\footnotesize(0.35)\\
Abitur    &     0.18&     0.18&     0.18&     0.17&     0.18&     0.18&     0.19&     0.19&     0.13\\
          &\footnotesize(0.39)&\footnotesize(0.39)&\footnotesize(0.39)&\footnotesize(0.38)&\footnotesize(0.39)&\footnotesize(0.39)&\footnotesize(0.39)&\footnotesize(0.39)&\footnotesize(0.34)\\
Education (years)&    11.69&    11.80&    11.52&    11.51&    11.52&    11.10&    11.57&    11.68&    10.73\\
          &\footnotesize(2.78)&\footnotesize(2.65)&\footnotesize(2.95)&\footnotesize(2.88)&\footnotesize(2.98)&\footnotesize(3.29)&\footnotesize(2.94)&\footnotesize(2.92)&\footnotesize(2.95)\\
Age (years)&    44.99&    45.08&    44.85&    45.16&    44.75&    45.25&    44.69&    45.15&    41.32\\
          &\footnotesize(18.91)&\footnotesize(19.68)&\footnotesize(17.70)&\footnotesize(18.36)&\footnotesize(17.47)&\footnotesize(21.10)&\footnotesize(17.01)&\footnotesize(17.05)&\footnotesize(16.37)\\
Employed  &     0.57&     0.57&     0.57&     0.56&     0.57&     0.37&     0.59&     0.60&     0.50\\
          &\footnotesize(0.50)&\footnotesize(0.50)&\footnotesize(0.50)&\footnotesize(0.50)&\footnotesize(0.50)&\footnotesize(0.48)&\footnotesize(0.49)&\footnotesize(0.49)&\footnotesize(0.50)\\
Gross monthly income (\euro)&  2647.93&  2507.90&  2863.50&  2727.80&  2908.87&  2943.38&  2906.49&  2911.46&  2863.03\\
          &\footnotesize(10963.00)&\footnotesize(13845.96)&\footnotesize(3163.77)&\footnotesize(2826.56)&\footnotesize(3267.59)&\footnotesize(4290.42)&\footnotesize(3184.88)&\footnotesize(3222.07)&\footnotesize(2839.52)\\
Year of first participation&     2006&     2002&     2013&     2012&     2013&     2013&     2013&     2013&     2014\\
          &\footnotesize(11.32)&\footnotesize(11.42)&\footnotesize(7.34)&\footnotesize(7.92)&\footnotesize(7.12)&\footnotesize(7.34)&\footnotesize(7.09)&\footnotesize(7.23)&\footnotesize(5.75)\\
SOEP-Core (share)&     0.90&     0.90&     0.90&     0.84&     0.92&     0.94&     0.92&     0.92&     0.95\\
          &\footnotesize(0.30)&\footnotesize(0.30)&\footnotesize(0.30)&\footnotesize(0.36)&\footnotesize(0.27)&\footnotesize(0.24)&\footnotesize(0.27)&\footnotesize(0.28)&\footnotesize(0.21)\\[0.3ex]
          \hline \\[-1.8ex]
            N  &   116'278&    70'933&    45'345&    11'612&    33'733&     3'380&    30'353&    26'686&     3'667\\
            Share of total (in \%) & 100 & 61 & 39 & 10 & 29 & 3 & 26 & 23 & 3 \\[0.3ex]
            \bottomrule
            \bottomrule
        \end{tabular} 
        \begin{tablenotes}
            \item \footnotesize{Table \ref{tab:consent_SOEPtotal} provides the estimated mean and standard deviation (in brackets) of observable characteristics for different subsamples of SOEP based on inclusion of the question of consent in the respective survey wave (\textit{Asked}), the latest available consent decision (\textit{Consented}), and the match type (\textit{Matched}). The group \textit{Other match} includes probabilistic and manual matches. All estimated statistics concerning time-varying variables are based on the last available observation for each individual. $N$ denotes the total number of cross-sectional units in a given subsample. The shares are relative to the total sample size ($116'278$). The shares may not add up to one due to rounding errors. Data: SOEP, SOEP-CMI-ADIAB.}
        \end{tablenotes}
    \end{threeparttable}
    \end{adjustbox}
\end{sidewaystable}

Columns $(4)$ and $(5)$ aim to capture selection effects arising from individual consent decisions. Females appear to be less likely to consent to linkage, which is a feature that carries over to our main sample used in subsequent analyses. Non-German citizens are less likely to agree with data linkage, which translates to a smaller share of individuals with migration background within the consenting group. Moreover, consenters are on average $0.41$ years younger and have completed the highest form of German secondary education, the \textit{Abitur}, in $18\%$ of cases compared to $17\%$ among the non-consenters. The gap in monthly earnings of \euro$181.07$ could be interpreted as a likely cause from this. Additionally, individuals with a slightly more recent entry to the survey appear to be more likely to belong to the group of consenters, as indicated by the more recent average year of first participation (2013 vs.\ 2012). Lastly, the consenting behavior appears to differ across SOEP modules. Namely, participants of the SOEP-IS are relatively well represented within the group of non-consenters. These considerations hint at the necessity of looking at the results for each module separately, which is done in Tables \ref{tab:consent_SOEPCore_unweighted} and \ref{tab:consent_SOEPIS_unweighted}.

Let us finish the discussion of Table \ref{tab:consent_SOEPtotal} by highlighting differences in the samples depending on the success of the matching procedure outlined above. Although most consenters were successfully matched to corresponding social security data observations, a few differences are apparent. Matching is more successful for German nationals, which may be related to a higher likelihood of misspelling uncommon names. The high share of persons with a migration background lacking an exact match ($71\%$ in column $(9)$) may be taken as additional evidence for this hypothesis. A similar observation holds for differences in average years of education, which is roughly one year higher among the group of individuals with an exact match. Matched individuals are younger, on average, and with $60\%$ a much larger share is employed compared to $37\%$ for the group without a successful match. Regarding the mode of matching (columns $(8)$ and $(9)$) it appears that difficulties arise for relatively young, non-German citizens residing in Western Germany, without \textit{Abitur} degree, and a slightly lower income compared to exactly matched individuals.

In summary, the resulting SOEP-CMI-ADIAB sample, which consists of $30'353$ successfully matched individuals, differs on average with respect to the overall SOEP sample in the following dimensions: age (younger), education (less), German citizenship (fewer), and year of first participation in the SOEP survey (later). If measurement error is correlated across these dimensions, the results in Section \ref{sec:analysis} characterizing the measurement error distribution may not be transferable to the overall population. Note that the interpretation of the results is based on point estimates without consideration of any inference procedure accounting for sampling error.

To get a better understanding of the origin of the variation in the covariates, Table \ref{tab:consent_SOEPtotal} is replicated for the subsamples of the SOEP-Core (Tables \ref{tab:consent_SOEPCore_unweighted} and \ref{tab:consent_SOEPCore_weighted}) and SOEP-IS (Tables \ref{tab:consent_SOEPIS_unweighted} and \ref{tab:consent_SOEPIS_weighted}). Since the two modules within SOEP are constructed as comprehensive surveys in their own right, we discuss summary statistics based on each module individually and compare them to each other column by column. In addition, we consider males and females separately in Tables \ref{tab:consent_SOEPtotal_women} and \ref{tab:consent_SOEPtotal_men}.

With only $10\%$ of the observations in SOEP arising from the innovation sample, the aggregate statistics are shaped mostly by observations from SOEP-Core. As the two modules have a slightly different purpose -- with SOEP-Core oversampling marginalized groups while the SOEP-IS strives to represent the entire German population --, their composition differs along many dimensions. More specifically, the average individual in SOEP-IS is more likely to be female and less likely to be single. The share of German nationals is $18$ pp.\ larger in SOEP-IS, which is also reflected in a smaller share with migration background of $24\%$ relative to $35\%$. Furthermore, the mean level of education surpasses that of SOEP-Core members, as emphasized by the larger fraction of individuals with \textit{Abitur} degree ($24\%$ vs.\ $18\%$) and more years of education ($12.39$ years vs.\ $11.60$ years). In addition, SOEP-Core covers individuals seven years younger than SOEP-IS participants on average. Although SOEP-IS initially rolled out in 1999, it was extended only in 2009 and subsequent years, which explains the rather recent average year of first participation (2013) compared to SOEP-Core (2005). Throughout the linkage procedure, most results from Table \ref{tab:consent_SOEPtotal} hold also for the module SOEP-Core. However, a couple of patterns in SOEP-IS stand out. Firstly, among the group of individuals asked in SOEP-IS, $59\%$ consented to linkage, which is $17$ pp.\ lower than the corresponding figure in SOEP-Core. This may be related to the differing sample composition, with an oversampling of high-income, well-educated German nationals in SOEP-IS relative to SOEP-Core. Secondly, the share of individuals living in Eastern Germany increases along the steps of linkage in SOEP-IS, while remaining relatively stable in SOEP-Core.

Using Tables \ref{tab:consent_SOEPCore_weighted} and \ref{tab:consent_SOEPIS_weighted} and by computing and interpreting weighted summary statistics, we again strive to assess the representative nature of the two modules, respectively. Although the weights refer to sampling probabilities in a cross-section of a given year, we proceed to consider all observations jointly by pooling across years. In SOEP-Core, the estimates in column $(1)$ of Table \ref{tab:consent_SOEPCore_weighted}, which are now constructed using sampling weights, exhibit a larger share of German nationals, which implies an undersampling of this part of the population by this module. The reverse holds for individuals from Eastern Germany, with a migration background, or of young age, which are oversampled in SOEP-Core. Moreover, the table suggests that there may be more participants with high incomes in the survey than in the population. When comparing summary statistics in columns $(7)$, we note that the group of matched consenters may underrepresent individuals with \textit{Abitur} degree or individuals in employment. 

Turning to SOEP-IS, the different aims of the two modules is reflected in different propensities to over- or underrepresent a certain part of the population residing within the boundaries of the German Federal Republic, as can be seen by the juxtaposition of Tables \ref{tab:consent_SOEPIS_unweighted} and \ref{tab:consent_SOEPIS_weighted}. Contrary to SOEP-Core, young individuals or persons with migration background appear to be undersampled in this module while German nationals are overrepresented in this module. In addition, the proportion of females (singles) is higher (lower) compared to the population. In regard to the final SOEP-CMI-ADIAB product, restricted now to observations from SOEP-IS, the share of Germans living in the East is disproportionately high whereas the share of employed individuals is too low relative to the population.

Finally, we present summary statistics by sex in Tables \ref{tab:consent_SOEPtotal_women} and \ref{tab:consent_SOEPtotal_men}, pooling across modules again. Samples are evenly split across steps of the linkage procedure, which is indicated by comparable sizes and shares of the cross-section by gender. On average, the share of single women is smaller compared to their male counterparts. Moreover, men appear to be slightly younger on average and drive the share of non-Germans upward. More importantly, labor market outcomes appear to be heterogeneous across gender, with men enjoying $66\%$ higher monthly salaries on average and a $11$ pp.\ higher share of employed individuals. These gaps, although slightly reduced, persist in the sample of matched consenters. If the income level is relevant for the quality of income reports, then considering the distributions by gender is crucial to capture the underlying heterogeneity in reporting correctly or not.

\subsection{Data processing}\label{sec:data_processing}

In the following, we will briefly explain the measures taken to retrieve the main sample to be used in subsequent analyses. Broadly, the procedure can be separated into two parts. The first part deals with finding the correct employment spell from the register data to be compared with the reported income from SOEP. In the second part, we impose some sample restrictions, which are based on \cite{schmillenetal2024} as well as own considerations. Table \ref{tab:sample_restrictions} tracks respective changes in the sample size.\footnote{This table was inspired by \cite{caliendoetal2024}.}

\begin{table}[!htbp] 
    \centering
    \begin{threeparttable}
        \caption{Sample restrictions and sample sizes}
        \label{tab:sample_restrictions}
        \begin{tabular}{llccc}
            \toprule 
            \toprule
            & & \multicolumn{3}{c}{No. of cross-sectional units} \\
            \multicolumn{2}{l}{} & \small{SOEP-Core} & \small{SOEP-IS} & \small{Total} \\[0.3ex]
            \hline \\[-1.8ex]
             & SOEP & 104'523 & 11'755 & 116'278 \\
            \multicolumn{2}{l}{\textit{Linkage}} \\
             & SOEP-CMI-ADIAB  & 27'929 & 2'424 & 30'353 \\
            \multicolumn{2}{l}{\textit{Matching spells}} \\
             & Unbalanced $(i,t)$-panel & 19'288 & 1'624 & 20'607 \\
            \multicolumn{2}{l}{\textit{Sample restrictions}} \\
            \multicolumn{1}{c}{\hspace{1em}1.} & Below assessment limit & 17'027 & 1'440 & 17'027 \\
            \multicolumn{1}{r}{2.} & Typical pay structures & 14'339 & 1'209 & 15'360 \\
            \multicolumn{1}{r}{3.} & $|u_{it}| \leq 150\%$ & 14'067 & 1'185 & 15'068 \\
            \multicolumn{1}{r}{4.} & Coinciding YoB & 14'018 & 1'184 & 15'020 \\
            \multicolumn{1}{r}{5.} & Working population & 13'871 & 1'167 & 14'863 \\
            \multicolumn{1}{r}{6.} & Within SSC limits & 13'335 & 1'108 & 14'286 \\
            \multicolumn{1}{r}{7.} & Income not imputed & 12'926 & 1'108 & 13'887 \\
            \multicolumn{2}{l}{\textit{Main sample}} \\
            & Main sample & 12'926 & 1'108 & 13'887 \\
            \multicolumn{2}{l}{\textit{Balancing}} \\
            & Weakly balanced & 12'926 & 1'108 & 13'887 \\
            & Strongly balanced & 4'667 & 527 & 5'118 \\
            [0.3ex]
            \bottomrule
            \bottomrule
        \end{tabular}  
        \begin{tablenotes}
            \item \footnotesize{Table \ref{tab:sample_restrictions} displays the number of unique cross-sectional units left after application of sample restrictions (in order). The naming convention of the sample restrictions reflects the condition to remain in the sample. That is, observations not fulfilling a requirement are dropped at the respective step. SOEP-CMI-ADIAB denotes the subsample of the SOEP for those cross-sectional units, who were asked, consented to linkage, and were successfully matched (c.f.\ Table \ref{tab:consent_SOEPtotal}). The tuple $(i,t)$ refers to individual $i$ and survey year $t$. Notation: SSC -- social security contribution(s); $u_{it}$ -- deviation of reported income from administrative income; YoB -- year of birth. The sum of cross-sectional units across modules may not add up to the last column due to survey participants switching modules over time. For more information, please refer to the main text. Data source: SOEP, SOEP-CMI-ADIAB.}
        \end{tablenotes}
    \end{threeparttable}
\end{table}

For our analysis we require our main sample to have a similar structure to the least granular data source. Although the register data at hand provides information on the entire employment history of an employed person throughout a year, possibly including several employment relationships for a single individual at a certain point in time, SOEP survey data is limited to information about the main source of labor income the month prior of the interviewing date. For the analysis of measurement error, the task thus lies in identifying the income of the corresponding spell in IEB data. We first select spells which cover the month prior to the month the SOEP interview took place. Secondly, we drop entries related to unemployment insurance, which leaves us with employment-related notices only. If more than one spell at different firms is available, the main job is selected, which we identify as the one with the highest pay. Moreover, we exclude one-time payments. In effect, this implies that we omit $9'746$ mostly unemployed individuals from the linked dataset, which do not allow for comparing incomes across the two data sources due to missing data in the first place.

By interpreting the SOEP query as earnings from the main (i.e., the highest paying) job, we deviate from the baseline sample restrictions in \cite{caliendoetal2024}. If revenues from multiple jobs are indicated in the register data for overlapping spells, the authors compute the sum of all earnings for juxtaposition with the survey income in a corresponding month. Admittedly, the phrasing of the survey question in isolation leaves room for interpretation. However, if seen as part of a sequence of questions, the question has to be considered jointly with the following statement preceding the question: \textit{"If you have more than one professional activity, please answer the following questions only for your current main professional activity."}\footnote{Original version in German: \textit{“Wenn Sie mehr als eine berufliche Tätigkeit ausüben, beantworten Sie die folgenden Fragen bitte nur für Ihre derzeitige berufliche Haupttätigkeit.”} (c.f.\ p.\ 35 in \citeauthor{SOEPsurvey2020}, \citeyear{SOEPsurvey2020})} We argue that the information elicited captures the concept described above and, thus, reflects earnings from the highest paying job. \cite{schmillenetal2024} appear to follow this line of argument in their derivations as well.

Sample restrictions are based on information provided by SOEP and mostly follow conventional procedures from the validation data literature. The aim is to limit the noise in the administrative data and to enable a comparison of two variables, which mirror the same concept. First, occupations not subject to social security contributions are discarded.\footnote{These include persons without employment, apprentices, trainees, interns, retirees, self-employed, civil servants, individuals in educational training, and individuals performing military or voluntary service.} Secondly, professions with untypical pay structures are eliminated from the sample. These include working at piecework rates, zero-hour contracts, and tipping.\footnote{E.g., cab drivers, cleaners, hotel and tourist industry staff, waiters, barkeepers, managers, clerks, hairdressers and tattooists, cosmeticians, artists, and performers.} These two restrictions are the most consequential, leading to a joint loss of $5'247$ units in the cross-section. 

Following \cite{schmillenetal2024}, we omit observations for which the absolute difference of the reported income and the register data exceeds $150\%$ relative to either the survey data or the administrative data. To further limit the extent of a false linkage of survey to register data, we exclude individuals with disagreeing year of birth across the two data sources or with responses from proxies. In addition, we restrict the sample to include members from the working-age population only, which we take to be given by the ages $18$ to $65$. We apply the annually changing threshold of the assessment limits supplied by \cite{dautheppelsheimer2020} and drop top-coded\footnote{In fact, we utilize $98\%$ of the respective assessment limit as the restricting threshold, following a recommendation by \cite{dautheppelsheimer2020}.} incomes from the sample. Since spells of marginal employment are covered reliably only from 1999 onwards, we omit any observations with monthly earnings below the corresponding marginal employment threshold before 1999. In addition, imputed incomes in the SOEP are excluded.\footnote{In the context of earnings volatility, \cite{ziliaketal2023} find that the inclusion of imputed income data introduces bias. Hence, we consider solely observations with non-imputed income.} The resulting main sample constitutes the basis for our analysis. We end up with a panel of $13'887$ unique individuals over the period of 1984 until 2021.

In order to characterize the dynamics of measurement errors over time, we rely on two balanced subsets of the main sample. In particular, we allow only for cross-sectional units interviewed and employed in consecutive survey years, i.e., individuals with non-missing earnings data from one year to another. The first observation of an individual $i$ with active employment spell serves as the initial observation, i.e.\ the temporal reference point. Observations following a gap in an individual's employment history are discarded. The resulting panel is then referred to as \textit{weakly balanced}, which remains to cover the $13'887$ units of the main sample. Note that this still results in an unbalanced panel in the sense that the number of observations differs across units due to gaps in the employment histories. In order to obtain the \textit{strongly balanced} panel, we further impose the requirement of consecutive employment for at least the number of time periods that we consider for the characterization of dynamics of measurement error and the true signal. Since the PASS dataset used by \cite{schmillenetal2024} includes four waves, the authors are only able to consider dynamics over a time horizon of $T=4$. The strongly balanced sample, whose relatively small size of $5'118$ cross-sectional units is indicated in Table \ref{tab:sample_restrictions}, is based on this time horizon.

Tables \ref{tab:main_summary_stats_all} and \ref{tab:main_summary_stats_all_IEB} present summary statistics for the cross-section by gender for selected subsamples: the unbalanced panel resulting from omitting observations without employment in a given month from all matched observations (columns $(1)$ and $(4)$), the main sample after applying the sample restrictions outlined above ($(2)$ and $(5)$), and the two balanced samples, which differ based on the periods of consecutive employment required ($T=1$ for columns $(3)$ and $(7)$, $T=4$ for columns $(4)$ and $(8)$). All sample averages are based on the respective underlying sample pooled across all combinations of $(i,t)$. The outcomes in the overall SOEP panel, which can be found in column $(1)$ of Table \ref{tab:consent_SOEPtotal}, will serve as a benchmark in the following discussion of Table \ref{tab:main_summary_stats_all}.

\begin{sidewaystable}
    \centering
    \begin{threeparttable}
        \caption{Main samples -- Summary statistics by gender (SOEP)}
        \label{tab:main_summary_stats_all}
        \begin{tabular}{lcccccccc}
            \toprule 
            \toprule
            & \multicolumn{4}{c}{Men} & \multicolumn{4}{c}{Women} \\ 
            &  \small{Unbalanced} & \small{Main} & \small{Weakly} & \small{Strongly} &  \small{Unbalanced} & \small{Main} & \small{Weakly} & \small{Strongly} \\
            & \small{$(i,t)$-panel} & \small{sample} & \small{balanced} & \small{balanced} & \small{$(i,t)$-panel} & \small{sample} & \small{balanced} & \small{balanced} \\
            & \footnotesize{(1)} & \footnotesize{(2)} & \footnotesize{(3)} & \footnotesize{(4)} & \footnotesize{(5)} & \footnotesize{(6)} & \footnotesize{(7)} & \footnotesize{(8)} \\[0.3ex]
            \hline \\[-1.8ex]
Single      &        0.37&        0.36&        0.38&        0.27&        0.32&        0.30&        0.32&        0.24\\
            &\footnotesize(0.47)&\footnotesize(0.46)&\footnotesize(0.48)&\footnotesize(0.42)&\footnotesize(0.45)&\footnotesize(0.44)&\footnotesize(0.46)&\footnotesize(0.41)\\
German      &        0.63&        0.68&        0.68&        0.89&        0.80&        0.84&        0.84&        0.93\\
            &\footnotesize(0.48)&\footnotesize(0.46)&\footnotesize(0.46)&\footnotesize(0.30)&\footnotesize(0.40)&\footnotesize(0.36)&\footnotesize(0.37)&\footnotesize(0.25)\\
Migration background&        0.48&        0.44&        0.44&        0.24&        0.34&        0.30&        0.30&        0.20\\
            &\footnotesize(0.50)&\footnotesize(0.50)&\footnotesize(0.50)&\footnotesize(0.43)&\footnotesize(0.47)&\footnotesize(0.46)&\footnotesize(0.46)&\footnotesize(0.40)\\
East        &        0.18&        0.19&        0.19&        0.23&        0.20&        0.21&        0.21&        0.23\\
            &\footnotesize(0.38)&\footnotesize(0.39)&\footnotesize(0.39)&\footnotesize(0.42)&\footnotesize(0.40)&\footnotesize(0.41)&\footnotesize(0.41)&\footnotesize(0.42)\\
Abitur      &        0.17&        0.17&        0.17&        0.18&        0.22&        0.25&        0.25&        0.24\\
            &\footnotesize(0.37)&\footnotesize(0.38)&\footnotesize(0.38)&\footnotesize(0.39)&\footnotesize(0.41)&\footnotesize(0.43)&\footnotesize(0.43)&\footnotesize(0.43)\\
Education (years)&       11.52&       11.67&       11.63&       12.14&       12.13&       12.50&       12.46&       12.65\\
            &\footnotesize(2.82)&\footnotesize(2.69)&\footnotesize(2.67)&\footnotesize(2.42)&\footnotesize(2.57)&\footnotesize(2.56)&\footnotesize(2.55)&\footnotesize(2.41)\\
Age (years) &       38.83&       38.89&       37.78&       41.72&       40.25&       40.66&       39.20&       42.96\\
            &\footnotesize(13.32)&\footnotesize(11.76)&\footnotesize(11.82)&\footnotesize(10.70)&\footnotesize(12.87)&\footnotesize(11.49)&\footnotesize(11.76)&\footnotesize(10.48)\\
Employed    &        0.90&        1.00&        1.00&        1.00&        0.89&        1.00&        1.00&        1.00\\
            &\footnotesize(0.25)&\footnotesize(0.00)&\footnotesize(0.00)&\footnotesize(0.00)&\footnotesize(0.25)&\footnotesize(0.00)&\footnotesize(0.00)&\footnotesize(0.00)\\
Gross monthly income in SOEP (\euro)&     2667.74&     2611.50&     2500.74&     2861.15&     1758.44&     1910.63&     1803.11&     2048.17\\
            &\footnotesize(2212.07)&\footnotesize(1254.39)&\footnotesize(1269.08)&\footnotesize(997.17)&\footnotesize(1326.80)&\footnotesize(1126.27)&\footnotesize(1138.34)&\footnotesize(1022.43)\\
Gross monthly income in IEB (\euro)&     2457.65&     2774.30&     2652.44&     3088.79&     1736.78&     2060.23&     1946.12&     2243.59\\
            &\footnotesize(1695.72)&\footnotesize(1393.53)&\footnotesize(1401.61)&\footnotesize(1123.91)&\footnotesize(1352.05)&\footnotesize(1269.28)&\footnotesize(1278.23)&\footnotesize(1171.12)\\
Year of first participation&        2013&        2012&        2012&        2008&        2011&        2010&        2010&        2008\\
            &\footnotesize(7.58)&\footnotesize(8.12)&\footnotesize(8.12)&\footnotesize(8.79)&\footnotesize(7.84)&\footnotesize(8.01)&\footnotesize(8.01)&\footnotesize(8.30)\\
SOEP-Core (share)&        0.93&        0.93&        0.93&        0.90&        0.92&        0.92&        0.92&        0.90\\
            &\footnotesize(0.25)&\footnotesize(0.25)&\footnotesize(0.25)&\footnotesize(0.30)&\footnotesize(0.27)&\footnotesize(0.27)&\footnotesize(0.27)&\footnotesize(0.29)\\[0.3ex]
            \hline \\[-1.8ex]
No.\ of $(i,t)$ observations           &       63'952&       36'730&       27'080&       20'050&       66'737&       38'788&       27'037&       20'084\\
N       &        10'787&        7'013&        7'013&  2'495&        9'820&        6'874& 6'874&        2'623 \\[0.3ex]
            \bottomrule
            \bottomrule
        \end{tabular}  
        \begin{tablenotes}
            \item \footnotesize{Table \ref{tab:main_summary_stats_all} presents summary statistics and sample sizes for selected observables from the survey data source in the underlying SOEP-CMI-ADIAB dataset, the main dataset used for pooled analyses, and two balanced subsamples restricted to consecutive employment across years used for the calculation of the variance-covariance matrix. The naming of the subsamples corresponds to the one used in Table \ref{tab:sample_restrictions}. For the time-varying variables, the summary statistics capture the empirical mean across individuals of individual-level average outcomes. Standard errors in parentheses. Data: SOEP-CMI-ADIAB (SOEP variables only).}
        \end{tablenotes}
    \end{threeparttable}
\end{sidewaystable}

\begin{sidewaystable}
    \centering
    \begin{threeparttable}
        \caption{Main samples -- Summary statistics by gender (IEB)}
        \label{tab:main_summary_stats_all_IEB}
        \begin{tabular}{lcccccccc}
            \toprule 
            \toprule
            & \multicolumn{4}{c}{Men} & \multicolumn{4}{c}{Women} \\ 
            &  \small{Unbalanced} & \small{Main} & \small{Weakly} & \small{Strongly} &  \small{Unbalanced} & \small{Main} & \small{Weakly} & \small{Strongly} \\
            & \small{$(i,t)$-panel} & \small{sample} & \small{balanced} & \small{balanced} & \small{$(i,t)$-panel} & \small{sample} & \small{balanced} & \small{balanced} \\
            & \footnotesize{(1)} & \footnotesize{(2)} & \footnotesize{(3)} & \footnotesize{(4)} & \footnotesize{(5)} & \footnotesize{(6)} & \footnotesize{(7)} & \footnotesize{(8)} \\[0.3ex]
            \hline \\[-1.8ex]
Marginal employment&        0.14&        0.06&        0.07&        0.02&        0.24&        0.13&        0.15&        0.07\\
            &\footnotesize(0.31)&\footnotesize(0.22)&\footnotesize(0.24)&\footnotesize(0.10)&\footnotesize(0.36)&\footnotesize(0.31)&\footnotesize(0.34)&\footnotesize(0.21)\\
Complex tasks&        0.24&        0.27&        0.26&        0.29&        0.20&        0.26&        0.26&        0.28\\
            &\footnotesize(0.40)&\footnotesize(0.42)&\footnotesize(0.43)&\footnotesize(0.43)&\footnotesize(0.36)&\footnotesize(0.41)&\footnotesize(0.42)&\footnotesize(0.42)\\
Flexible hours&        0.09&        0.09&        0.12&        0.12&        0.13&        0.13&        0.16&        0.15\\
            &\footnotesize(0.23)&\footnotesize(0.25)&\footnotesize(0.31)&\footnotesize(0.29)&\footnotesize(0.27)&\footnotesize(0.28)&\footnotesize(0.34)&\footnotesize(0.30)\\
Temporary employment&        0.26&        0.22&        0.22&        0.11&        0.25&        0.22&        0.22&        0.13\\
            &\footnotesize(0.39)&\footnotesize(0.38)&\footnotesize(0.39)&\footnotesize(0.27)&\footnotesize(0.37)&\footnotesize(0.37)&\footnotesize(0.39)&\footnotesize(0.29)\\
Share w.\ multiple spells&        0.10&        0.09&        0.09&        0.08&        0.11&        0.11&        0.11&        0.10\\
            &\footnotesize(0.23)&\footnotesize(0.24)&\footnotesize(0.26)&\footnotesize(0.20)&\footnotesize(0.24)&\footnotesize(0.26)&\footnotesize(0.28)&\footnotesize(0.23)\\
Year establishment first observed in IEB&        1996&        1995&        1994&        1991&        1994&        1993&        1993&        1991\\
            &\footnotesize(13.59)&\footnotesize(13.67)&\footnotesize(14.00)&\footnotesize(12.28)&\footnotesize(12.70)&\footnotesize(13.05)&\footnotesize(13.39)&\footnotesize(12.42)\\
No.\ of employees at establishment&      838.84&      842.15&      839.47&      997.52&      600.48&      652.12&      642.71&      650.31\\
            &\footnotesize(3819.03)&\footnotesize(3742.20)&\footnotesize(3828.13)&\footnotesize(4562.67)&\footnotesize(2604.63)&\footnotesize(2605.65)&\footnotesize(2594.92)&\footnotesize(2618.67)\\
No.\ of employees working full time&      621.56&      627.04&      628.51&      775.31&      361.55&      395.10&      390.79&      407.47\\
            &\footnotesize(3165.04)&\footnotesize(3101.94)&\footnotesize(3191.09)&\footnotesize(3829.04)&\footnotesize(2052.75)&\footnotesize(2009.42)&\footnotesize(2018.50)&\footnotesize(2105.40)\\
No.\ of employees in marginal employment&       41.41&       37.88&       39.47&       29.26&       51.58&       47.52&       51.56&       36.40\\
            &\footnotesize(206.69)&\footnotesize(216.50)&\footnotesize(241.78)&\footnotesize(163.27)&\footnotesize(212.42)&\footnotesize(231.28)&\footnotesize(252.17)&\footnotesize(170.59)\\
Notification: \textit{Annual report}&        0.70&        0.73&        0.71&        0.81&        0.68&        0.72&        0.69&        0.81\\
            &\footnotesize(0.33)&\footnotesize(0.33)&\footnotesize(0.37)&\footnotesize(0.18)&\footnotesize(0.31)&\footnotesize(0.32)&\footnotesize(0.37)&\footnotesize(0.18)\\
Notification: \textit{Employment termination}&        0.20&        0.17&        0.18&        0.09&        0.20&        0.17&        0.19&        0.09\\
            &\footnotesize(0.30)&\footnotesize(0.30)&\footnotesize(0.33)&\footnotesize(0.14)&\footnotesize(0.28)&\footnotesize(0.28)&\footnotesize(0.33)&\footnotesize(0.14)\\[0.3ex]
            \hline \\[-1.8ex]
No.\ of $(i,t)$ observations           &       63'952&       36'730&       27'080&       20'050&       66'737&       38'788&       27'037&       20'084\\
N       &        10'787&        7'013&        7'013&  2'495&        9'820&        6'874& 6'874&        2'623 \\[0.3ex]
            \bottomrule
            \bottomrule
        \end{tabular}  
        \begin{tablenotes}
            \item \footnotesize{Table \ref{tab:main_summary_stats_all_IEB} presents summary statistics and sample sizes for selected observables made available through the administrative source in the underlying SOEP-CMI-ADIAB dataset, the main dataset used for pooled analyses, and two balanced subsamples restricted to consecutive employment across years used for the calculation of the variance-covariance matrix. The naming corresponds to the one used in Table \ref{tab:sample_restrictions}. For the time-varying variables, the summary statistics capture the empirical mean across individuals of individual-level average outcomes. Standard errors in parentheses. Notation: No.\ -- number; w.\ -- with; imp.\ -- imputed. Data: SOEP-CMI-ADIAB (IEB variables only).}
        \end{tablenotes}
    \end{threeparttable}
\end{sidewaystable}

Let us first analyze variables provided by the survey dataset. In the first unbalanced subsample of SOEP-CMI-ADIAB, the median sex is male. However, the share of observations associated with a female individual is equal to $51\%$, which is closer to the SOEP benchmark of $50\%$. Moreover, the average individual is younger, more likely to be from Eastern Germany, and the share of singles is larger. The latter is driven largely by male individuals. The heterogeneity across gender carries over to the share of German nationals (individuals with a migration background), which is particularly low (high) among men. In the case of education, the mean of SOEP is surpassed by the \textit{unbalanced $(i,t)$-panel} sample, although there is heterogeneity across gender with females driving this statistic. Since this first subsample was generated by mostly omitting observations without active employment, the average employment status across individuals ($90\%$) is much higher than in SOEP ($57\%$). The positive gender gap in incomes persists, which is arguably the result of women being more likely to work part-time than men.

The restrictions imposed to yield the main sample are non-random as hinted at by the obtained summary statistics. The share of people with migration background drops by $4$ pp.\ across gender with the average still staying slightly above the overall share in SOEP ($37\%$ vs.\ $34\%$). Further, the average education level rises by $0.26$ years compared to the first subsample, which is largely driven by observations associated with women. This is also reflected in the share of individuals with an \textit{Abitur} degree, which only increases among women. Moreover, the share of spells linked to unemployment drops to zero -- as required -- and the average income level increases. In general, omitting occupations with untypical pay structures as well as top-coded observations appear to be responsible for these changes in the summary statistics across subsamples.

For the \textit{weakly balanced} panel, which similarly requires at least one existing spell of employment, summary statistics with respect to time-invariant characteristics in the first seven rows are roughly comparable to columns $(2)$ and $(6)$. Any changes in the calculated statistics stems from dropping all observations following a gap in the employment history on an individual level. Thus, the number of individuals remains the same while the number of observations per person decreases. Muting the income growth channel for such individuals, this sample restriction affects average incomes, where we observe a drop in levels for both income-related variables across gender. Naturally, we also observe a lower average number of years of age, which may also be reflected in larger share of singles.

Finally, the subsample requiring consecutive employment for a time horizon of $T=4$ since the first period of employment indication exhibits vastly different summary statistics than the previous subsample. This hints at more severe selection effects stemming from this requirement. The majority of remaining respondents are German nationals, without migration background, more likely living in the East. The average age is four years above the average age of the previous subsample, thus getting closer to the average age in SOEP ($44.99$). The absence of individual breaks in the employment history is associated with higher incomes and a larger number of years of education enjoyed. Since all linked individuals must have participated in SOEP survey in 2019, 2020, or 2021, the longer time horizon also implies that relatively recent additions to the pool of respondents are omitted from this subsample. This is also reflected in the lower average year of first participation in SOEP. The relatively large average number of observations per cross-sectional unit ($7.8$) mirrors the longer time horizon as well. Moreover, the share of respondents from the SOEP-Core module drops to $90\%$, hinting at the fact that -- due to the distinct sampling scheme with a disproportionate focus on economically poor -- individuals sampled in that module are less likely to fulfill the imposed requirements.

A comparison of average gross monthly income across data sources, i.e.\ administrative and survey-based variables, reveals that the structure of the measurement error may not be uniform across created subsamples. In columns $(1)$ and $(5)$, the average reported income is larger than the average true signal. This relationship is reversed once we impose additional sample restrictions as well as the requirement of consecutive employment. The same holds for the variances. Most likely, this sign reversal of the difference is due to the top coding of administrative income data, which is not the case for the survey data. In turn, once we restrict ourselves to labor earnings within the assessment limits -- as done in columns $(2)$ to $(4)$ and $(6)$ to $(8)$--, administrative earnings are larger on average (c.f.\ Figure \ref{fig:SOEP_vs_IEB}).

Based on the summary statistics in table \ref{tab:main_summary_stats_all_IEB}, we mimic the exercise from above relying on variables provided by the register data. In the \textit{Unbalanced $(i,t)$-panel} (columns $(1)$ and $(5)$), the average income level of $17\%$ of the individuals lies below the marginal assessment limit, which classifies them as marginally employed. This form of employment is more prevalent among women, where $24\%$ of the observations are associated with marginal employment spells. The underlying jobs for women involve complex tasks to a lesser extent than for men ($20\%$ vs.\ $24\%$). Interestingly, temporary employment spells appear to be equally frequent across gender, with an average of $25\%$. The same holds for the average share of observations with multiple employment spells applying to roughly $10\%$ of all observations. In terms of the number of coworkers, men tend to work at larger establishments. Roughly $90\%$ of all employer-based notices making up the administrative data were filed as part of the mandatory annual report or indicating the end of employment.

Narrowing the range of incomes considered and excluding certain occupations leads to a sharp decrease in the share of marginally employed or temporarily employed individuals. Moreover, the task complexity gap across gender vanishes. Omitting observations after a gap in the respective individual's employment history does not alter the estimates by much, as can be seen by comparing columns $(2)$ and $(6)$ to columns $(3)$ and $(7)$. However, there is an increase in the share of employment spells with flexible hours from $11\%$ to $14\%$ in the aggregate. Requiring consecutive employment for at least four years is the most consequential in terms of observables, as can be deduced from columns $(4)$ and $(8)$. While still being more frequently observed among women, the share of observations associated with marginal employment drops to $4\%$. More stable employment appears to be positively associated with task complexity, which increases to over $28\%$. Moreover, the employer associated with a position is more established (in terms of age) and larger (in terms of establishment size), with an average year first observed of $1991$ and an average size of $819.41$ employees at an establishment.

In summary, the sample restrictions imposed have a large impact on the constitution of the resulting subsample in terms of observable characteristics. This calls for caution when aiming to extrapolate the results of this paper to the general population.

The comparison of administrative and survey income data will be based on monthly earnings before taxes. Since the IEB data provides the average daily income of the corresponding employment spell, a conversion scheme is required. We follow \cite{schmillenetal2024} and compute gross monthly administrative earnings by multiplying the daily income with the average number of days per month $\left(\frac{7}{12} \cdot 31 + \frac{4}{12} \cdot 30 + \frac{1}{12}\cdot 28.75 \approx 30.48\right)$. Measurement error $u_{it}$ is then defined as the difference between (natural) log-transformed survey income data $Y_{it}$ and log-transformed administrative income data $Y_{it}^*$ for individual $i$ in year $t$:
\begin{equation*}
u_{it} = log(Y_{it}) - log(Y_{it}^{*}) = log\left(\frac{ Y_{it}}{ Y_{it}^*}\right).
\end{equation*}
This choice of notation using the asterisks relates directly to the introductory discussion on measurement error models in Section \ref{sec:ME_linear_model}. Namely, throughout our analysis, we take administrative data to represent the true gross monthly income level of a given observation. In that sense, we interpret resulting deviations of SOEP incomes from the true signal as reporting errors in the survey data.

Note that we impose the additive measurement error model not on nominal earnings but their log-transformed counterparts. In many empirical studies, this appears to be the purported variable of interest as opposed to untransformed nominal incomes. Exploiting the properties of the logarithmic function, the resulting measurement error variable can be interpreted as the deviation of nominal reported income relative to its administrative counterpart in percentage terms. We provide two additional notions of reporting error to be used in the graphical analysis, i.e.\ a \textit{nominal} measurement error $Y_{it} - Y_{it}^{*}$ (in units of \euro) and a \textit{relative} measurement error $\frac{Y_{it} - Y_{it}^{*}}{Y_{it}^{*}}$. The latter allows for a direct interpretation of the measurement error in terms of shares of the true signal, which is only approximated by the baseline measurement error definition using the logarithm.

\section{Empirical analysis}\label{sec:analysis}

In the following, the results of the empirical analysis are presented, which are largely based on \cite{schmillenetal2024}. These include Mincer-type regressions of both available earnings measures as well as regressions capturing the relationship between measurement error and observable characteristics in the pooled sample. Moreover, the distribution of measurement errors in SOEP is characterized by providing summary statistics and empirical density functions. For the subsample of consecutive employment, we estimate pairwise correlation coefficients and covariances as well as the variance-covariance matrix for a time horizon of $T=4$, which are then used for the calculation of reliability ratios assuming a classical error structure. We rely on OLS estimation regressing administrative income data on survey income data to obtain reliability ratios for the non-classical case. Robustness of the regression-based reliability ratios to the applied data processing steps is assessed by considering varying subsamples of SOEP-CMI-ADIAB.

\subsection{Measurement error distribution in SOEP}\label{sec:ME_distribution}

In order to get a first idea by how much the income distributions of $Y$ and $Y^*$ differ, histograms based on the unweighted pooled observations across years by income measure are plotted in Figure \ref{fig:SOEP_vs_IEB}.\footnote{Graphical year-by-year comparisons yielded very similar results.} Firstly, both variables are observed over similar supports, which is the result of top-coding in particular. Compared to the survey data, the empirical density distribution of IEB income is shifted towards the right in a uniform manner, which is reflected in the empirical mean of the respective distribution. This suggests underreporting of gross monthly labor income across the entire income distribution. In particular, the fact that individuals at the lower end of the income distribution underreport their income is not consistent with the hypothesis of mean-reversion, which would have implied a shift of density mass towards the mean or median for the IEB earnings distribution relative to the survey data distribution. In terms of the cumulative density functions (cdf's), the presence of mean reversion would imply the two cdf's to intersect at some point of the support. As can be seen from the right panel of Figure \ref{fig:SOEP_vs_IEB}, this is not the case as the empirical cdf of survey income dominates the cdf of the administrative income variable. 

\begin{figure}[H]
    \centering
    \caption{Gross monthly earnings in survey and administrative data}
    \includegraphics[scale=0.55]{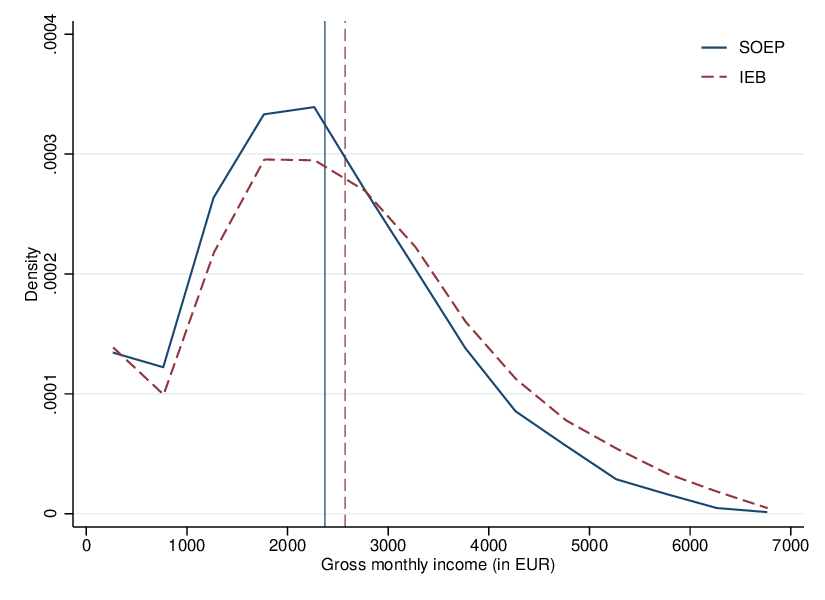}
    \includegraphics[scale=0.55]{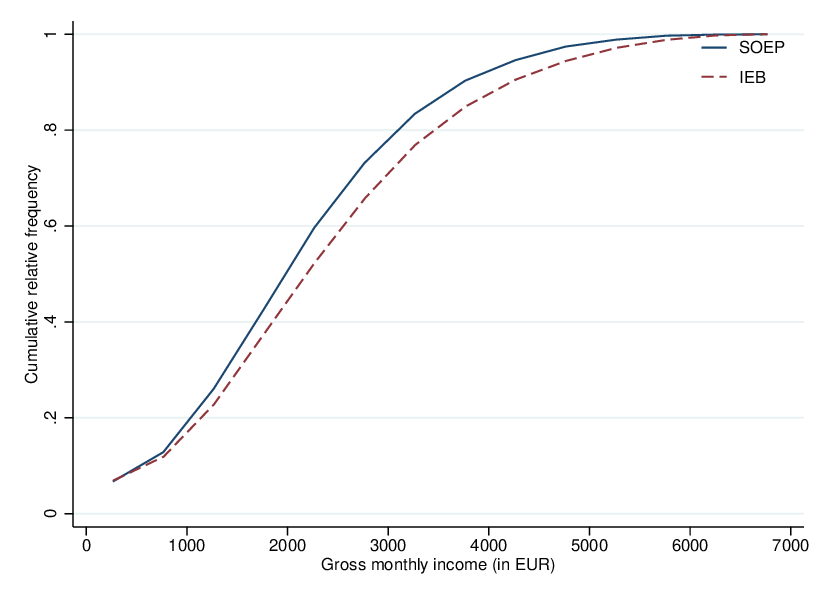}    
    \label{fig:SOEP_vs_IEB}
    \begin{minipage}{\textwidth}
        \begin{singlespace}
            \footnotesize{Figure \ref{fig:SOEP_vs_IEB} plots the unweighted empirical (right panel: cumulative) density functions of survey (SOEP) and administrative (IEB) income data, respectively, estimated by histograms with $14$ bins of size \euro$500$ in between \euro$15$ and \euro$7100$ pooling all observations across years. The vertical lines denote the empirical mean of the respective variable from SOEP (\euro$2'371.43$) and from IEB (\euro$2'572.66$). Data: SOEP-CMI-ADIAB (main sample).}
        \end{singlespace}
    \end{minipage}
\end{figure}

This reasoning, however, may be misleading since we do not compare the reported and the true income level for each observation directly but look at differences across the overall, i.e.\ aggregate, distributions. This is improved upon in Figure \ref{fig:distrib_MElog}, which now sketches the distribution of the measurement error $u_{it}$ for the main sample (upper panel) and by gender (lower panel). Most prominently, all three plots lie to the left of the vertical rule at $u_{it} = 0$, which suggests a non-zero estimated mean of $\bar{u} = \frac{1}{\textrm{no.\ of }(i,t)\textrm{ obs.}} \sum_{(i,t)} u_{it}= -0.07$. Translating this into a more interpretable number: the average measurement error corresponds to an average underreporting of monthly earnings by $7\%$ relative to the true signal, which is a common way to interpret log-transformed ratios.\footnote{Here, the term \textit{average} refers to the geometric mean. Let $\tilde{N}$ denote the number of observations in the main (unbalanced) sample. Then $\bar{u} = log\left(\underset{(i,t)}{\Pi}\frac{Y_{it}}{Y_{it}^*}\right)^{\frac{1}{\tilde{N}}} = -0.07 \implies e^{\bar{u}} = \left(\underset{(i,t)}{\Pi} \frac{Y_{it}}{Y_{it}^*}\right)^{\frac{1}{\tilde{N}}} = 0.93$, i.e., the relative size of reported incomes is equal to $93\%$ of the true signal, on average.} Since the standard deviation provided for each graph, respectively, is only slightly larger than the interquartile range ($0.15$), for at least $73\%$ of the underlying observations the reported income is smaller than the administrative income, which results in negative $u_{it}$. Hence, although there is evidence for widespread underreporting, roughly a quarter of observations do exhibit overreporting to a varying degree. For comparison, the range of $u_{it}$ in Figure \ref{fig:distrib_MElog} is similar to that of \cite{schmillenetal2024}. The measurement error distribution has more mass around its center than the normal distribution while having slightly larger tails at the same time. However, unlike those authors, we do not make out substantial differences across gender.

\begin{figure}[H]
    \centering
    \caption{Measurement error distribution}
    \includegraphics[scale=0.9]{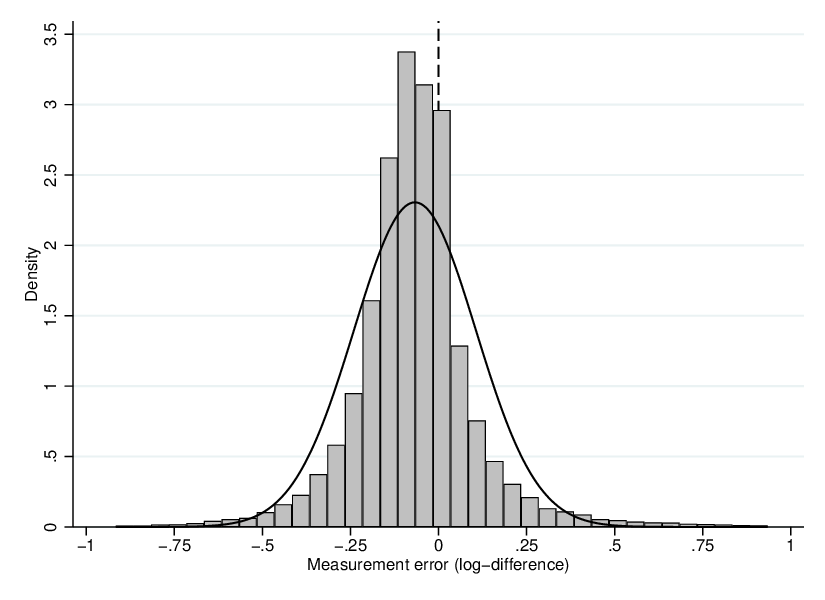}
    \includegraphics[scale=0.55]{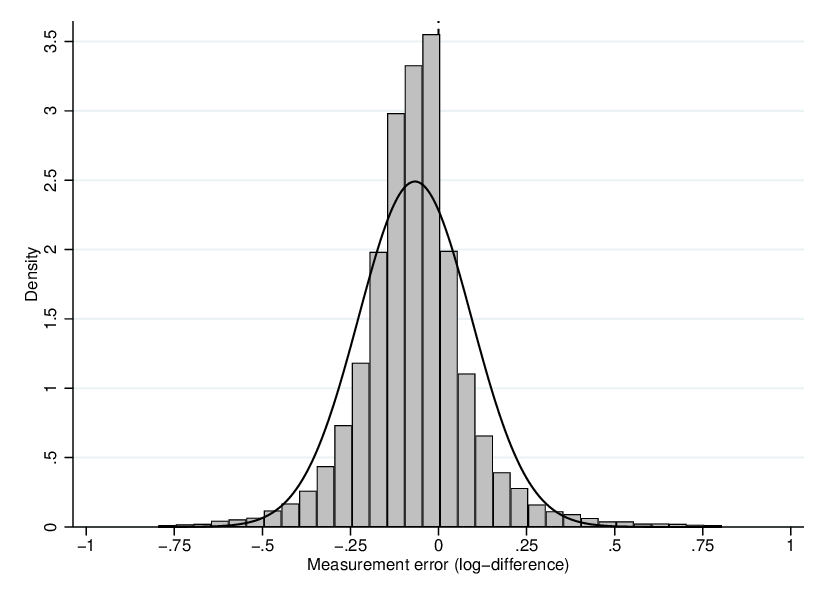}
    \includegraphics[scale=0.55]{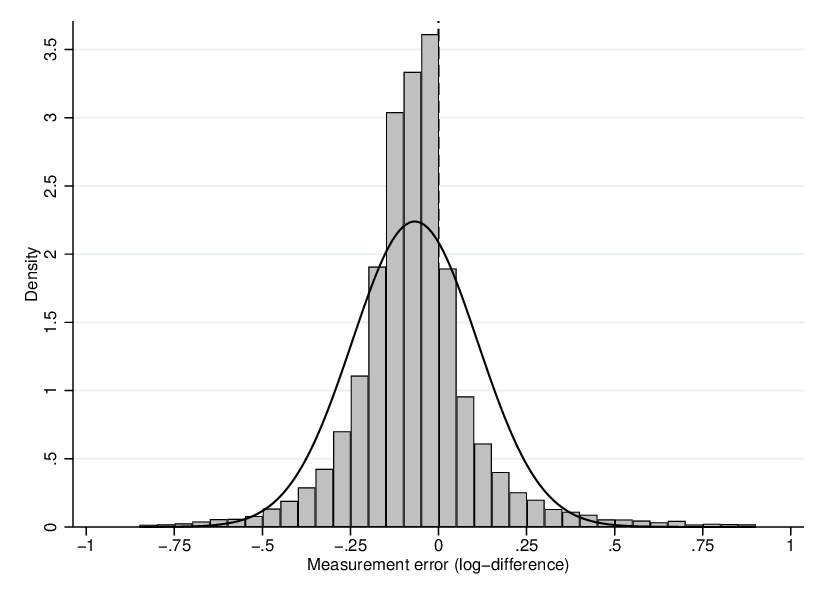}
    \label{fig:distrib_MElog}
    \begin{minipage}{\textwidth}
        \begin{singlespace}
            \footnotesize{Figure \ref{fig:distrib_MElog} presents the empirical distributions of measurement error using histograms for the main subsample (upper panel) as well as separately by gender (lower panels: men on the left, women on the right) with a bin width of $0.05$. A total of $0$ ($76$, $44$) observations in the tails of the income distribution of the main sample (men, women) were omitted due to data privacy concerns. The normal distribution with mean and standard deviation corresponding to the underlying subsample is displayed by the continuous bell-shaped curve and the vertical dashed line goes through the origin. The measurement error is defined as the differences of log-transformed gross monthly survey earnings and log-transformed gross monthly administrative earnings. Mean: $-0.07$, $-0.07$, $-0.07$; standard deviation: $0.17$, $0.16$, $0.18$ (counter-clockwise, starting at the upper panel). Data: SOEP-CMI-ADIAB (main sample).}
        \end{singlespace}
    \end{minipage}
\end{figure}

Figure \ref{fig:distrib_MEraw} repeats the same exercise for nominal measurement error. The above remarks with respect to the mean (negative) and the share of observations with negative reporting errors ($73\%$) apply in this context as well. However, the effect of applying a logarithmic transformation on variables with a right-skewed distribution, as is the case for earnings, becomes apparent as the distribution of nominal measurement error is no longer symmetric but exhibits negative skewness. The measurement error distribution of women has a smaller support, which is corroborated by a smaller estimated variance. This hints at the fact that misreporting among women may not be as pronounced, also indicated by the smaller mean of the distribution. \cite{caliendoetal2024} provide a similar figure for data up to 2019, which appears to be comparable to ours in shape and range with the exception of some more mass in the tails in their graph (c.f.\ Figure \textit{1b} therein). Importantly, this similarity also suggests that allowing for multiple jobs when calculating gross monthly earnings may not apply to many observations in the final sample and, thus, does not introduce or reduce noise attributed to reporting error.

\begin{figure}[H]
    \centering
    \caption{Distribution of nominal measurement error}
    \includegraphics[scale=0.9]{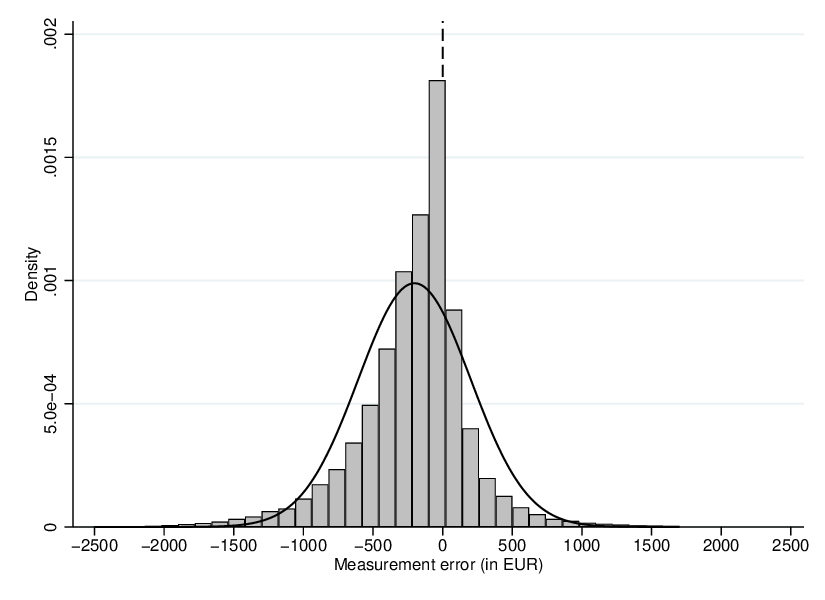}
    \includegraphics[scale=0.55]{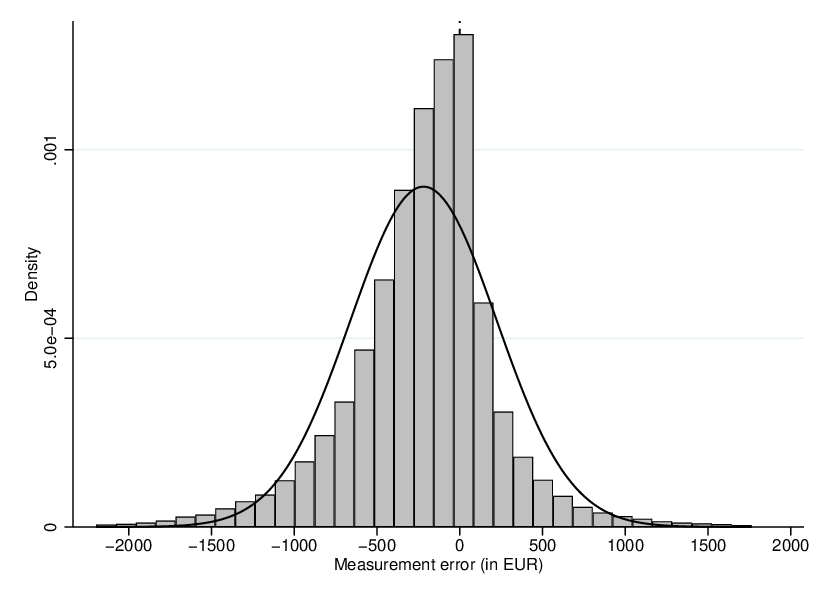}
    \includegraphics[scale=0.55]{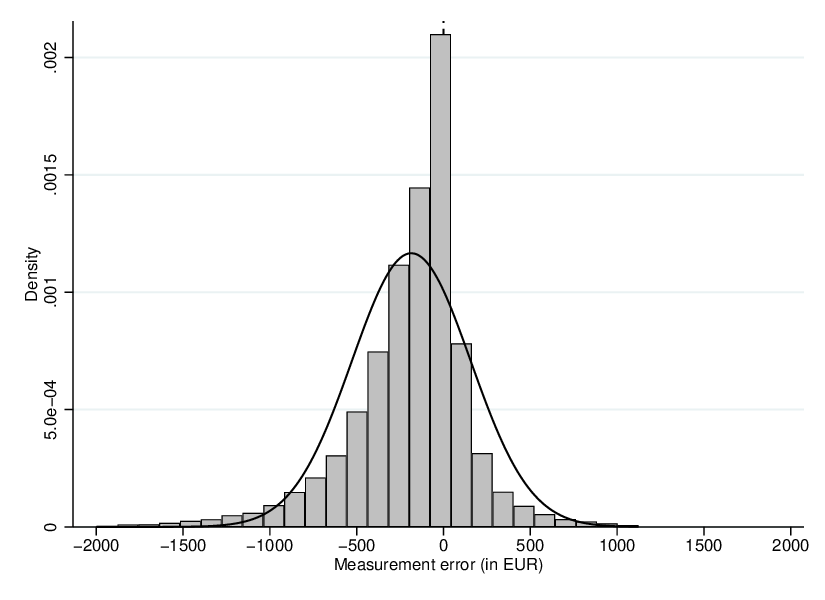}
    \label{fig:distrib_MEraw}
    \begin{minipage}{\textwidth}
        \begin{singlespace}
            \footnotesize{Figure \ref{fig:distrib_MEraw} presents the empirical distributions of the raw difference between survey and administrative data using histograms for the main subsample (upper panel) as well as separately by gender (lower panels, men on the left) with a bin width of \euro$120$. $142$ ($126$, $189$) observations in the tails of the income distribution of the main sample (men, women) were omitted due to data privacy concerns. The normal distribution with mean and standard deviation corresponding to the underlying sample is displayed by the continuous bell-shaped curve and the vertical dashed line goes through the origin. Mean: $-201.23$, $-220.09$, $-183.36$; standard deviation: $416.38$, $463.12$, $365.78$ (counter-clockwise, starting at the upper panel). Data: SOEP-CMI-ADIAB (main sample).}
        \end{singlespace}
    \end{minipage}
\end{figure}

To address the issue of mean reversion in survey income data, Figure \ref{fig:ME_by_wagequant} plots the measurement error by earnings quantile in a given survey year, with nominal measurement error on the left and relative measurement error on the right. In nominal terms, the median, mean, and quartiles are decreasing in the earnings quantile, which suggests a negative correlation between the measurement error and the income level. On average, all income quantiles, but the first two, seem to understate their their income. Moreover, the nominal measurement error is heteroskedastic with respect to the income level, exhibiting a monotonically increasing variance across earnings quantiles. Normalizing this by the true income level allows for adjustment by this factor and yields (relative) measurement errors, for which the variation across earnings quantiles is similar, judging by the constant interquartile ranges for earnings quantiles $3$ to $20$. The lowest two income groups are an exception with a larger variation in the size of relative measurement error. Moreover, the plot suggests that the average individual in the lowest two income quantiles overstates her true gross monthly income level by approx.\ $10\%$. The remaining groups tend to underreport on average, with the severity slowly increasing in the income level. Again, for reference, \cite{caliendoetal2024} do not appear to account for the position in the income distribution of a given year but construct quantiles based on the pooled dataset. However, we argue that the position of an individual with respect to his or her income in the earnings distribution in a given survey year is relevant when intending to evaluate the hypothesis of mean reversion. 

\begin{figure}[H]
    \centering
    \caption{Measurement error distribution by earnings quantile}
    \includegraphics[scale=0.55]{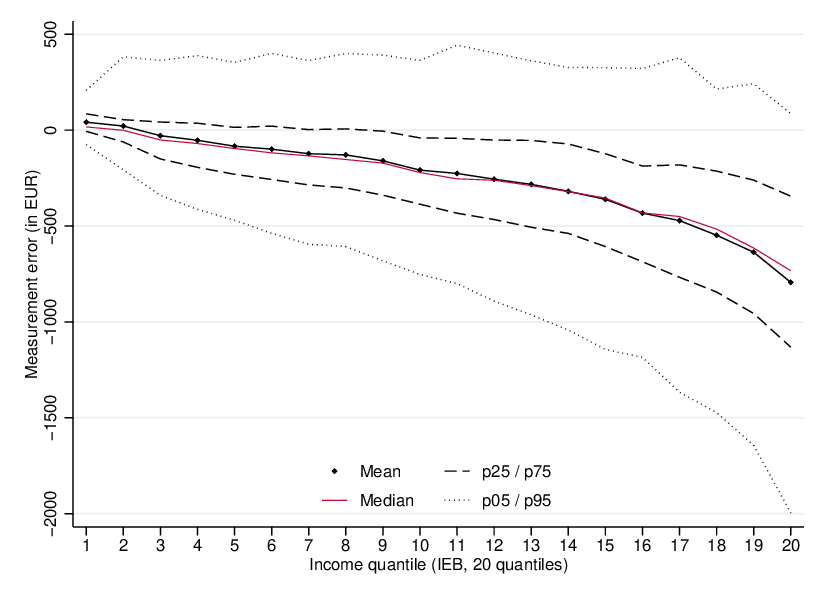}
    \includegraphics[scale=0.55]{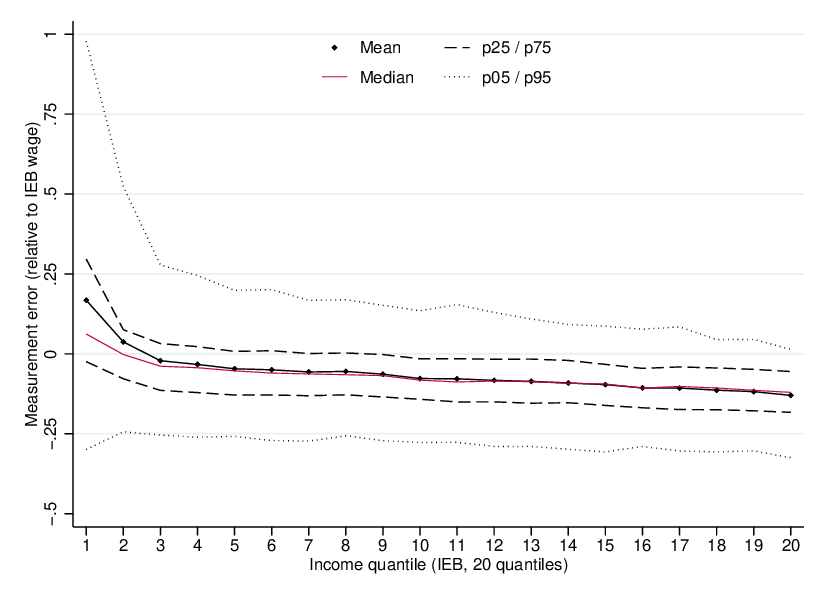}
    \label{fig:ME_by_wagequant}
    \begin{minipage}{\textwidth}
        \begin{singlespace}
            \footnotesize{Figure \ref{fig:ME_by_wagequant} displays key statistics (mean, median, $5^{th}$ percentile, $95^{th}$ percentile, quartiles) of the 20 quantile-specific measurement error distributions for (i) nominal differences of survey and administrative gross monthly incomes (left panel), and (ii) the nominal difference relative to the administrative income level (right panel) for a given $(i,t)$ observation. The classification of an observation to a given quantile is based on the IEB earnings distribution in a given year. Data: SOEP-CMI-ADIAB (main sample).}
        \end{singlespace}
    \end{minipage}
\end{figure}

\subsection{$Cov(u,X)$ and Mincer-type regressions}\label{sec:mincer}

In a first step, we explore possible correlations of time-invariant observable characteristics $X$ with the measurement error $u$. This is important as measurement error covarying with other covariates included in a regression causes inconsistent estimates in the linear model, as discussed in Section \ref{sec:ME_linear_model}. Using the pooled sample and OLS, we estimate a model for $u_{it}$, which is linear in the covariates. The results are presented in the first three columns of Table \ref{tab:mincer_corr}.

\begin{sidewaystable}
    \centering
    \begin{threeparttable}
        \caption{Correlation with observables and Mincer-type regressions}
        \label{tab:mincer_corr}
        \begin{tabular}{lccccccccc}
            \toprule 
            \toprule
            Dependent variable & \multicolumn{3}{c}{$u_{it}$} & \multicolumn{3}{c}{$log(Y_{it})$} & \multicolumn{3}{c}{$log(Y_{it}^*)$} \\
            & \small{All} & \small{Men} & \small{Women} & \small{All} & \small{Men} & \small{Women} & \small{All} & \small{Men} & \small{Women} \\
            & \footnotesize{(1)} & \footnotesize{(2)} & \footnotesize{(3)} & \footnotesize{(4)} & \footnotesize{(5)} & \footnotesize{(6)} & \footnotesize{(7)} & \footnotesize{(8)} & \footnotesize{(9)} \\[0.3ex]
            \hline \\[-1.8ex]
Female      &     -0.0126&    --        &    --        &     -0.3149&     --       &      --      &     -0.3023&      --      &       --     \\
            &\footnotesize(0.0012)&            &            &\footnotesize(0.0030)&            &            &\footnotesize(0.0030)&            &            \\
Single      &      0.0064&      0.0078&      0.0053&      0.0305&     -0.0784&      0.1653&      0.0241&     -0.0862&      0.1600\\
            &\footnotesize(0.0016)&\footnotesize(0.0022)&\footnotesize(0.0023)&\footnotesize(0.0037)&\footnotesize(0.0043)&\footnotesize(0.0059)&\footnotesize(0.0037)&\footnotesize(0.0042)&\footnotesize(0.0059)\\
German      &     -0.0196&     -0.0276&     -0.0083&      0.0025&      0.0275&     -0.0373&      0.0221&      0.0551&     -0.0289\\
            &\footnotesize(0.0028)&\footnotesize(0.0038)&\footnotesize(0.0041)&\footnotesize(0.0061)&\footnotesize(0.0069)&\footnotesize(0.0105)&\footnotesize(0.0061)&\footnotesize(0.0067)&\footnotesize(0.0105)\\
Migration background&      0.0126&      0.0119&      0.0135&      0.0277&     -0.0122&      0.0557&      0.0152&     -0.0241&      0.0423\\
            &\footnotesize(0.0021)&\footnotesize(0.0029)&\footnotesize(0.0029)&\footnotesize(0.0048)&\footnotesize(0.0054)&\footnotesize(0.0077)&\footnotesize(0.0049)&\footnotesize(0.0052)&\footnotesize(0.0079)\\
East        &     -0.0055&     -0.0108&     -0.0006&     -0.0537&     -0.1435&      0.0368&     -0.0482&     -0.1328&      0.0375\\
            &\footnotesize(0.0015)&\footnotesize(0.0022)&\footnotesize(0.0021)&\footnotesize(0.0045)&\footnotesize(0.0047)&\footnotesize(0.0066)&\footnotesize(0.0047)&\footnotesize(0.0047)&\footnotesize(0.0070)\\
Abitur      &      0.0035&      0.0028&      0.0050&     -0.1502&     -0.1553&     -0.1500&     -0.1536&     -0.1580&     -0.1551\\
            &\footnotesize(0.0025)&\footnotesize(0.0036)&\footnotesize(0.0035)&\footnotesize(0.0065)&\footnotesize(0.0084)&\footnotesize(0.0095)&\footnotesize(0.0065)&\footnotesize(0.0084)&\footnotesize(0.0095)\\
Education (years)&      0.0013&      0.0025&      0.0002&      0.0433&      0.0319&      0.0565&      0.0420&      0.0293&      0.0563\\
            &\footnotesize(0.0005)&\footnotesize(0.0006)&\footnotesize(0.0007)&\footnotesize(0.0013)&\footnotesize(0.0014)&\footnotesize(0.0021)&\footnotesize(0.0013)&\footnotesize(0.0014)&\footnotesize(0.0022)\\
Age (years) &      0.0002&      0.0001&      0.0003&      0.0045&      0.0035&      0.0060&      0.0043&      0.0034&      0.0057\\
            &\footnotesize(0.0001)&\footnotesize(0.0001)&\footnotesize(0.0001)&\footnotesize(0.0002)&\footnotesize(0.0002)&\footnotesize(0.0003)&\footnotesize(0.0002)&\footnotesize(0.0002)&\footnotesize(0.0003)\\
Marginal employment&      0.1467&      0.1298&      0.1524&     -1.6585&     -1.9218&     -1.5327&     -1.8052&     -2.0516&     -1.6851\\
            &\footnotesize(0.0049)&\footnotesize(0.0099)&\footnotesize(0.0056)&\footnotesize(0.0096)&\footnotesize(0.0186)&\footnotesize(0.0110)&\footnotesize(0.0091)&\footnotesize(0.0169)&\footnotesize(0.0105)\\
Complex task&      0.0006&      0.0004&      0.0006&      0.1568&      0.1497&      0.1517&      0.1563&      0.1492&      0.1511\\
            &\footnotesize(0.0015)&\footnotesize(0.0021)&\footnotesize(0.0021)&\footnotesize(0.0040)&\footnotesize(0.0045)&\footnotesize(0.0059)&\footnotesize(0.0041)&\footnotesize(0.0045)&\footnotesize(0.0060)\\
Notification: \textit{Annual report}&     -0.0192&     -0.0177&     -0.0209&     -0.0042&      0.0050&     -0.0117&      0.0150&      0.0226&      0.0092\\
            &\footnotesize(0.0020)&\footnotesize(0.0028)&\footnotesize(0.0029)&\footnotesize(0.0048)&\footnotesize(0.0058)&\footnotesize(0.0072)&\footnotesize(0.0048)&\footnotesize(0.0057)&\footnotesize(0.0073)\\
Notification: \textit{Employment termination}&      0.0091&      0.0085&      0.0095&     -0.1014&     -0.0909&     -0.1119&     -0.1106&     -0.0994&     -0.1215\\
            &\footnotesize(0.0030)&\footnotesize(0.0041)&\footnotesize(0.0043)&\footnotesize(0.0074)&\footnotesize(0.0086)&\footnotesize(0.0114)&\footnotesize(0.0076)&\footnotesize(0.0086)&\footnotesize(0.0116)\\
Multiple empl.\ spells&      0.0163&      0.0089&      0.0221&     -0.0521&     -0.0166&     -0.0766&     -0.0685&     -0.0255&     -0.0986\\
            &\footnotesize(0.0026)&\footnotesize(0.0037)&\footnotesize(0.0037)&\footnotesize(0.0055)&\footnotesize(0.0068)&\footnotesize(0.0080)&\footnotesize(0.0055)&\footnotesize(0.0066)&\footnotesize(0.0081)\\
No.\ of coworkers&     -0.0000&     -0.0000&     -0.0000&      0.0000&     -0.0000&      0.0000&      0.0000&     -0.0000&      0.0000\\
            &\footnotesize(0.0000)&\footnotesize(0.0000)&\footnotesize(0.0000)&\footnotesize(0.0000)&\footnotesize(0.0000)&\footnotesize(0.0000)&\footnotesize(0.0000)&\footnotesize(0.0000)&\footnotesize(0.0000)\\
Avg.\ wage of coworkers&     -0.0006&     -0.0007&     -0.0006&      0.0048&      0.0048&      0.0046&      0.0054&      0.0055&      0.0052\\
            &\footnotesize(0.0000)&\footnotesize(0.0000)&\footnotesize(0.0000)&\footnotesize(0.0001)&\footnotesize(0.0001)&\footnotesize(0.0002)&\footnotesize(0.0001)&\footnotesize(0.0001)&\footnotesize(0.0003)\\
Constant    &      0.0008&      0.0028&     -0.0179&      6.6297&      6.8426&      6.0795&      6.6289&      6.8399&      6.0974\\
            &\footnotesize(0.0067)&\footnotesize(0.0090)&\footnotesize(0.0100)&\footnotesize(0.0163)&\footnotesize(0.0194)&\footnotesize(0.0260)&\footnotesize(0.0163)&\footnotesize(0.0193)&\footnotesize(0.0258)\\[0.3ex]
            \hline \\[-1.8ex]
F-statistic           &      188.14&      111.74&      110.07 & -- & --& --& --& --& -- \\
Time-FE        &        No&        No&        No&        No&        No&        No&        No&        No&        No\\
$R^2$          &        0.08&        0.07&        0.09&        0.63&        0.64&        0.59&        0.67&        0.69&        0.63\\
No.\ of obs.\          &       70'567&       34'996&       35'571&       70'567&       34'996&       35'571&       70'567&       34'996&       35'571\\[0.3ex]
            \bottomrule
            \bottomrule
        \end{tabular}  
        \begin{tablenotes}
            \item \footnotesize{Table \ref{tab:mincer_corr} presents the regression results of a linear model, in which the outcome variable is regressed on the covariates indicated and a constant. C.f.\ Table \ref{tab:mincer_corr_FE} for results of a similar model including survey year indicators (time-FE). Robust standard errors in parentheses; $R^2$ is unadjusted; F-statistic is based on the null hypothesis $H_0: \beta = 0$ (excl.\ the constant). Notation: No.\ -- number; empl.\ -- employment; avg.\ -- average. Data: SOEP-CMI-ADIAB (main sample).}
        \end{tablenotes}
    \end{threeparttable}
\end{sidewaystable}

The majority of estimated coefficients in columns $(1)$ through $(3)$ are (individually) statistically significant at conventional significance levels but mostly small in magnitude relative to an average measurement error of $-0.07$ in the overall sample (c.f.\ Figure \ref{fig:distrib_MElog}). An exception to this are \textit{Age}, \textit{Abitur}, and \textit{Complex task}, which do not appear to be related to measurement error. Judging from the $R^2$ statistics, the proposed linear model is able to capture $8\%$ of the variation in the measurement error. Ceteris paribus (c.p.), the largest association arises for marginal employment spells, which exhibit an increase by $15$ pp. Holding a German passport comes in a distant second, which involves an average decrease in the ratio of survey to administrative incomes by $1.96$ pp. A similar observation can be made with respect to the education level, where another year of education is associated with a $0.13$ pp.\ decrease of SOEP relative to IEB incomes. Being single, having a history of migration, residing in Western Germany, or working multiple jobs at once: all these characteristics are on average associated with larger measurement error and, thus -- since the average measurement error is negative --, with slightly smaller deviations of survey to administrative data, ranging from increases of $0.13$ pp.\ (\textit{Education}) to $1.63$ pp.\ (\textit{Multiple empl.\ spells}).

Running these regressions for each sex separately reveals substantial heterogeneity, hinting at a more complex relationship of measurement error and time-invariant observable characteristics. Despite this heterogeneity, the signs of the individual coefficients coincide across gender. However, whereas measurement error of men does not correlate with age, the correlation for women is weak but statistically significant. Moreover, the covariation with respect to East German residency (negative) and years of education completed (positive) appear to be driven by men while the corresponding coefficients based on female observations only are not statistically significant.

In a second step, we want to provide more direct evidence for the effects of measurement error in the linear model. The results in columns $(1)$ to $(3)$ give rise to doubts regarding the assumption of a non-differential measurement error, which does not allow for a dependence with covariates. This can be interpreted as evidence for non-classical measurement error as well. As discussed above, only non-classical measurement error in the dependent variable to be modeled leads to biased estimates. Thus, if measurement error is classical, estimated coefficients should largely coincide. For that purpose, we run Mincer-type regressions with reported earnings (columns $(4)$ -- $(6)$ in Table \ref{tab:mincer_corr}) and administrative earnings (columns $(7)$ -- $(9)$) as dependent variables and compare estimates across regressions. Although the focus is not so much on the individual estimates, i.e.\ the association of the respective characteristics with the outcome, we briefly interpret the sign and magnitude of the estimated coefficients in column $(7)$. It appears useful to see if estimates have the expected sign. Moreover, estimates across the two specifications differing in the dependent variable's data source feature agreeing signs. Hence, the interpretation of columns $(7)$-$(9)$ carries over to the results in columns $(4)$-$(6)$.

Some common themes from studies aiming to explain the income level are corroborated in the main sample. Firstly, the estimates imply a payoff from more education. On the margin, an additional year of education is associated with a $4.2\%$ earning's increase. Moreover, the gender pay gap disadvantaging women is sizeable with $30.2\%$. Respondents in East Germany make $4.8\%$ less on average based on administrative income data. Although small in magnitude, an additional year in age, which arguably corresponds to more experience on average, is associated with larger labor earnings. The estimated coefficient associated with \textit{Complex task} suggest a sort of rent-sharing between the employer and the employee, in the sense that part of the value-added accrues to the employee. Since the classification of \textit{Marginal employment} is directly tied to a low income level, the negative coefficient is not surprising. Similarly, multiple employment spells are more frequently encountered among individuals with low earnings, which finds its expression in an estimate of negative sign. A prediction that is at odds with common beliefs about determinants of labor market outcomes is with respect to the quality of secondary education. The estimate implies that \textit{Abitur} degree holders make $15.4\%$ less on average in our sample. We hypothesize that this is due to the sampling restriction associated with the upper assessment limit. Since we omit roughly $10\%$ of top earners, who are more likely to have an \textit{Abitur} degree, we may end up with a selection of “underperformers” of that group in our main sample. This in turn may explain the negative correlation. The sign and magnitude of estimates associated with the remaining variables is harder to rationalize. C.p., singles, German nationals, and individuals with migration background have close to $2\%$ higher earnings on average. 

There is substantial heterogeneity by gender. Women appear to be “hurt” from being in a relationship in terms of income ($-16\%$), which can be attributed to the unequal allocation of childcare related tasks within a household. The opposite appears to be the case for men, although the estimate is only half of the magnitude of the one in the females-only sample. Moreover, the average increase in income associated with being a German national is largely driven by men (with the female point estimate being negative), whereas higher incomes for women with migration background drive the corresponding estimate in the overall sample. Only males' incomes are negatively correlated with East German residency ($-13.3\%$), while the same estimate is positive for women ($3.8\%$). Moreover, the positive association between the notification type \textit{Annual report} and income in the aggregate is largely driven by male observations. Relative to measurement error as the dependent variable, the linear model based on time-invariant covariates is able to explain $63\%$ to $69\%$ of the variation in log-earnings. However, this appears to work better in explaining male outcomes ($R^2 = 24\%$) than female outcomes ($R^2 = 17\%$).

Turning to survey data, the overall direction of the correlations between observable characteristics and earnings mostly persists. The change in the estimated coefficients compared to the baseline specification with administrative income as the dependent variable is proportional to the sign and magnitude of the estimated parameters in columns $(1)$ through $(3)$: If the relation between a covariate and the measurement error is negative, this results in a downward bias in columns $(4)$, $(5)$, and $(6)$, respectively (and vice versa). Moreover, the larger the association of measurement error with a covariate, the larger the bias in absolute terms. This also implies that the coefficients are not merely attenuated, i.e.\ drawn towards zero, but that the bias can go into either direction.

Estimates based on reported income as the dependent variable imply a slightly larger gender gap and larger absolute effects from being single, having a migration background, or living in East Germany. Most remaining estimates are virtually unchanged or shrunk towards zero, where the most sizeable drop (in terms of absolute deviations from the baseline coefficient) of more than $12\%$ is associated with returns from labor from being a German national. This change appears to be driven by male respondents, for which the downward bias due to falsely reported earnings amounts to $49\%$. The attenuation bias finds its extreme in the case of \textit{German} and \textit{Annual report}, yielding estimates not significantly different from zero. Regarding regional differences, there appears to be heterogeneity across genders in terms of bias due to mismeasured income data. Namely, survey data implies a larger average earnings difference of East German males while a decrease in the positive regional gap for women relative to the estimates based on administrative data is apparent. Moreover, the share of the variance explained by the linear model is slightly smaller for the specification using survey income data, which could be attributed to the non-differential noise introduced by measurement error.

We can also test the null hypothesis that the estimated coefficients from the two regressions differing only in the chosen dependent variable coincide (misreported survey income vs.\ true signal). As already discussed, this would be the case under the classical measurement error assumption. More formally, let $\hat{\beta}_{OLS}^{Y^*X}$ denote the OLS estimator of a regression using the administrative income variable as the regressand and $\hat{\beta}_{OLS}^{YX}$ its counterpart based on survey income as the dependent variable. Thus, the hypothesis to be tested can be written as $H_0: \hat{\beta}_{OLS}^{Y^*X}  = \hat{\beta}_{OLS}^{YX}$. Together with the linear measurement error model imposed, it turns out that this is equivalent to $H_0: \hat{\beta}_{OLS}^{uX}  = 0$ and, thus, amounts to testing hypothesis that the estimated OLS coefficients of a regression of measurement error $u$ on selected covariates $X$ are jointly different from zero (excl.\ the intercept). The relatively large F-statistics ($>100$) provided in table \ref{tab:mincer_corr} allow us to reject $H_0$. This suggests that measurement error in the SOEP income data is of non-classical nature.

Since the evolution of incomes is usually upward sloping over time and none of the covariates is able to account for these kinds of dynamics, Table \ref{tab:mincer_corr_FE} presents the results of regressions, which include indicators for each survey year from 1984 to 2021. Moreover, we may be worried about survey year-specific differences in reporting behavior and, thus, measurement error. Although the magnitude of the estimated coefficients in the Mincer-type income models change substantially, the differences across specifications when using the reported income $log(Y_{it})$ or the true signal $log(Y_{it}^*)$ as dependent variables, respectively, are minor. Hence, these results allow for a similar interpretation as of those in Table \ref{tab:mincer_corr}, namely that the bias from reporting error is rather limited. The association between the indicator of being single and measurement error $u_{it}$ is now negative, which results in a downward bias of the corresponding coefficient in the income regression. A similar observation holds for the variable \textit{Age}. The peculiarities of a positive association of the migration background indicator with earnings and a negative payoff from having an \textit{Abitur} degree are now even more pronounced. Interestingly, the $R^2$ statistics of Mincer-type regressions are quite large -- one could even say “too large” --, since the numbers imply that the model is able to predict log-incomes almost perfectly. This observation is robust to solely considering observations from 2010 to 2021, implying that a large share of the variance of log-incomes in our main sample can be described by time-fixed effects. Although to a lesser extent, this holds for the measurement error as well, although here the explanatory power of the model is limited to a maximum of $25\%$.

\subsection{Reliability ratios}\label{sec:reliab_ratio}

The reliability ratio is a measure for the degree of attenuation bias introduced by measurement error in independent variables in the linear model. Its interpretation is very similar to the $R^2$ statistic frequently presented in the context of OLS estimation in that it indicates the share of variation in the mismeasured variable that can be attributed to variation of the true signal. Although this measure arguably has its conceptual origin in the context of classical measurement error, it has been extended by previous validation studies to cover the non-classical case, which allows for correlation of the true signal with the measurement error. In either case, the basic components underlying the reliability ratio are variances and covariances of the true signal and the measurement error. These can be estimated from the data.

By itself, the reliability ratio was not envisioned for a panel setting. That is, this statistic usually makes a statement about a cross-section. With the panel at hand, we are able to calculate reliability ratios for each time period $t$ and to compare the evolution of reliability ratios across time. To estimate the average dynamics of income variables and of measurement error with the aim of calculating reliability ratios, we require the existence of a balanced panel such that individuals' outcomes can be observed in consecutive years. Previous studies have been restricted to a short panel of pre-determined length, i.e.\ $t \in \{1, \cdots, T\}$ for some $T \in \mathbb{N}_{\geq 1}$ (\citeauthor{boundkrueger1991}, \citeyear{boundkrueger1991}; \citeauthor{boundetal1994}, \citeyear{boundetal1994}; \citeauthor{pischke1995}, \citeyear{pischke1995}; \citeauthor{schmillenetal2024}, \citeyear{schmillenetal2024}). This has the advantage that the initial period is also pre-determined. For SOEP-CMI-ADIAB, the record linkage process can be described as “backdating”: a positive consent decision of an individual allows for linkage of all prior observations in the survey data. This requires the researcher to determine an initial period $t=1$. 

We proceed as follows: The first observation for each cross-sectional unit $i$ refers to the first observation with active employment and, thus, strictly positive labor income. The resulting variable captures variation in the very first reported income level across units. In order to consider the resulting sample as identically distributed, I assume that the income and measurement error distributions to not vary over survey years but with years spent as a participant in the survey. For $T>1$, I do not allow for gaps in the employment history, i.e., solely individuals with consecutive employment for the given time-horizon are considered, starting from $t=1$. 

In that sense, the variation in a thereby constructed variable over time captures variation due to increasing seniority in SOEP. The resulting sample is weakly balanced in the sense that cross-sections in $t$ are affected by irreversible attrition from the cross-section in $t-1$. The adopted approach has the benefit that it is flexible with respect to the time horizon of interest. Namely, extending the time horizon $T$ does not alter the constitution of past cross-sections dating back to the initial period. Moreover, the weakly balanced nature allows me to exploit a larger number of observations for small $t$. For robustness, I repeat some analyses using a strongly balanced sample for consecutively employed individuals given the specific time horizon $T=4$. I discuss alternative strategies closer to previous work and issues associated with these in Section \ref{sec:discussion}.

A key ingredient for calculating reliability ratios are estimated population covariances and variances. Further, a characterization of dynamics of measurement error as well as its correlation with the true signal provides useful evidence to assess the assumption of classical measurement error in SOEP. In Table \ref{tab:varcov}, we present estimated variance terms as well as pairwise covariation and corresponding correlation coefficients of the true signal and measurement error for a time horizon of $T=4$. Since estimated variance-covariance matrices are usually based on the same number of observations in each estimated cell, I provide similar statistics based on the strongly balanced panel with consecutive employment for four periods in Table \ref{tab:varcov_balanced}.

\begin{table}[!htbp] 
    \centering
    \begin{threeparttable}
        \caption{Pairwise variances, covariances, and correlations}
        \label{tab:varcov}
        \begin{tabular}{llcccccccc}
            \toprule 
            \toprule
            \multicolumn{2}{l}{All} & \scriptsize{[13887]} & \scriptsize{[8867]} & \scriptsize{[6583]} & \scriptsize{[5118]} & \scriptsize{[13887]} & \scriptsize{[8867]} & \scriptsize{[6583]} & \scriptsize{[5118]} \\
            & & $log(Y^*_{1})$ & $log(Y^*_{2})$ & $log(Y^*_{3})$ & $log(Y^*_{4})$ & $u_{1}$ & $u_{2}$ & $u_{3}$ & $u_{4}$ \\[0.3ex]
            \hline \\[-1.8ex]
\multicolumn{2}{l}{$ log(Y_{1}^*) $}     &       0.744&       0.927&       0.892&       0.863&      -0.320&      -0.274&      -0.227&      -0.209\\
\multicolumn{2}{l}{$ log(Y_{2}^*) $}     &       0.590&       0.545&       0.950&       0.915&      -0.266&      -0.329&      -0.249&      -0.228\\
\multicolumn{2}{l}{$ log(Y_{3}^*) $}     &       0.522&       0.476&       0.460&       0.951&      -0.239&      -0.293&      -0.307&      -0.237\\
\multicolumn{2}{l}{$ log(Y_{4}^*) $}     &       0.467&       0.424&       0.405&       0.393&      -0.216&      -0.271&      -0.258&      -0.286\\
\multicolumn{2}{l}{$ u_{1} $}     &      -0.057&      -0.041&      -0.037&      -0.033&       0.043&       0.403&       0.356&       0.338\\
\multicolumn{2}{l}{$ u_{2} $}     &      -0.042&      -0.044&      -0.035&      -0.033&       0.015&       0.032&       0.430&       0.401\\
\multicolumn{2}{l}{$ u_{3} $}     &      -0.032&      -0.030&      -0.034&      -0.028&       0.012&       0.013&       0.027&       0.443\\
\multicolumn{2}{l}{$ u_{4} $}     &      -0.029&      -0.027&      -0.026&      -0.029&       0.011&       0.012&       0.012&       0.026\\[0.3ex]
            \hline \\[-1.8ex]
            \multicolumn{2}{l}{Men} & \scriptsize{[7013]} & \scriptsize{[4321]} & \scriptsize{[3181]} & \scriptsize{[2495]} & \scriptsize{[7013]} & \scriptsize{[4321]} & \scriptsize{[3181]} & \scriptsize{[2495]} \\
            & & $log(Y^*_{1})$ & $log(Y^*_{2})$ & $log(Y^*_{3})$ & $log(Y^*_{4})$ & $u_{1}$ & $u_{2}$ & $u_{3}$ & $u_{4}$ \\[0.3ex]
            \hline \\[-1.8ex]
\multicolumn{2}{l}{$ log(Y_{1}^*) $}     &       0.587&       0.915&       0.872&       0.836&      -0.299&      -0.230&      -0.199&      -0.209\\
\multicolumn{2}{l}{$ log(Y_{2}^*) $}     &       0.428&       0.372&       0.935&       0.883&      -0.248&      -0.283&      -0.211&      -0.220\\
\multicolumn{2}{l}{$ log(Y_{3}^*) $}     &       0.360&       0.307&       0.290&       0.929&      -0.213&      -0.221&      -0.263&      -0.227\\
\multicolumn{2}{l}{$ log(Y_{4}^*) $}      &       0.302&       0.254&       0.236&       0.222&      -0.169&      -0.189&      -0.181&      -0.292\\
\multicolumn{2}{l}{$ u_{1} $}     &      -0.045&      -0.032&      -0.028&      -0.022&       0.038&       0.378&       0.350&       0.361\\
\multicolumn{2}{l}{$ u_{2} $}     &      -0.030&      -0.029&      -0.021&      -0.018&       0.013&       0.029&       0.403&       0.393\\
\multicolumn{2}{l}{$ u_{3} $}     &      -0.023&      -0.020&      -0.022&      -0.015&       0.011&       0.011&       0.024&       0.452\\
\multicolumn{2}{l}{$ u_{4} $}     &      -0.025&      -0.021&      -0.019&      -0.021&       0.011&       0.010&       0.011&       0.024\\[0.3ex]
            \hline \\[-1.8ex]
            \multicolumn{2}{l}{Women} & \scriptsize{[6874]} & \scriptsize{[4546]} & \scriptsize{[3402]} & \scriptsize{[2623]} & \scriptsize{[6874]} & \scriptsize{[4546]} & \scriptsize{[3402]} & \scriptsize{[2623]} \\
            & & $log(Y^*_{1})$ & $log(Y^*_{2})$ & $log(Y^*_{3})$ & $log(Y^*_{4})$ & $u_{1}$ & $u_{2}$ & $u_{3}$ & $u_{4}$ \\[0.3ex]
            \hline \\[-1.8ex]
\multicolumn{2}{l}{$ log(Y_{1}^*) $}      &       0.810&       0.924&       0.886&       0.850&      -0.365&      -0.322&      -0.258&      -0.245\\
\multicolumn{2}{l}{$ log(Y_{2}^*) $}        &       0.657&       0.623&       0.950&       0.915&      -0.312&      -0.382&      -0.289&      -0.270\\
\multicolumn{2}{l}{$ log(Y_{3}^*) $}     &       0.585&       0.550&       0.538&       0.952&      -0.290&      -0.358&      -0.356&      -0.281\\
\multicolumn{2}{l}{$ log(Y_{4}^*) $}     &       0.526&       0.497&       0.480&       0.473&      -0.285&      -0.345&      -0.329&      -0.325\\
\multicolumn{2}{l}{$ u_{1} $}     &      -0.073&      -0.053&      -0.049&      -0.048&       0.049&       0.423&       0.362&       0.320\\
\multicolumn{2}{l}{$ u_{2} $}     &      -0.054&      -0.056&      -0.049&      -0.047&       0.017&       0.035&       0.449&       0.408\\
\multicolumn{2}{l}{$ u_{3} $}     &      -0.040&      -0.039&      -0.045&      -0.040&       0.014&       0.014&       0.030&       0.435\\
\multicolumn{2}{l}{$ u_{4} $}     &      -0.036&      -0.035&      -0.034&      -0.037&       0.012&       0.013&       0.012&       0.027\\[0.3ex]
            \hline \\[-1.8ex]
            \bottomrule
            \bottomrule
        \end{tabular}  
        \begin{tablenotes}
            \item \footnotesize{Table \ref{tab:varcov} presents pairwise correlation coefficients (upper half), estimated covariances (lower half), and estimated variances (diagonal) for the entire dataset (panel 1) and by gender (panels 2 and 3). $Y^*_1 \coloneq Y^*_{i1}$ denotes earnings in the first available period with employment for individual $i$ (similarly for $u_t \coloneq u_{it}$). Since pairwise correlations are considered, the estimates may be based on different sample sizes, which are declining in the number of periods $t \in \{1, 2, 3, 4\}$. I.e., if correlations across different periods are considered, the number of observations underlying this statistic is at most equal to the minimum available number of observations for the variables to be compared. The effective number of observations for each individual variable can be found in square brackets in the row above each matrix. Data: SOEP-CMI-ADIAB (weakly balanced sample).}
        \end{tablenotes}
    \end{threeparttable}
\end{table}

Log-incomes appear to be highly time persistent with a large positive correlation for as many as $4$ periods in the past of $0.863$. Comparing correlation coefficients in adjacent years across time reveals an increasing degree of correlation in $t$. This is most likely caused by requiring consecutive employment for later periods. Measurement errors behave similar over time. However, the variation amounts to $3$ - $11\%$ of the variation in $Y^*_t$'s. Moreover, $u_t$ exhibits a weaker degree of autocorrelation, which does not appear vanish over time either. Some heterogeneity is present across gender: Variation of earnings and measurement error of male observations is lower compared to their female counterparts. Females, in particular, exhibit a larger variation in true log-incomes relative to their male counterparts. This decreases the importance of measurement error for these observations, as the variation in measurement error is comparable across gender. Moreover, the degree of autocorrelation within the measurement error and the true signal, respectively, is slightly larger for women.

Although the sample size for future periods is much smaller, the variance of both the measurement error $u_t$ and the true signal $Y^*_t$ decreases in $t$, which could be due to requiring consecutive employment as well. Previous validation studies of income variables have attributed measurement error to the transitory component of income. Arguably, the share of permanent income compared to transitory income is larger for consecutively employed than for those with frequent unemployment breaks, which provides an explanation for the decreasing variance over time $t$. However, estimates of the variance-covariance matrix using a strongly balanced panel as depicted in Table \ref{tab:varcov_balanced} feature comparable -- if anything, slightly more attenuated -- dynamics. Thus, this may be evidence of the fact that more experience in the survey process increases the accuracy of reported income. This explanation, however, does not appear sensible in explaining the declining trend in earnings variation.

The standardized measures of association between the true signal and measurement error as provided by the correlation coefficients provide insights into the nature of measurement error. Together with the previous distributional observation of a non-zero expected value, the results suggest that measurement error in SOEP should be considered non-classical. Firstly, the sign of the correlation coefficients indicate a negative relationship between measurement error and the true signal, which can be interpreted as some form of mean-reversion. Again, the degree of the negative association is decreasing across time periods in the weakly balanced panel, which in this case is not robust to the consecutive employment requirement judging from the corresponding estimates in Table \ref{tab:varcov_balanced}. Administrative earnings are weakly correlated with past and future measurement error (and vice versa), although the correlation with measurement error in past periods appears to be stronger than with future measurement error. The reverse holds for correlation of measurement error with past and future true incomes, where earnings one period ahead appear to be slightly more correlated with present measurement error than earnings from the previous period. These observations carry over to the subsample considering men and women separately. However, there do not exist any salient patterns across gender.

Assuming classical measurement error, these estimates allow us to calculate reliability ratios $\lambda_t$ for survey income $Y_{it}$ as well as first differences $\Delta Y_{is}$ in the weakly balanced sample in each time period $t \in \{1, 2, 3, 4\}$ and $s \in \{2, 3, 4\}$, which are presented in the first three columns of Table \ref{tab:reliab_ratio}. These are based on considerations made with respect to the attenuation bias in a univariate regression with a mismeasured independent variable (c.f.\ Equation \ref{eq:bias_class_ME}) and amount to 
\begin{equation}\label{eq:reliab_ratio_class}
\lambda(Y_{it}) = \frac{\sigma^2_{Y^*_t}}{\sigma^2_{Y^*_t} +\sigma^2_{u_t}}
\end{equation}
\noindent for nominal log-income, and to 
\begin{equation}\label{eq:reliab_ratio_class_diff}
\lambda(\Delta Y_{it}) = \frac{\sigma^2_{\Delta Y^*_t}}{\sigma^2_{\Delta Y^*_t} + \sigma^2_{\Delta u_t}} = \frac{\sigma^2_{Y^*_t} + \sigma^2_{Y^*_{t-1}} - 2\sigma_{Y^*_tY^*_{t-1}}}{\sigma^2_{Y^*_t} + \sigma^2_{Y^*_{t-1}} + \sigma^2_{u_t} + \sigma^2_{u_{t-1}} - 2\sigma_{Y^*_tY^*_{t-1}}}
\end{equation}
\noindent for first differences of log-income. Note that we allow the reliability ratios to differ with $t$.

To obtain the reliability ratio in the non-classical measurement error context, we run regressions of the true signal on the mismeasured variable for each time period $t$, i.e.\
\begin{equation}\label{eq:reliab_ratio_nonclass}
Y_{it}^* = \alpha + b_{Y_tY^*_t} Y_{it} + \eta_{it}.
\end{equation}
A similar expression is used for first differences, where $Y_{it}^*$ ($Y_{it}$) is replaced by $\Delta Y_{it}^*$ ($\Delta Y_{it}$). The results of this exercise can be found in Table \ref{tab:reliab_ratio}. Assuming a classical measurement error structure, the reliability ratios for each period are close to $94\%$. Disaggregation by gender reveals a larger reliability ratio among females. This implies that regressions with reported earnings among the covariates based on a cross-section of men will be subject to larger attenuation bias. As expected, the effect of measurement error accumulates when calculating first differences, with substantially lower reliability ratios of less than $59\%$. Most of this effect can be explained by the large time persistence of incomes, which is not allowed for in the classical measurement error case. Unlike undifferenced log-income, the reliability ratios decline with $t$, which indicates a more severe contamination of log-income changes by measurement error for later reports. This decline appears to be more pronounced for males, for which the reliability ratio drops from $61\%$ in $t=2$ to $46\%$ in $t=4$. Table \ref{tab:reliab_ratio_class_II}, which presents the results based on the strongly balanced panel, shows that these results are robust to the imbalanced sampling structure for females only. For males, the share of measurement error among reported incomes is now larger. Further, the reliability ratios for the first-differenced variables have decreased as well, indicating an even larger attenuation bias for analyses relying on the strongly balanced sample.

\begin{table}[!htbp] 
    \centering
    \begin{threeparttable}
        \caption{Reliability ratios}
        \label{tab:reliab_ratio}
        \begin{tabular}{lcccccc}
            \toprule 
            \toprule
            & \multicolumn{3}{c}{Classical} & \multicolumn{3}{c}{Non-classical} \\
            & All & Men & Women & All & Men & Women \\
            & \footnotesize{(1)} & \footnotesize{(2)} & \footnotesize{(3)} & \footnotesize{(4)} & \footnotesize{(5)} & \footnotesize{(6)} \\[0.3ex]
            \hline \\[-1.8ex]
$ log(Y_{i1}) $     &       0.945&       0.939&       0.943 &       1.021&       1.012&       1.033\\
$ log(Y_{i2}) $     &       0.944&       0.928&       0.947 &       1.023&       1.001&       1.039\\
$ log(Y_{i3}) $     &       0.945&       0.925&       0.948 &       1.018&       0.993&       1.032\\
$ log(Y_{i4}) $     &       0.939&       0.902&       0.946 &       1.008&       0.986&       1.023\\
$ \Delta log(Y_{i2}) $&       0.590&       0.607&       0.590 &       0.716&       0.681&       0.741\\
$ \Delta log(Y_{i3}) $&       0.477&       0.475&       0.487 &       0.627&       0.586&       0.661\\
$ \Delta log(Y_{i4}) $&       0.460&       0.460&       0.472 &       0.616&       0.589&       0.634\\[0.3ex]
            \bottomrule
            \bottomrule
        \end{tabular}  
        \begin{tablenotes}
            \item \footnotesize{Table \ref{tab:reliab_ratio} presents the reliability ratios for log-incomes and first differences of log-incomes for a time horizon of $T=4$. In the classical measurement error context, the statistics are based on Equations \ref{eq:reliab_ratio_class} and \ref{eq:reliab_ratio_class_diff} and the (pairwise) estimated variance-covariance matrix displayed in Table \ref{tab:varcov}. In the non-classical measurement error context, the estimates are based on a univariate regression following Equation \ref{eq:reliab_ratio_nonclass}. The standard error is smaller than $0.07$ for all regression-based estimated coefficients. Data source: SOEP-CMI-ADIAB (weakly balanced sample).}
        \end{tablenotes}
    \end{threeparttable}
\end{table}

Columns $(4)$ through $(6)$ provide yet another indication that the assumption of classical measurement error may not be justified. As in the classical case, the reliability ratios of nominal log-incomes are relatively stable over time. However, when allowing for non-classical measurement error, the level is approximately $7$-$8$ pp.\ larger. This is due to the negative correlation between the true signal and measurement error, which diminishes the variation of the mismeasured variable $Y_{it}$ by more than the covariance between $Y_{it}$ and the true signal $Y_{it}^*$. This implies a small positive bias due to measurement error in univariate regressions with log-income as the independent variable ($0.8$-$2.3\%$). The increase in reliability ratios holds for each sex separately as well. However, the disaggregation by gender further reveals that earnings reported by women exhibit a large positive bias from $2.3$ (in $t=4$) up to $3.9\%$ (in $t=2$) while the results for males suggest that estimates of a univariate model for males may also suffer from attenuation of ($1.4\%$). 

Concerning first differences in log-income, the autocorrelated nature of the measurement error benefits the reliability ratios, which increase by $7.4$-$17.4$ pp.\ compared to the classical case in columns $(1)$ through $(3)$. The dynamics by gender detected in the classical case carry over to these statistics as well: The drop across time for first-differences is again largely driven by observations associated with male survey participants, yielding reliability ratios as low as $46\%$ (in $t=4$).

In Table \ref{tab:reliab_robust}, we assess whether the figures obtained for the non-classical measurement error context are particular to the constructed sample or whether they are robust to that choice. With a standard error smaller than $0.042$ all estimates individually are statistically different from zero. Column $(1)$ restates the calculated reliability ratios for the weakly balanced sample (c.f.\ column $(4)$ in Table \ref{tab:reliab_ratio}). As in the classical measurement error case, we rerun regressions using the strongly balanced panel, which affects estimation of variables in earlier time periods. The corresponding results are presented in column $(2)$. Reliability ratios for nominal log-incomes are slightly lower than the estimates based on the weakly balanced panel but remain larger than calculated reliability ratios restricted to classical measurement error. For first differences, the decrease across $t$ persists. Here, we also observe a slightly lower reliability ratios compared to using the weakly balanced data. Another source for bias from measurement error stems from the ambiguity of the survey question eliciting monthly incomes paired with an individual potentially having multiple jobs. In such a case, the calculated measurement error could reflect a mismatch between the employment spells being compared. In order to evaluate the influence the effect of these \textit{false comparisons} on the calculated statistics, we exclude observations with multiple employment spells. As only a few observations are affected, the point estimates hardly change. Once again, this indicates that the varying sample preparation steps compared to \cite{caliendoetal2024} should only have a minor effect on estimated reliability ratios. In that sense, the results in this paper complement the results in the aforementioned one.

Column $(4)$ explores whether there is variation in reliability ratios across early (1984-2009) and late periods of reporting in SOEP by restricting the sample to observations from 2010 to 2021. Indeed, regression-based reliability ratios of $Y_{it}$ in more recent years are slightly larger, although the increase is limited to at most $0.7$ pp. The change in reliability for first differences is slightly larger, but also positive. This exercise allows us to shed some light on variation of reporting errors and administrative income across time. Since less than $2\%$ of observations are associated with reports the past millennium, the calculated baseline statistics in column $(1)$ are mostly based on recent reports. More specifically, at least $75\%$ of observations in the main sample, the weakly balanced sample, and the strongly balanced sample, respectively, refer to the years after 2009. This arguably limits the amount of noise due to changes in reporting behavior across time, which is reflected in small differences relative to the baseline estimates.

\begin{table}[!htbp] 
    \centering
    \begin{threeparttable}
        \caption{Reliability ratios -- Robustness to sample variation}
        \label{tab:reliab_robust}
        \begin{tabular}{lccccccc}
            \toprule 
            \toprule
            & \small{Baseline} & \small{Strongly} & \small{Excl.} & \small{2010} & \small{Incl.} & \small{Incl.} & \small{Incl. untypical}  \\
            & \small{(pairwise)} & \small{balanced} & \small{mult.\ spells} & \small{to 2021} & \small{imputed obs.\ } & \small{$|u_{it}|>150\%$} & \small{occupations} \\
            & \footnotesize{(1)} & \footnotesize{(2)} & \footnotesize{(3)} & \footnotesize{(4)} & \footnotesize{(5)} & \footnotesize{(6)} & \footnotesize{(7)} \\[0.3ex]
            \hline \\[-1.8ex]
$ log(Y_{i1}) $      &       1.021&       1.005&       1.019&       1.024&       1.016&       0.982&       1.023\\[-1.0ex]
            &\scriptsize{[13887]}&\scriptsize{[5118]}&\scriptsize{[12446]}&\scriptsize{[12303]}&\scriptsize{[14286]}&\scriptsize{[14048]}&\scriptsize{[16509]}\\
$ log(Y_{i2}) $     &       1.023&       1.015&       1.022&       1.028&       1.022&       0.989&       1.029\\[-1.0ex]
            &\scriptsize{[8867]}&\scriptsize{[5118]}&\scriptsize{[8016]}&\scriptsize{[7754]}&\scriptsize{[9380]}&\scriptsize{[9000]}&\scriptsize{[10701]}\\
$ log(Y_{i3}) $      &       1.018&       1.013&       1.017&       1.023&       1.012&       0.984&       1.027\\[-1.0ex]
            &\scriptsize{[6583]}&\scriptsize{[5118]}&\scriptsize{[6004]}&\scriptsize{[5668]}&\scriptsize{[7058]}&\scriptsize{[6693]}&\scriptsize{[7965]}\\
$ log(Y_{i4}) $      &       1.008&       1.008&       1.007&       1.015&       1.009&       0.989&       1.014\\[-1.0ex]
            &\scriptsize{[5118]}&\scriptsize{[5118]}&\scriptsize{[4675]}&\scriptsize{[4357]}&\scriptsize{[5567]}&\scriptsize{[5203]}&\scriptsize{[6209]}\\
$ \Delta log(Y_{i2}) $ &       0.716&       0.693&       0.700&       0.727&       0.673&       0.553&       0.762\\[-1.0ex]
            &\scriptsize{[8867]}&\scriptsize{[5118]}&\scriptsize{[8016]}&\scriptsize{[7754]}&\scriptsize{[9380]}&\scriptsize{[9000]}&\scriptsize{[10701]}\\
$ \Delta log(Y_{i3}) $ &       0.627&       0.612&       0.635&       0.644&       0.595&       0.523&       0.707\\[-1.0ex]
            &\scriptsize{[6583]}&\scriptsize{[5118]}&\scriptsize{[6004]}&\scriptsize{[5668]}&\scriptsize{[7058]}&\scriptsize{[6693]}&\scriptsize{[7965]}\\
$ \Delta log(Y_{i4}) $ &       0.616&       0.616&       0.592&       0.643&       0.574&       0.514&       0.653\\[-1.0ex]
            &\scriptsize{[5118]}&\scriptsize{[5118]}&\scriptsize{[4675]}&\scriptsize{[4357]}&\scriptsize{[5567]}&\scriptsize{[5203]}&\scriptsize{[6209]}\\[0.3ex]
            \bottomrule
            \bottomrule
        \end{tabular}  
        \begin{tablenotes}
            \item \footnotesize{Table \ref{tab:reliab_robust} presents regression based reliability ratios (c.f.\ Equation \ref{eq:reliab_ratio_nonclass}) for varying sample restrictions, with the baseline results copied from Table \ref{tab:reliab_ratio} (column $(4)$). The effective number of observations used in each regression is printed in square brackets below the corresponding estimate. Results in column $(2)$ are based on the strongly balanced sample. Column $(3)$ is based on observations, which are not characterized by additional employment spells other than the highest-paying job. In column $(4)$, the weakly balanced sample is restricted to observations from 2010 to 2021. The sample used for columns $(5)$ and $(6)$ include observations imputed in SOEP and observations with absolute measurement error exceeding $150\%$ of administrative income, respectively. In column $(7)$, observations associated with occupations with untypical pay structures are admitted to the weakly balanced sample. The standard deviation is smaller than $0.042$ for all estimates presented. Data: SOEP-CMI-ADIAB.}
        \end{tablenotes}
    \end{threeparttable}
\end{table}

In order to single out the “pure” reporting error, we have applied restrictions on the sample. To see if this influences the degree of attenuation, reliability ratios for non-classical measurement error are calculated for samples, which include the observations left out in the previous steps (c.f.\ Table \ref{tab:sample_restrictions}). Columns $(5)$ and $(6)$ corroborate our decision to exclude imputed observations ($5\%$ of the weakly balanced sample) as well as those with measurement error exceeding $150\%$ (approx.\ $1\%$ of the weakly balanced sample). In both cases, the resulting reliability ratios of nominal log-income and first differences are lower. In particular, the results concerning first-differenced when including imputed earnings for the analyses question the quality of imputation procedures adopted by IAB. When including extreme observations, the deterioration is as large as reversing the sign of the bias (attenuation). In light of the rather small number of observations which exhibit large absolute differences in reported earnings relative to the register earnings, this effect seems quite large. Lastly, reliability ratios based on a sample, which includes occupations with untypical pay structures ($19\%$ of the weakly balanced panel), are slightly larger compared to the baseline results. This small change is indeed surprising as our motive for excluding this group of observations was that we expected a larger degree of measurement error among this group of workers, which was then taken to have a large effect on reliability ratio estimates. Except for including observations with very large absolute measurement error, the reliability ratios remain stable across specifications.

A brief comparison to previous results from the validation study literature is in order. For nominal log-income, the obtained reliability ratios denote the upper bound of the set established by previous literature. Our reliability ratios for first differences correspond to the average value found in the literature. In the classical measurement error context, \cite{boundkrueger1991} find lower reliability ratios among men in the range of $0.815$ to $0.844$ whereas -- similar to our results -- the statistics for women are larger. However, with reliability ratios of $0.924$ to $0.939$, these figures are in line with ours from Table \ref{tab:reliab_ratio} (column $(3)$). Further, the authors find larger values for first differences ($0.625$; $0.708$ for men, $0.814$ for women). Using data from the 1980s, reliability ratios obtained by \cite{boundetal1994}, who pool the data across gender, can be related in a similar fashion. These range from $0.698$ to $0.867$ for static log-income and from $0.685$ to $0.708$ for first differences. \cite{schmillenetal2024}'s ratios are smaller for nominal ($0.878$ -- $0.893$) as well as first-differenced log-income ($0.289$ -- $0.407$). However, in their case male log-income exhibits larger reliability ratios, which is the opposite of what we observe in our results.

For non-classical measurement error, reliability ratios in \cite{boundkrueger1991} and \cite{boundetal1994} increase due to the negative correlation between measurement error and the true signal. The general theme that our figures are larger persists.\footnote{The same applies to \cite{pischke1995}, who uses the same data as \cite{boundetal1994}.} From the parametric validation study literature (\citeauthor{kapteynypma2007}, \citeyear{kapteynypma2007}; \citeauthor{jenkinsriosavila2023}, \citeyear{jenkinsriosavila2023}; \citeauthor{meijeretal2012}, \citeyear{meijeretal2012}), we obtain lower values of the reliability associated with survey data than ours ($0.690$ -- $0.822$).

\section{Discussion}\label{sec:discussion}
In this section, we will briefly discuss our results by highlighting limitations, applications, and extensions. Moreover, this part presents thoughts collected in the course of our project.

The calculation of reliability ratios for consecutive time periods required us to construct a balanced panel. We proceeded to take the first observation for each unit $i$ with available income data as the initial period ($t=1$). Alternatively, for fixed $T$, we could have used the temporal dimension in SOEP directly and accounted for this time horizon in the past relative to 2021. For $T=4$, this would have amounted to $t=1$ in 2018, $t=2$ in 2019, etc. For nominal log-incomes, this approach allows one to use the cross-section of a given year and, thus, is not limited to a balanced panel. However, when changes in nominal log-income are of interest, this requires restricting to a balanced panel once again. Thus far, the same considerations hold in the context of our approach. The issue arises when wanting to consider a larger time horizon $T$.\footnote{In principle, the SOEP allows us to consider every time horizon up until $T=35$.} Suppose $T=5$. Then, the alternative approach now yields $t=1$ in 2017, $t=2$ in 2018, etc. Due to the varying composition of SOEP, the calculated reliability ratios for $t=1$ may differ across $T$. Our approach, on the other hand, leaves the reliability ratios based on the length of the baseline time horizon unchanged.

Our results can be interpreted to imply a large reliability of SOEP data with respect to reported incomes. The hope is that bias from measurement error is rather small relative to the size of the true parameter of interest. However, the metric used to reach this verdict -- the reliability ratio -- is based on the closed form expression of the attenuation bias in the linear model. Reliability ratios in the non-classical measurement error setting allow for non-zero covariances among the measurement error and the true signal, which is a measure of linear dependence. It is not clear whether our claim holds true in other structural or even nonparametric settings. Hence, the development of more general statistics about the reliability of mismeasured data and the severity of measurement error is necessary. This could be the subject of future research.

In this paper, we offer a rather general characterization of measurement error in SOEP. The question remains which repercussions we can expect to arise from the presence of measurement error in applied topics. In addition to modelling assumptions, complications arising from measurement error are probably specific to a particular context. For example, mean-reverting measurement error implies a smaller degree of income inequality when considering statistics based on the income distribution, since both tails of the mismeasured income distribution are shifted towards the center. In this context, establishing the “most robust” income inequality measure appears particularly valuable. Methodologically speaking, an emphasis could be placed on commonly used income inequality statistics such as, e.g., the Gini coefficient, Theil index, or top/bottom shares, and to study the effect of different forms of measurement error therein. The non-linear nature of most of these statistics make this effort particularly interesting as classical measurement error may not simply attenuate but amplify the parameter estimate of interest, i.e.\ the bias' sign is unknown ex-ante.

Statistical analyses are usually not based on one explanatory variable but allow for additional information through observables. In the linear model, one can show that the bias from measurement error propagates to other estimates if the underlying covariates are correlated with the measurement error and the true signal \citep{pischke2007}. Our characterization reveals that measurement error in SOEP is indeed related to observable characteristics. If the entire variation in measurement error was explained by a model of observables, we may be able to reduce the bias. This idea is related to two-stage least squares estimation in the context of instrumental variables in that we reduce the variance stemming from measurement error in a first step while using variation solely associated with the true signal in a second step. However, as Tables \ref{tab:mincer_corr} and \ref{tab:mincer_corr_FE} show, the information available to us does not suffice to explain a sufficient share in the variance of measurement error. On the contrary, it appears that a much larger share of the true signal is explained by these covariates, which is the opposite of what we would like for this approach to work.

The results of our analysis suggest that measurement error is serially correlated over time. This may be able to explain a share of the estimated dynamics in income inequality based on mismeasured data. Taking the stance that administrative data provides the pure signal, a disproportionate increase in income inequality based on the survey data may merely reflect a decreasing degree of mean-reversion over time. \cite{stueberetal2023} claim that diverging results with respect to income inequality in Germany over time across data sources result from non-comparable samples, which can be resolved through proper harmonization. However, the results in this paper as well as of previous validation studies argue for a more systematic discrepancy between survey and administrative data, which cannot be overcome by simply making samples “more comparable”.

An issue with the dataset based on social security contributions, which is particularly prevalent in the context of income inequality, is top-coding. Methods modeling the latency of high incomes, which applies to 10-13\% of all full-time employment spells in Germany, have been envisaged and are provided by IAB as well (\citeauthor{gartner2005}, \citeyear{gartner2005}; \citeauthor{drechsleretal2023}, \citeyear{drechsleretal2023}). The value of the validation dataset at hand lies in its ability to provide some (potentially noisy) information on income of individuals pertaining to this group. The estimated measurement error distribution on the observable part of the support of the earnings distribution could be extrapolated to “correct” the survey data for measurement error and to, thus, assess the quality of high incomes as predicted by latent variable models.

All of our analyses build on the assumption that the administrative data is without error, thus providing the \textit{true} gross monthly income. A more sincere approach is to treat the income data provided by survey and administrative sources, respectively, as repeated measurements of the same underlying concept: gross labor income. The idea is to gather more information about the distribution of the true signal by imposing assumptions on how the measurement errors of the two variables relate.\footnote{In general, this approach is of particular importance when dealing with more abstract concepts as skills, beliefs, or valuations, where -- unlike administrative data -- the plausibility of the existence of a variable that encodes the true signal is limited.}  Using linked survey and administrative data from New Zealand, \cite{hysloptownsend2020} allow for measurement error in register data as well. Based on a model of earnings dynamics fitted to the data using minimum distance estimation methods, the authors find that the share of measurement error in first differences of log-earnings decreases relative to taking register data as the true signal. When having two variables measuring the same concept at hand, \cite{abowdstinson2013} proffer that progress can be made by placing a prior on the likelihood of which variable resembles the true signal more. Implicitly, our procedure assigned a probability of 1 to the event that register data reflects the true signal, which may be too strong of an assumption to impose. The authors devise a method, which allows the researcher to bound reliabilities of each variable by imposing different priors. This could be exploited in future research.


\section{Conclusion}\label{sec:conclusion}
This paper studies measurement error of reported income in the German \textit{Sozio-ökonomisches Panel} (SOEP) survey dataset using a recent linkage to register data. We visualize and apply recent advances with respect to the analysis of validation datasets, mostly following considerations by \cite{schmillenetal2024}. The process of linking administrative and survey data requires formal consent by each respondent. We find that this process is selective, yielding a non-random subsample of SOEP. Considering administrative data as the true signal, our results suggest the presence of non-classical measurement error: It is negatively correlated with the true signal, exhibits autocorrelation, and is not equal to zero in expectation. A large share of observations is associated with negative measurement error, which implies underreporting to a degree varying across the income distribution. Mincer-type regressions using both reported and true earnings data, respectively, show a rather small bias due to measurement error, whose sign is proportional to the relationship of the corresponding covariate with the measurement error. This is corroborated by high estimated reliability ratios of nominal log-income close to $1$. For first differences, reliability ratios are mostly below $70\%$ mirroring the increasing severity of measurement error when using a combination of mismeasured variables.


\newpage
\begin{singlespace}
\renewcommand{\bibsection}{\section*{References}\addcontentsline{toc}{section}{References}}
\bibliography{bib}
\end{singlespace}

\endgroup

\newpage
\section*{\appendixname}\addcontentsline{toc}{section}{\appendixname}\label{sec:appendix}
\normalsize

\subsection*{Additional figures and tables}

\begin{figure}[H]
    \centering
    \caption{Representativeness of SOEP -- Main sample}
    \includegraphics[scale=0.55]{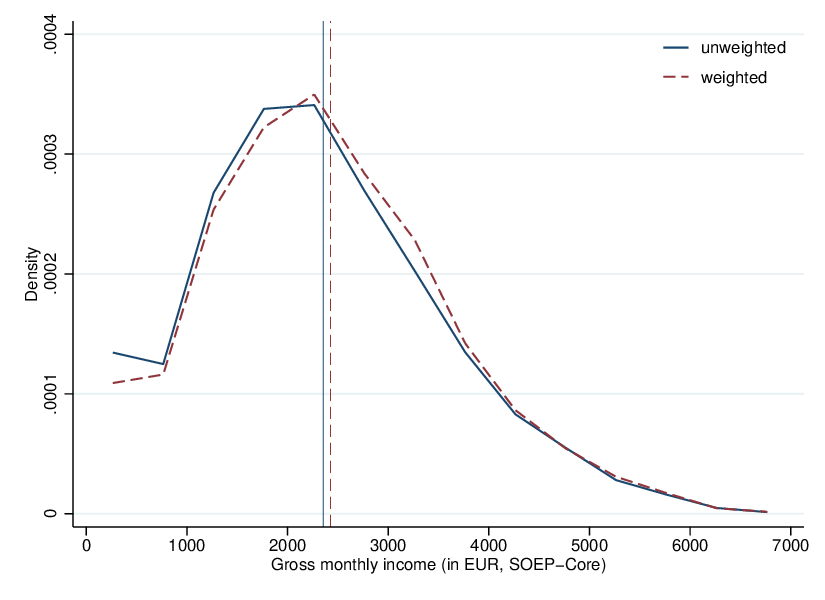}
    \includegraphics[scale=0.55]{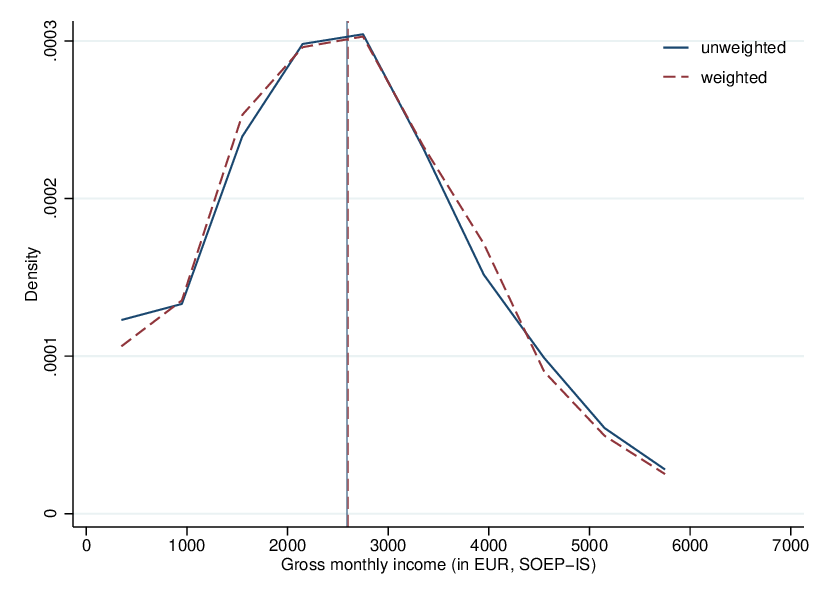}
    \label{fig:SOEP_sampling_main}
    \begin{minipage}{\textwidth}
        \begin{singlespace}
            \footnotesize{Figure \ref{fig:SOEP_sampling_main} shows the unweighted and weighted empirical density functions of gross monthly income in the SOEP-Core (left panel) and SOEP-IS (right panel) module, respectively, using a histogram with $15$ (SOEP-IS: $12$) bins of size \euro$500$ (SOEP-IS: \euro$600$) for incomes up to \euro$7'000$ (SOEP-IS: \euro$6'200$) for individuals in the main sample, where, in each figure respectively, less than $0.26\%$ of all observations located in the upper tail of the respective distribution were omitted due to data privacy concerns. Mean: \euro$2'354.54$ (\euro$2'591.00$), \euro$2'426.30$ (\euro$2'601.45$); standard deviation: \euro$1'219.98$ (\euro$1'283.38$), \euro$1'205.51$ (\euro$1'256.53$) (left to right, weighted statistics in brackets). Data: SOEP-CMI-ADIAB (main sample).}
        \end{singlespace}
    \end{minipage}
\end{figure}

\begin{sidewaystable}[!hbtp]
        \centering
        \begin{threeparttable}
            \caption{Correlation with observables and Mincer-type regressions -- Survey year-FE}
            \label{tab:mincer_corr_FE}
            \begin{tabular}{@{}lccccccccc@{}}
                \toprule 
                \toprule
                Dep.\ var.\ & \multicolumn{3}{c}{$u_{it}$} & \multicolumn{3}{c}{$log(Y_{it})$} & \multicolumn{3}{c}{$log(Y_{it}^*)$} \\
                & \small{All} & \small{Men} & \small{Women} & \small{All} & \small{Men} & \small{Women} & \small{All} & \small{Men} & \small{Women} \\[0.3ex]
                \hline \\[-1.8ex]
Female      &     -0.0144&       --     &      --      &     -0.3259&       --     &      --      &     -0.3115&      --      &     --       \\
            &\footnotesize(0.0012)&            &            &\footnotesize(0.0037)&            &            &\footnotesize(0.0037)&            &            \\
Single      &     -0.0019&     -0.0012&     -0.0027&      0.0786&     -0.0155&      0.1932&      0.0805&     -0.0143&      0.1959\\
            &\footnotesize(0.0015)&\footnotesize(0.0021)&\footnotesize(0.0022)&\footnotesize(0.0066)&\footnotesize(0.0093)&\footnotesize(0.0087)&\footnotesize(0.0066)&\footnotesize(0.0093)&\footnotesize(0.0087)\\
German      &     -0.0147&     -0.0204&     -0.0081&      0.1473&      0.1795&      0.0897&      0.1620&      0.2000&      0.0978\\
            &\footnotesize(0.0027)&\footnotesize(0.0037)&\footnotesize(0.0041)&\footnotesize(0.0115)&\footnotesize(0.0144)&\footnotesize(0.0185)&\footnotesize(0.0116)&\footnotesize(0.0144)&\footnotesize(0.0186)\\
Migration background&      0.0027&     -0.0004&      0.0051&      0.1383&      0.1250&      0.1317&      0.1355&      0.1254&      0.1266\\
            &\footnotesize(0.0020)&\footnotesize(0.0028)&\footnotesize(0.0029)&\footnotesize(0.0107)&\footnotesize(0.0154)&\footnotesize(0.0139)&\footnotesize(0.0107)&\footnotesize(0.0155)&\footnotesize(0.0140)\\
East        &     -0.0076&     -0.0158&     -0.0005&     -0.0785&     -0.1779&      0.0204&     -0.0709&     -0.1620&      0.0209\\
            &\footnotesize(0.0016)&\footnotesize(0.0022)&\footnotesize(0.0022)&\footnotesize(0.0046)&\footnotesize(0.0054)&\footnotesize(0.0065)&\footnotesize(0.0049)&\footnotesize(0.0055)&\footnotesize(0.0069)\\
Abitur      &      0.0015&      0.0014&      0.0040&     -0.4001&     -0.4625&     -0.3329&     -0.4016&     -0.4639&     -0.3370\\
            &\footnotesize(0.0024)&\footnotesize(0.0034)&\footnotesize(0.0035)&\footnotesize(0.0182)&\footnotesize(0.0271)&\footnotesize(0.0230)&\footnotesize(0.0183)&\footnotesize(0.0273)&\footnotesize(0.0231)\\
Education (years)&      0.0009&      0.0024&     -0.0008&      0.1018&      0.1030&      0.1004&      0.1008&      0.1006&      0.1012\\
            &\footnotesize(0.0005)&\footnotesize(0.0006)&\footnotesize(0.0007)&\footnotesize(0.0043)&\footnotesize(0.0062)&\footnotesize(0.0057)&\footnotesize(0.0043)&\footnotesize(0.0062)&\footnotesize(0.0057)\\
Age (years) &     -0.0003&     -0.0004&     -0.0003&      0.0080&      0.0079&      0.0084&      0.0084&      0.0084&      0.0086\\
            &\footnotesize(0.0001)&\footnotesize(0.0001)&\footnotesize(0.0001)&\footnotesize(0.0004)&\footnotesize(0.0005)&\footnotesize(0.0005)&\footnotesize(0.0004)&\footnotesize(0.0005)&\footnotesize(0.0005)\\
Marginal employment&      0.1378&      0.1184&      0.1445&     -1.6368&     -1.9029&     -1.5059&     -1.7746&     -2.0213&     -1.6504\\
            &\footnotesize(0.0049)&\footnotesize(0.0099)&\footnotesize(0.0057)&\footnotesize(0.0104)&\footnotesize(0.0202)&\footnotesize(0.0120)&\footnotesize(0.0100)&\footnotesize(0.0188)&\footnotesize(0.0117)\\
Complex task&      0.0043&      0.0034&      0.0052&      0.1021&      0.0762&      0.1156&      0.0978&      0.0727&      0.1104\\
            &\footnotesize(0.0015)&\footnotesize(0.0021)&\footnotesize(0.0021)&\footnotesize(0.0062)&\footnotesize(0.0089)&\footnotesize(0.0083)&\footnotesize(0.0064)&\footnotesize(0.0090)&\footnotesize(0.0085)\\
Notification: \textit{Annual report}&     -0.0210&     -0.0195&     -0.0228&      0.1087&      0.1580&      0.0602&      0.1296&      0.1775&      0.0831\\
            &\footnotesize(0.0020)&\footnotesize(0.0028)&\footnotesize(0.0029)&\footnotesize(0.0096)&\footnotesize(0.0147)&\footnotesize(0.0117)&\footnotesize(0.0096)&\footnotesize(0.0148)&\footnotesize(0.0118)\\
Notification: \textit{Employment termination}&      0.0064&      0.0058&      0.0065&      0.0205&      0.0720&     -0.0344&      0.0140&      0.0661&     -0.0410\\
            &\footnotesize(0.0030)&\footnotesize(0.0040)&\footnotesize(0.0044)&\footnotesize(0.0120)&\footnotesize(0.0175)&\footnotesize(0.0153)&\footnotesize(0.0122)&\footnotesize(0.0176)&\footnotesize(0.0157)\\
Multiple empl.\ spells&      0.0104&      0.0035&      0.0157&     -0.0565&     -0.0148&     -0.0839&     -0.0669&     -0.0183&     -0.0996\\
            &\footnotesize(0.0026)&\footnotesize(0.0037)&\footnotesize(0.0037)&\footnotesize(0.0060)&\footnotesize(0.0083)&\footnotesize(0.0082)&\footnotesize(0.0060)&\footnotesize(0.0082)&\footnotesize(0.0084)\\
No.\ of coworkers&     -0.0000&     -0.0000&     -0.0000&      0.0000&      0.0000&      0.0000&      0.0000&      0.0000&      0.0000\\
            &\footnotesize(0.0000)&\footnotesize(0.0000)&\footnotesize(0.0000)&\footnotesize(0.0000)&\footnotesize(0.0000)&\footnotesize(0.0000)&\footnotesize(0.0000)&\footnotesize(0.0000)&\footnotesize(0.0000)\\
Avg.\ wage of coworkers&     -0.0009&     -0.0010&     -0.0008&      0.0042&      0.0042&      0.0040&      0.0051&      0.0052&      0.0048\\
            &\footnotesize(0.0000)&\footnotesize(0.0000)&\footnotesize(0.0001)&\footnotesize(0.0001)&\footnotesize(0.0001)&\footnotesize(0.0002)&\footnotesize(0.0002)&\footnotesize(0.0001)&\footnotesize(0.0003)\\[0.3ex]
                \hline \\[-1.8ex]
    F statistic  &      437.27&      222.01&      241.27 & -- & --& --& --& --& -- \\
    Survey year-FE        &        Yes&        Yes&        Yes&        Yes&        Yes&        Yes&        Yes&        Yes&        Yes\\
    $R^2$          &        0.24&        0.24&        0.25&        1.00&        1.00&        1.00&        1.00&        1.00&        1.00\\
    No.\ of obs.\           &       70'567&       34'996&       35'571&       70'567&       34'996&       35'571&       70'567&       34'996&       35'571\\[0.3ex]
                \bottomrule
                \bottomrule
            \end{tabular}  
            \begin{tablenotes}
                \item \footnotesize{Table \ref{tab:mincer_corr_FE} presents estimates of a linear model using OLS with the indicated dependent and independent variables. Contrary to Table \ref{tab:mincer_corr}, the regressions include survey year-fixed effects. Robust standard errors in parentheses; $R^2$ is unadjusted; F-statistic is based on the null hypothesis $H_0: \beta = 0$ (excl.\ the constant). Notation: No.\ -- number; empl.\ -- employment; avg.\ -- average. Data: SOEP-CMI-ADIAB (main sample).}
            \end{tablenotes}
        \end{threeparttable}
\end{sidewaystable}

\begin{table}[!htbp] 
    \centering
    \begin{threeparttable}
        \caption{Variance-covariance matrix}
        \label{tab:varcov_balanced}
        \begin{tabular}{llcccccccc}
            \toprule 
            \toprule
            \multicolumn{2}{l}{{All}} \\
            & & $log(Y^*_1)$ & $log(Y^*_2)$ & $log(Y^*_3)$ & $log(Y^*_4)$ & $u_1$ & $u_2$ & $u_3$ & $u_4$ \\[0.3ex]
            \hline \\[-1.8ex]
\multicolumn{2}{l}{$ log(Y^*_1) $}     &       0.467&       0.931&       0.893&       0.863&      -0.280&      -0.241&      -0.209&      -0.209\\
\multicolumn{2}{l}{$ log(Y^*_2) $}     &       0.418&       0.431&       0.952&       0.915&      -0.233&      -0.305&      -0.233&      -0.228\\
\multicolumn{2}{l}{$ log(Y^*_3) $}     &       0.390&       0.399&       0.408&       0.951&      -0.217&      -0.279&      -0.293&      -0.237\\
\multicolumn{2}{l}{$ log(Y^*_4) $}     &       0.370&       0.377&       0.381&       0.393&      -0.216&      -0.271&      -0.258&      -0.286\\
\multicolumn{2}{l}{$ u_1 $}     &      -0.034&      -0.028&      -0.025&      -0.024&       0.032&       0.421&       0.352&       0.338\\
\multicolumn{2}{l}{$ u_2 $}     &      -0.027&      -0.033&      -0.029&      -0.028&       0.013&       0.027&       0.438&       0.401\\
\multicolumn{2}{l}{$ u_3 $}     &      -0.022&      -0.024&      -0.029&      -0.025&       0.010&       0.011&       0.024&       0.443\\
\multicolumn{2}{l}{$ u_4 $}     &      -0.023&      -0.024&      -0.024&      -0.029&       0.010&       0.011&       0.011&       0.026\\[0.3ex]
            \hline \\[-1.8ex]
            \multicolumn{2}{l}{{Men}} \\
            & & $log(Y^*_1)$ & $log(Y^*_2)$ & $log(Y^*_3)$ & $log(Y^*_4)$ & $u_1$ & $u_2$ & $u_3$ & $u_4$ \\[0.3ex]
            \hline \\[-1.8ex]
\multicolumn{2}{l}{$ log(Y^*_1) $}     &       0.280&       0.938&       0.878&       0.836&      -0.254&      -0.184&      -0.155&      -0.209\\
\multicolumn{2}{l}{$ log(Y^*_2) $}     &       0.249&       0.253&       0.931&       0.883&      -0.204&      -0.230&      -0.168&      -0.220\\
\multicolumn{2}{l}{$ log(Y^*_3) $}     &       0.225&       0.227&       0.234&       0.929&      -0.169&      -0.195&      -0.224&      -0.227\\
\multicolumn{2}{l}{$ log(Y^*_4) $}     &       0.208&       0.209&       0.212&       0.222&      -0.169&      -0.189&      -0.181&      -0.292\\
\multicolumn{2}{l}{$ u_1 $}     &      -0.023&      -0.017&      -0.014&      -0.013&       0.028&       0.418&       0.360&       0.361\\
\multicolumn{2}{l}{$ u_2 $}     &      -0.015&      -0.018&      -0.015&      -0.014&       0.011&       0.024&       0.422&       0.393\\
\multicolumn{2}{l}{$ u_3 $}     &      -0.012&      -0.012&      -0.016&      -0.013&       0.009&       0.010&       0.022&       0.452\\
\multicolumn{2}{l}{$ u_4 $}     &      -0.017&      -0.017&      -0.017&      -0.021&       0.009&       0.010&       0.010&       0.024\\[0.3ex]
            \hline \\[-1.8ex]
            \multicolumn{2}{l}{{Women}} \\
            & & $log(Y^*_1)$ & $log(Y^*_2)$ & $log(Y^*_3)$ & $log(Y^*_4)$ & $u_1$ & $u_2$ & $u_3$ & $u_4$ \\[0.3ex]
            \hline \\[-1.8ex]
\multicolumn{2}{l}{$ log(Y^*_1) $}     &       0.540&       0.915&       0.881&       0.850&      -0.346&      -0.300&      -0.262&      -0.245\\
\multicolumn{2}{l}{$ log(Y^*_2) $}     &       0.479&       0.506&       0.953&       0.915&      -0.296&      -0.380&      -0.294&      -0.270\\
\multicolumn{2}{l}{$ log(Y^*_3) $}     &       0.451&       0.472&       0.485&       0.952&      -0.288&      -0.356&      -0.363&      -0.281\\
\multicolumn{2}{l}{$ log(Y^*_4) $}     &       0.430&       0.448&       0.456&       0.473&      -0.285&      -0.345&      -0.329&      -0.325\\
\multicolumn{2}{l}{$ u_1 $}     &      -0.048&      -0.040&      -0.038&      -0.037&       0.036&       0.423&       0.347&       0.320\\
\multicolumn{2}{l}{$ u_2 $}     &      -0.038&      -0.047&      -0.043&      -0.041&       0.014&       0.030&       0.450&       0.408\\
\multicolumn{2}{l}{$ u_3 $}   &      -0.032&      -0.034&      -0.042&      -0.037&       0.011&       0.013&       0.027&       0.435\\
\multicolumn{2}{l}{$ u_4 $}     &      -0.030&      -0.032&      -0.032&      -0.037&       0.010&       0.012&       0.012&       0.027\\[0.3ex]
            \hline \\[-1.8ex]
            \bottomrule
            \bottomrule
        \end{tabular}  
        \begin{tablenotes}
            \item \footnotesize{Table \ref{tab:varcov_balanced} presents the variance-covariance matrices of the indicated random variables for the entire dataset (panel 1) and by gender (panels 2 and 3), where the covariance terms on the upper half have been replaced by corresponding correlation coefficients. $Y^*_1 \coloneq Y^*_{i1}$ denotes the income in the first available period with employment for individual $i$ (similarly for $u_t \coloneq u_{it}$ and $\forall t \in \{1, 2, 3, 4\}$). Since all estimates within a matrix are based on the same amount of observations (All: $5118$; Men: $2495$; Women: $2623$), we refrain from displaying the effective number of observations underlying each matrix. Data: SOEP-CMI-ADIAB (strongly balanced sample).}
        \end{tablenotes}
    \end{threeparttable}
\end{table}

\begin{table}[!htbp] 
    \centering
    \begin{threeparttable}
        \caption{Reliability ratios -- Strongly balanced sample}
        \label{tab:reliab_ratio_class_II}
        \begin{tabular}{lcccccc}
            \toprule 
            \toprule
            & \multicolumn{3}{c}{Classical} & \multicolumn{3}{c}{Non-classical} \\
            & All & Men & Women & All & Men & Women \\
            & \footnotesize{(1)} & \footnotesize{(2)} & \footnotesize{(3)} & \footnotesize{(4)} & \footnotesize{(5)} & \footnotesize{(6)} \\[0.3ex]
            \hline \\[-1.8ex]
$ Y_1 $     &       0.935&       0.908&       0.937 &       1.005&       0.978&       1.025\\
$ Y_2 $     &       0.940&       0.912&       0.944 &       1.015&       0.974&       1.038\\
$ Y_3 $     &       0.944&       0.915&       0.947 &       1.013&       0.975&       1.034\\
$ Y_4 $     &       0.939&       0.902&       0.946 &       1.008&       0.986&       1.023\\
$ \Delta Y_2 $&       0.511&       0.390&       0.574 &  0.693&       0.532&       0.772\\
$ \Delta Y_3 $&       0.440&       0.423&       0.453 &       0.612&       0.575&       0.640\\
$ \Delta Y_4 $&       0.442&       0.416&       0.461 &       0.616&       0.589&       0.634\\[0.3ex]
            \bottomrule
            \bottomrule
        \end{tabular}  
        \begin{tablenotes}
            \item \footnotesize{Table \ref{tab:reliab_ratio_class_II} presents calculated reliability ratios for the classical measurement error case (based on Equation \ref{eq:reliab_ratio_class} and the estimated variance-covariance matrix in Table \ref{tab:varcov_balanced}) and for the non-classical measurement error case (regression-based, c.f.\ Equation \ref{eq:reliab_ratio_nonclass}) using the strongly balanced sample. The standard error is smaller than $0.07$ for all estimated coefficients. Data source: SOEP-CMI-ADIAB (strongly balanced sample).}
        \end{tablenotes}
    \end{threeparttable}
\end{table}

\begin{sidewaystable}
    \centering
    \begin{threeparttable}
        \caption{Consent decision -- Unweighted summary statistics (SOEP-Core)}
        \label{tab:consent_SOEPCore_unweighted}
        \begin{tabular}{lccccccccc}
            \toprule 
            \toprule
            & \multicolumn{9}{c}{\small{Subsamples of the SOEP-Core}} \\
            & \small{Total} & \small{Not asked} & \small{Asked} & \small{Not consented} & \small{Consented} & \small{Not matched} & \small{Matched} & \small{Exact match} & \small{Other match} \\
            & \footnotesize{(1)} & \footnotesize{(2)} & \footnotesize{(3)} & \footnotesize{(4)} & \footnotesize{(5)} & \footnotesize{(6)} & \footnotesize{(7)} & \footnotesize{(8)} & \footnotesize{(9)} \\[0.3ex]
            \hline \\[-1.8ex]
Female    &     0.50&     0.51&     0.49&     0.50&     0.49&     0.50&     0.49&     0.49&     0.47\\
          &\footnotesize(0.50)&\footnotesize(0.50)&\footnotesize(0.50)&\footnotesize(0.50)&\footnotesize(0.50)&\footnotesize(0.50)&\footnotesize(0.50)&\footnotesize(0.50)&\footnotesize(0.50)\\
Single    &     0.30&     0.31&     0.30&     0.31&     0.29&     0.30&     0.29&     0.29&     0.31\\
          &\footnotesize(0.46)&\footnotesize(0.46)&\footnotesize(0.46)&\footnotesize(0.46)&\footnotesize(0.46)&\footnotesize(0.46)&\footnotesize(0.46)&\footnotesize(0.45)&\footnotesize(0.46)\\
German    &     0.76&     0.87&     0.60&     0.59&     0.60&     0.47&     0.62&     0.65&     0.38\\
          &\footnotesize(0.43)&\footnotesize(0.34)&\footnotesize(0.49)&\footnotesize(0.49)&\footnotesize(0.49)&\footnotesize(0.50)&\footnotesize(0.49)&\footnotesize(0.48)&\footnotesize(0.48)\\
Migration background&     0.35&     0.24&     0.53&     0.57&     0.51&     0.64&     0.50&     0.46&     0.73\\
          &\footnotesize(0.48)&\footnotesize(0.43)&\footnotesize(0.50)&\footnotesize(0.49)&\footnotesize(0.50)&\footnotesize(0.48)&\footnotesize(0.50)&\footnotesize(0.50)&\footnotesize(0.45)\\
East      &     0.18&     0.19&     0.17&     0.15&     0.18&     0.15&     0.18&     0.19&     0.15\\
          &\footnotesize(0.39)&\footnotesize(0.39)&\footnotesize(0.38)&\footnotesize(0.35)&\footnotesize(0.38)&\footnotesize(0.35)&\footnotesize(0.39)&\footnotesize(0.39)&\footnotesize(0.35)\\
Abitur    &     0.18&     0.18&     0.18&     0.16&     0.18&     0.18&     0.18&     0.19&     0.13\\
          &\footnotesize(0.38)&\footnotesize(0.38)&\footnotesize(0.38)&\footnotesize(0.37)&\footnotesize(0.38)&\footnotesize(0.38)&\footnotesize(0.38)&\footnotesize(0.39)&\footnotesize(0.33)\\
Education (years)&    11.60&    11.74&    11.40&    11.29&    11.43&    11.00&    11.48&    11.60&    10.64\\
          &\footnotesize(2.77)&\footnotesize(2.64)&\footnotesize(2.95)&\footnotesize(2.85)&\footnotesize(2.98)&\footnotesize(3.28)&\footnotesize(2.94)&\footnotesize(2.93)&\footnotesize(2.94)\\
Age (years)&    44.27&    44.64&    43.74&    43.27&    43.89&    43.69&    43.91&    44.37&    40.68\\
          &\footnotesize(18.74)&\footnotesize(19.70)&\footnotesize(17.25)&\footnotesize(17.72)&\footnotesize(17.09)&\footnotesize(20.23)&\footnotesize(16.70)&\footnotesize(16.74)&\footnotesize(15.99)\\
Employed  &     0.57&     0.57&     0.57&     0.57&     0.57&     0.38&     0.59&     0.61&     0.50\\
          &\footnotesize(0.50)&\footnotesize(0.50)&\footnotesize(0.49)&\footnotesize(0.50)&\footnotesize(0.49)&\footnotesize(0.49)&\footnotesize(0.49)&\footnotesize(0.49)&\footnotesize(0.50)\\
Gross monthly income (\euro)&  2628.52&  2485.58&  2846.32&  2685.84&  2896.33&  2919.50&  2894.65&  2902.27&  2830.26\\
          &\footnotesize(11459.08)&\footnotesize(14511.71)&\footnotesize(3230.48)&\footnotesize(2912.68)&\footnotesize(3321.79)&\footnotesize(4314.52)&\footnotesize(3238.14)&\footnotesize(3282.66)&\footnotesize(2834.37)\\
Year of first participation&     2005&     2001&     2013&     2012&     2013&     2013&     2013&     2013&     2014\\
          &\footnotesize(11.60)&\footnotesize(11.22)&\footnotesize(7.61)&\footnotesize(8.45)&\footnotesize(7.30)&\footnotesize(7.46)&\footnotesize(7.29)&\footnotesize(7.45)&\footnotesize(5.81)\\
SOEP-Core (share)&     1.00&     1.00&     1.00&     1.00&     1.00&     1.00&     1.00&     1.00&     1.00\\
          &\footnotesize(0.00)&\footnotesize(0.00)&\footnotesize(0.00)&\footnotesize(0.00)&\footnotesize(0.00)&\footnotesize(0.00)&\footnotesize(0.00)&\footnotesize(0.00)&\footnotesize(0.00)\\[0.3ex]
            \hline \\[-1.8ex]
            N &   104'523&    63'624&    40'899&     9'789&    31'110&     3'181&    27'929&    24'428&     3'501\\
            Share of total (in \%) & 100 & 61 & 39 & 9 & 30 & 3 & 27 & 23 & 3 \\[0.3ex]
            \bottomrule
            \bottomrule
        \end{tabular}  
        \begin{tablenotes}
            \item \footnotesize{Table \ref{tab:consent_SOEPCore_unweighted} provides the estimated mean and standard deviation (in brackets) of observable characteristics for different subsamples of SOEP-Core based on inclusion of the question of consent in the respective survey wave (\textit{Asked}), the latest available consent decision (\textit{Consented}), and the match type (\textit{Matched}; \textit{Other match}, includes probabilistic and manual matches). All estimated statistics concerning time-varying variables are based on the last available observation for each individual. $N$ denotes the total number of cross-sectional units in a given subsample. The shares are relative to the total sample size ($104'523$). The shares may not add up due to rounding errors. Data: SOEP, SOEP-CMI-ADIAB (SOEP-Core only).}
        \end{tablenotes}
    \end{threeparttable}
\end{sidewaystable}

\begin{sidewaystable}
    \centering
    \begin{threeparttable}
        \caption{Consent decision -- Unweighted summary statistics (SOEP-IS)}
        \label{tab:consent_SOEPIS_unweighted}
        \begin{tabular}{lccccccc}
            \toprule 
            \toprule
            & \multicolumn{7}{c}{\small{Subsamples of the SOEP-IS}} \\
            & \small{Total} & \small{Not asked} & \small{Asked} & \small{Not consented} & \small{Consented} & \small{Matched} & \small{Exact match} \\
            & \footnotesize{(1)} & \footnotesize{(2)} & \footnotesize{(3)} & \footnotesize{(4)} & \footnotesize{(5)} & \footnotesize{(7)} & \footnotesize{(8)} \\[0.3ex]
            \hline \\[-1.8ex]
Female    &     0.53&     0.52&     0.53&     0.54&     0.53&     0.53&     0.53\\
          &\footnotesize(0.50)&\footnotesize(0.50)&\footnotesize(0.50)&\footnotesize(0.50)&\footnotesize(0.50)&\footnotesize(0.50)&\footnotesize(0.50)\\
Single    &     0.26&     0.28&     0.23&     0.24&     0.22&     0.23&     0.23\\
          &\footnotesize(0.44)&\footnotesize(0.45)&\footnotesize(0.42)&\footnotesize(0.43)&\footnotesize(0.42)&\footnotesize(0.42)&\footnotesize(0.42)\\
German    &     0.94&     0.93&     0.96&     0.95&     0.97&     0.97&     0.97\\
          &\footnotesize(0.24)&\footnotesize(0.26)&\footnotesize(0.20)&\footnotesize(0.22)&\footnotesize(0.18)&\footnotesize(0.18)&\footnotesize(0.17)\\
Migration background&     0.24&     0.28&     0.19&     0.21&     0.18&     0.18&     0.17\\
          &\footnotesize(0.43)&\footnotesize(0.45)&\footnotesize(0.39)&\footnotesize(0.41)&\footnotesize(0.38)&\footnotesize(0.39)&\footnotesize(0.38)\\
East      &     0.18&     0.16&     0.20&     0.18&     0.22&     0.22&     0.23\\
          &\footnotesize(0.38)&\footnotesize(0.37)&\footnotesize(0.40)&\footnotesize(0.38)&\footnotesize(0.41)&\footnotesize(0.42)&\footnotesize(0.42)\\
Abitur    &     0.24&     0.24&     0.25&     0.25&     0.24&     0.24&     0.24\\
          &\footnotesize(0.43)&\footnotesize(0.42)&\footnotesize(0.43)&\footnotesize(0.43)&\footnotesize(0.43)&\footnotesize(0.43)&\footnotesize(0.43)\\
Education (years)&    12.39&    12.31&    12.53&    12.57&    12.49&    12.49&    12.49\\
          &\footnotesize(2.71)&\footnotesize(2.68)&\footnotesize(2.76)&\footnotesize(2.78)&\footnotesize(2.74)&\footnotesize(2.71)&\footnotesize(2.71)\\
Age (years)&    51.09&    48.68&    55.05&    55.30&    54.88&    53.64&    53.56\\
          &\footnotesize(19.19)&\footnotesize(19.18)&\footnotesize(18.54)&\footnotesize(18.41)&\footnotesize(18.63)&\footnotesize(18.04)&\footnotesize(18.02)\\
Employed  &     0.55&     0.57&     0.52&     0.52&     0.52&     0.54&     0.55\\
          &\footnotesize(0.50)&\footnotesize(0.49)&\footnotesize(0.50)&\footnotesize(0.50)&\footnotesize(0.50)&\footnotesize(0.50)&\footnotesize(0.50)\\
Gross monthly income (\euro)&  2848.66&  2728.58&  3054.50&  3008.01&  3084.94&  3065.16&  3028.52\\
          &\footnotesize(2259.07)&\footnotesize(2234.50)&\footnotesize(2286.48)&\footnotesize(2145.98)&\footnotesize(2374.33)&\footnotesize(2352.98)&\footnotesize(2313.00)\\
Year of first participation&     2013&     2014&     2012&     2012&     2012&     2012&     2012\\
          &\footnotesize(4.22)&\footnotesize(4.20)&\footnotesize(4.17)&\footnotesize(3.96)&\footnotesize(4.30)&\footnotesize(4.24)&\footnotesize(4.26)\\
SOEP-Core (share)&     0.00&     0.00&     0.00&     0.00&     0.00&     0.00&     0.00\\
          &\footnotesize(0.00)&\footnotesize(0.00)&\footnotesize(0.00)&\footnotesize(0.00)&\footnotesize(0.00)&\footnotesize(0.00)&\footnotesize(0.00)\\[0.3ex]
            \hline \\[-1.8ex]
            N &    11'755&     7'309&     4'446&     1'823&     2'623&     2'424&     2'258\\
            Share of total (in \%) & 100 & 62 & 38 & 16 & 22 & 21 & 19 \\[0.3ex]
            \bottomrule
            \bottomrule
        \end{tabular}  
        \begin{tablenotes}
            \item \footnotesize{Table \ref{tab:consent_SOEPIS_unweighted} provides the estimated mean and standard deviation (in brackets) of observable characteristics for different subsamples of SOEP-IS based on inclusion of the question of consent in the respective survey wave (\textit{Asked}), the latest available consent decision (\textit{Consented}), and the match type (\textit{Matched}). All estimated statistics concerning time-varying variables are based on the last available observation for each individual. $N$ denotes the total number of cross-sectional units in a given subsample. The shares are relative to the total sample size ($11'755$). The shares may not add up due to rounding errors. Data: SOEP, SOEP-CMI-ADIAB (SOEP-IS only).}
        \end{tablenotes}
    \end{threeparttable}
\end{sidewaystable}

\begin{sidewaystable}
    \centering
    \begin{threeparttable}
        \caption{Consent decision -- Weighted summary statistics (SOEP-Core)}
        \label{tab:consent_SOEPCore_weighted}
        \begin{tabular}{lccccccccc}
            \toprule 
            \toprule
            & \multicolumn{9}{c}{\small{Subsamples of the SOEP-Core}} \\
            & \small{Total} & \small{Not asked} & \small{Asked} & \small{Not consented} & \small{Consented} & \small{Not matched} & \small{Matched} & \small{Exact match} & \small{Other match}\\
            & \footnotesize{(1)} & \footnotesize{(2)} & \footnotesize{(3)} & \footnotesize{(4)} & \footnotesize{(5)} & \footnotesize{(6)} & \footnotesize{(7)} & \footnotesize{(8)} & \footnotesize{(9)} \\[0.3ex]
            \hline \\[-1.8ex]
Female      &        0.50&        0.50&        0.50&        0.50&        0.50&        0.54&        0.50&        0.50&        0.48\\
            &\footnotesize(0.50)&\footnotesize(0.50)&\footnotesize(0.50)&\footnotesize(0.50)&\footnotesize(0.50)&\footnotesize(0.50)&\footnotesize(0.50)&\footnotesize(0.50)&\footnotesize(0.50)\\
Single      &        0.31&        0.31&        0.32&        0.32&        0.32&        0.28&        0.32&        0.32&        0.34\\
            &\footnotesize(0.46)&\footnotesize(0.46)&\footnotesize(0.47)&\footnotesize(0.47)&\footnotesize(0.47)&\footnotesize(0.45)&\footnotesize(0.47)&\footnotesize(0.47)&\footnotesize(0.47)\\
German      &        0.87&        0.90&        0.82&        0.76&        0.84&        0.79&        0.84&        0.86&        0.64\\
            &\footnotesize(0.33)&\footnotesize(0.30)&\footnotesize(0.39)&\footnotesize(0.43)&\footnotesize(0.37)&\footnotesize(0.41)&\footnotesize(0.36)&\footnotesize(0.34)&\footnotesize(0.48)\\
Migration background&        0.22&        0.18&        0.30&        0.38&        0.27&        0.32&        0.27&        0.24&        0.49\\
            &\footnotesize(0.41)&\footnotesize(0.39)&\footnotesize(0.46)&\footnotesize(0.49)&\footnotesize(0.44)&\footnotesize(0.47)&\footnotesize(0.44)&\footnotesize(0.43)&\footnotesize(0.50)\\
East        &        0.15&        0.15&        0.16&        0.13&        0.17&        0.13&        0.17&        0.18&        0.13\\
            &\footnotesize(0.36)&\footnotesize(0.36)&\footnotesize(0.37)&\footnotesize(0.33)&\footnotesize(0.37)&\footnotesize(0.34)&\footnotesize(0.38)&\footnotesize(0.38)&\footnotesize(0.34)\\
Abitur      &        0.18&        0.16&        0.23&        0.19&        0.24&        0.29&        0.23&        0.24&        0.22\\
            &\footnotesize(0.38)&\footnotesize(0.36)&\footnotesize(0.42)&\footnotesize(0.40)&\footnotesize(0.43)&\footnotesize(0.46)&\footnotesize(0.42)&\footnotesize(0.42)&\footnotesize(0.41)\\
Education (years)&       11.76&       11.59&       12.14&       11.85&       12.22&       12.35&       12.21&       12.24&       11.91\\
            &\footnotesize(2.58)&\footnotesize(2.49)&\footnotesize(2.73)&\footnotesize(2.67)&\footnotesize(2.75)&\footnotesize(3.12)&\footnotesize(2.71)&\footnotesize(2.70)&\footnotesize(2.82)\\
Age (years) &       48.65&       48.42&       49.14&       48.75&       49.26&       55.31&       48.73&       48.95&       46.55\\
            &\footnotesize(20.45)&\footnotesize(21.06)&\footnotesize(19.07)&\footnotesize(19.77)&\footnotesize(18.85)&\footnotesize(23.43)&\footnotesize(18.30)&\footnotesize(18.23)&\footnotesize(18.88)\\
Employed    &        0.55&        0.52&        0.61&        0.57&        0.62&        0.42&        0.64&        0.64&        0.64\\
            &\footnotesize(0.50)&\footnotesize(0.50)&\footnotesize(0.49)&\footnotesize(0.49)&\footnotesize(0.49)&\footnotesize(0.49)&\footnotesize(0.48)&\footnotesize(0.48)&\footnotesize(0.48)\\
Gross monthly income (\euro)&     2520.22&     2282.87&     2975.68&     2752.37&     3040.45&     3313.61&     3024.93&     3016.96&     3104.70\\
            &\footnotesize(7043.44)&\footnotesize(8462.21)&\footnotesize(2661.34)&\footnotesize(2600.27)&\footnotesize(2675.37)&\footnotesize(3449.27)&\footnotesize(2623.83)&\footnotesize(2649.29)&\footnotesize(2352.89)\\
Year of first participation&        2001&        1997&        2010&        2009&        2010&        2009&        2011&        2010&        2012\\
            &\footnotesize(11.73)&\footnotesize(10.74)&\footnotesize(8.45)&\footnotesize(9.50)&\footnotesize(8.06)&\footnotesize(8.40)&\footnotesize(8.01)&\footnotesize(7.99)&\footnotesize(8.12)\\
SOEP-Core (share)&        1.00&        1.00&        1.00&        1.00&        1.00&        1.00&        1.00&        1.00&        1.00\\
            &\footnotesize(0.00)&\footnotesize(0.00)&\footnotesize(0.00)&\footnotesize(0.00)&\footnotesize(0.00)&\footnotesize(0.00)&\footnotesize(0.00)&\footnotesize(0.00)&\footnotesize(0.00)\\[0.3ex]
            \hline \\[-1.8ex]
            N &      100'883&       60'675&       40'208&        9'625&       30'583&        3'145&       27'438&       23'983&        3'455\\
            Share of total (in \%) & 100 & 60 & 40 & 10 & 30 & 3 & 27 & 24 & 3 \\[0.3ex]
            \bottomrule
            \bottomrule
        \end{tabular}  
        \begin{tablenotes}
            \item \footnotesize{Table \ref{tab:consent_SOEPCore_weighted} provides the weighted estimated mean and standard deviation (in brackets) of observable characteristics for different subsamples of SOEP-Core based on inclusion of the question of consent in the respective survey wave (\textit{Asked}), the latest available consent decision (\textit{Consented}), and the match type (\textit{Matched}; \textit{Other match}, includes probabilistic and manual matches). All estimated statistics concerning time-varying variables are based on the last available observation for each individual. The weighting is based on the sample weights \textit{phrf} as provided by SOEP-Core. $N$ denotes the total number of cross-sectional units in a given subsample. The shares are relative to the total sample size ($100'883$). The shares may not add up due to rounding errors. Data: SOEP, SOEP-CMI-ADIAB (SOEP-Core only).}
        \end{tablenotes}
    \end{threeparttable}
\end{sidewaystable}

\begin{sidewaystable}
    \centering
    \begin{threeparttable}
        \caption{Consent decision -- Weighted summary statistics (SOEP-IS)}
        \label{tab:consent_SOEPIS_weighted}
        \begin{tabular}{lccccccc}
            \toprule 
            \toprule
            & \multicolumn{7}{c}{\small{Subsamples of the SOEP-IS}} \\
            & \small{Total} & \small{Not asked} & \small{Asked} & \small{Not consented} & \small{Consented} & \small{Matched} & \small{Exact match} \\
            & \footnotesize{(1)} & \footnotesize{(2)} & \footnotesize{(3)} & \footnotesize{(4)} & \footnotesize{(5)} & \footnotesize{(7)} & \footnotesize{(8)} \\[0.3ex]
            \hline \\[-1.8ex]
Female    &     0.51&     0.50&     0.52&     0.52&     0.51&     0.51&     0.50\\
          &\footnotesize(0.50)&\footnotesize(0.50)&\footnotesize(0.50)&\footnotesize(0.50)&\footnotesize(0.50)&\footnotesize(0.50)&\footnotesize(0.50)\\
Single    &     0.31&     0.34&     0.27&     0.28&     0.26&     0.27&     0.28\\
          &\footnotesize(0.46)&\footnotesize(0.47)&\footnotesize(0.45)&\footnotesize(0.45)&\footnotesize(0.44)&\footnotesize(0.44)&\footnotesize(0.45)\\
German    &     0.87&     0.88&     0.85&     0.81&     0.88&     0.87&     0.89\\
          &\footnotesize(0.34)&\footnotesize(0.32)&\footnotesize(0.36)&\footnotesize(0.39)&\footnotesize(0.33)&\footnotesize(0.33)&\footnotesize(0.31)\\
Migration background&     0.32&     0.31&     0.32&     0.37&     0.29&     0.30&     0.28\\
          &\footnotesize(0.47)&\footnotesize(0.46)&\footnotesize(0.47)&\footnotesize(0.48)&\footnotesize(0.45)&\footnotesize(0.46)&\footnotesize(0.45)\\
East      &     0.17&     0.17&     0.17&     0.16&     0.17&     0.17&     0.18\\
          &\footnotesize(0.37)&\footnotesize(0.37)&\footnotesize(0.37)&\footnotesize(0.37)&\footnotesize(0.38)&\footnotesize(0.38)&\footnotesize(0.39)\\
Abitur    &     0.22&     0.22&     0.23&     0.23&     0.23&     0.23&     0.24\\
          &\footnotesize(0.42)&\footnotesize(0.41)&\footnotesize(0.42)&\footnotesize(0.42)&\footnotesize(0.42)&\footnotesize(0.42)&\footnotesize(0.43)\\
Education (years)&    12.22&    12.13&    12.35&    12.38&    12.33&    12.32&    12.35\\
          &\footnotesize(2.68)&\footnotesize(2.60)&\footnotesize(2.78)&\footnotesize(2.82)&\footnotesize(2.75)&\footnotesize(2.73)&\footnotesize(2.74)\\
Age (years)&    48.02&    46.35&    50.41&    50.21&    50.56&    49.59&    49.41\\
          &\footnotesize(19.38)&\footnotesize(19.87)&\footnotesize(18.41)&\footnotesize(18.46)&\footnotesize(18.37)&\footnotesize(17.67)&\footnotesize(17.64)\\
Employed  &     0.58&     0.58&     0.59&     0.58&     0.59&     0.61&     0.62\\
          &\footnotesize(0.49)&\footnotesize(0.49)&\footnotesize(0.49)&\footnotesize(0.49)&\footnotesize(0.49)&\footnotesize(0.49)&\footnotesize(0.49)\\
Gross monthly income (\euro)&  2747.20&  2534.35&  3024.31&  3031.24&  3019.23&  3007.43&  2977.49\\
          &\footnotesize(2135.77)&\footnotesize(1984.55)&\footnotesize(2288.90)&\footnotesize(2304.10)&\footnotesize(2278.68)&\footnotesize(2252.59)&\footnotesize(2186.87)\\
Year of first participation&     2013&     2013&     2013&     2014&     2013&     2013&     2013\\
          &\footnotesize(3.95)&\footnotesize(4.19)&\footnotesize(3.58)&\footnotesize(3.38)&\footnotesize(3.72)&\footnotesize(3.65)&\footnotesize(3.64)\\
SOEP-Core (share)&     0.00&     0.00&     0.00&     0.00&     0.00&     0.00&     0.00\\
          &\footnotesize(0.00)&\footnotesize(0.00)&\footnotesize(0.00)&\footnotesize(0.00)&\footnotesize(0.00)&\footnotesize(0.00)&\footnotesize(0.00)\\[0.3ex]
            \hline \\[-1.8ex]
            N &    10'691&     6'247&     4'444&     1'822&     2'622&     2'423&     2'257\\
            Share of total (in \%) & 100 & 88 & 12 & 6 & 6 & 5 & 5 \\[0.3ex]
            \bottomrule
            \bottomrule
        \end{tabular}  
        \begin{tablenotes}
            \item \footnotesize{Table \ref{tab:consent_SOEPIS_weighted} provides the weighted estimated mean and standard deviation (in brackets) of observable characteristics for different subsamples of SOEP-IS based on inclusion of the question of consent in the respective survey wave (\textit{Asked}), the latest available consent decision (\textit{Consented}), and the match type (\textit{Matched}). All estimated statistics concerning time-varying variables are based on the last available observation for each individual. The weighting is based on the sample weights \textit{phrf} as provided by SOEP-IS. $N$ denotes the total number of cross-sectional units in a given subsample. The shares are relative to the total sample size ($10'691$). The shares may not add up due to rounding errors. Data: SOEP, SOEP-CMI-ADIAB (SOEP-IS only).}
        \end{tablenotes}
    \end{threeparttable}
\end{sidewaystable}

\begin{sidewaystable}
    \centering
    \begin{threeparttable}
        \caption{Consent decision -- Unweighted summary statistics (Women)}
        \label{tab:consent_SOEPtotal_women}
        \begin{tabular}{lccccccccc}
            \toprule 
            \toprule
            & \multicolumn{9}{c}{\small{Subsamples of the SOEP}} \\
            & \small{Total} & \small{Not asked} & \small{Asked} & \small{Not consented} & \small{Consented} & \small{Not matched} & \small{Matched} & \small{Exact match} & \small{Other match}\\
            & \footnotesize{(1)} & \footnotesize{(2)} & \footnotesize{(3)} & \footnotesize{(4)} & \footnotesize{(5)} & \footnotesize{(6)} & \footnotesize{(7)} & \footnotesize{(8)} & \footnotesize{(9)} \\[0.3ex]
            \hline \\[-1.8ex]
Female      &        1.00&        1.00&        1.00&        1.00&        1.00&        1.00&        1.00&        1.00&        1.00\\
            &\footnotesize(0.00)&\footnotesize(0.00)&\footnotesize(0.00)&\footnotesize(0.00)&\footnotesize(0.00)&\footnotesize(0.00)&\footnotesize(0.00)&\footnotesize(0.00)&\footnotesize(0.00)\\
Single      &        0.27&        0.29&        0.25&        0.26&        0.24&        0.23&        0.24&        0.25&        0.24\\
            &\footnotesize(0.44)&\footnotesize(0.45)&\footnotesize(0.43)&\footnotesize(0.44)&\footnotesize(0.43)&\footnotesize(0.42)&\footnotesize(0.43)&\footnotesize(0.43)&\footnotesize(0.42)\\
German      &        0.80&        0.88&        0.67&        0.67&        0.67&        0.51&        0.69&        0.73&        0.43\\
            &\footnotesize(0.40)&\footnotesize(0.33)&\footnotesize(0.47)&\footnotesize(0.47)&\footnotesize(0.47)&\footnotesize(0.50)&\footnotesize(0.46)&\footnotesize(0.45)&\footnotesize(0.50)\\
Migration background&        0.33&        0.24&        0.46&        0.50&        0.45&        0.60&        0.43&        0.40&        0.68\\
            &\footnotesize(0.47)&\footnotesize(0.43)&\footnotesize(0.50)&\footnotesize(0.50)&\footnotesize(0.50)&\footnotesize(0.49)&\footnotesize(0.50)&\footnotesize(0.49)&\footnotesize(0.46)\\
East        &        0.18&        0.18&        0.18&        0.15&        0.19&        0.15&        0.19&        0.20&        0.15\\
            &\footnotesize(0.39)&\footnotesize(0.39)&\footnotesize(0.38)&\footnotesize(0.36)&\footnotesize(0.39)&\footnotesize(0.36)&\footnotesize(0.39)&\footnotesize(0.40)&\footnotesize(0.36)\\
Abitur      &        0.18&        0.18&        0.19&        0.18&        0.20&        0.18&        0.20&        0.21&        0.14\\
            &\footnotesize(0.39)&\footnotesize(0.38)&\footnotesize(0.39)&\footnotesize(0.38)&\footnotesize(0.40)&\footnotesize(0.39)&\footnotesize(0.40)&\footnotesize(0.40)&\footnotesize(0.35)\\
Education (years)&       11.63&       11.65&       11.59&       11.53&       11.61&       11.03&       11.68&       11.78&       10.83\\
            &\footnotesize(2.73)&\footnotesize(2.61)&\footnotesize(2.90)&\footnotesize(2.82)&\footnotesize(2.93)&\footnotesize(3.24)&\footnotesize(2.88)&\footnotesize(2.86)&\footnotesize(2.96)\\
Age (years) &       45.40&       45.37&       45.45&       45.70&       45.37&       45.96&       45.30&       45.75&       41.84\\
            &\footnotesize(19.06)&\footnotesize(19.98)&\footnotesize(17.59)&\footnotesize(18.32)&\footnotesize(17.32)&\footnotesize(21.47)&\footnotesize(16.79)&\footnotesize(16.76)&\footnotesize(16.62)\\
Employed    &        0.51&        0.51&        0.52&        0.51&        0.52&        0.30&        0.55&        0.56&        0.43\\
            &\footnotesize(0.50)&\footnotesize(0.50)&\footnotesize(0.50)&\footnotesize(0.50)&\footnotesize(0.50)&\footnotesize(0.46)&\footnotesize(0.50)&\footnotesize(0.50)&\footnotesize(0.50)\\
Gross monthly income (\euro)&     1947.93&     1755.18&     2244.81&     2078.09&     2301.65&     2196.10&     2308.23&     2305.56&     2335.21\\
            &\footnotesize(1716.46)&\footnotesize(1621.82)&\footnotesize(1813.21)&\footnotesize(1749.27)&\footnotesize(1831.13)&\footnotesize(2607.98)&\footnotesize(1771.42)&\footnotesize(1729.25)&\footnotesize(2151.73)\\
Year of first participation&        2006&        2002&        2012&        2012&        2012&        2012&        2012&        2012&        2014\\
            &\footnotesize(11.21)&\footnotesize(11.35)&\footnotesize(7.45)&\footnotesize(7.97)&\footnotesize(7.25)&\footnotesize(7.63)&\footnotesize(7.20)&\footnotesize(7.31)&\footnotesize(6.09)\\
Share in SOEP-Core&        0.89&        0.89&        0.90&        0.83&        0.92&        0.94&        0.91&        0.91&        0.95\\
            &\footnotesize(0.31)&\footnotesize(0.31)&\footnotesize(0.31)&\footnotesize(0.37)&\footnotesize(0.28)&\footnotesize(0.23)&\footnotesize(0.28)&\footnotesize(0.29)&\footnotesize(0.22)\\[0.3ex]
            \hline \\[-1.8ex]
            N  &       58'628&       36'126&       22'502&        5'909&       16'593&        1'673&       14'920&       13'201&        1'719\\
            Share of total (in \%) & 100 & 62 & 38 & 10 & 28 & 3 & 25 & 23 & 3 \\[0.3ex]
            \bottomrule
            \bottomrule
        \end{tabular}  
        \begin{tablenotes}
            \item \footnotesize{Table \ref{tab:consent_SOEPtotal_women} provides the estimated mean and standard deviation (in brackets) of observable characteristics for different subsamples of SOEP based on inclusion of the question of consent in the respective survey wave (\textit{Asked}), the latest available consent decision (\textit{Consented}), and the match type (\textit{Matched}; \textit{Other match}, includes probabilistic and manual matches). All estimated statistics concerning time-varying variables are based on the last available observation for each female individual. $N$ denotes the total number of cross-sectional units in a given subsample. The shares are relative to the total sample size ($58'628$). The shares may not add up due to rounding errors. Data: SOEP, SOEP-CMI-ADIAB.}
        \end{tablenotes}
    \end{threeparttable}
\end{sidewaystable}

\begin{sidewaystable}
    \centering
    \begin{threeparttable}
        \caption{Consent decision -- Unweighted summary statistics (Men)}
        \label{tab:consent_SOEPtotal_men}
        \begin{tabular}{lccccccccc}
            \toprule 
            \toprule
            & \multicolumn{9}{c}{\small{Subsamples of the SOEP}} \\
            & \small{Total} & \small{Not asked} & \small{Asked} & \small{Not consented} & \small{Consented} & \small{Not matched} & \small{Matched} & \small{Exact match} & \small{Other match}\\
            & \footnotesize{(1)} & \footnotesize{(2)} & \footnotesize{(3)} & \footnotesize{(4)} & \footnotesize{(5)} & \footnotesize{(6)} & \footnotesize{(7)} & \footnotesize{(8)} & \footnotesize{(9)} \\[0.3ex]
            \hline \\[-1.8ex]
Female      &        0.00&        0.00&        0.00&        0.00&        0.00&        0.00&        0.00&        0.00&        0.00\\
            &\footnotesize(0.00)&\footnotesize(0.00)&\footnotesize(0.00)&\footnotesize(0.00)&\footnotesize(0.00)&\footnotesize(0.00)&\footnotesize(0.00)&\footnotesize(0.00)&\footnotesize(0.00)\\
Single      &        0.33&        0.33&        0.33&        0.34&        0.33&        0.35&        0.33&        0.33&        0.36\\
            &\footnotesize(0.47)&\footnotesize(0.47)&\footnotesize(0.47)&\footnotesize(0.47)&\footnotesize(0.47)&\footnotesize(0.48)&\footnotesize(0.47)&\footnotesize(0.47)&\footnotesize(0.48)\\
German      &        0.76&        0.87&        0.59&        0.62&        0.59&        0.49&        0.60&        0.63&        0.37\\
            &\footnotesize(0.43)&\footnotesize(0.34)&\footnotesize(0.49)&\footnotesize(0.48)&\footnotesize(0.49)&\footnotesize(0.50)&\footnotesize(0.49)&\footnotesize(0.48)&\footnotesize(0.48)\\
Migration background&        0.36&        0.25&        0.52&        0.54&        0.52&        0.62&        0.51&        0.48&        0.73\\
            &\footnotesize(0.48)&\footnotesize(0.43)&\footnotesize(0.50)&\footnotesize(0.50)&\footnotesize(0.50)&\footnotesize(0.49)&\footnotesize(0.50)&\footnotesize(0.50)&\footnotesize(0.44)\\
East        &        0.18&        0.19&        0.17&        0.15&        0.18&        0.14&        0.18&        0.19&        0.14\\
            &\footnotesize(0.39)&\footnotesize(0.39)&\footnotesize(0.38)&\footnotesize(0.36)&\footnotesize(0.38)&\footnotesize(0.35)&\footnotesize(0.39)&\footnotesize(0.39)&\footnotesize(0.35)\\
Abitur      &        0.18&        0.19&        0.17&        0.17&        0.17&        0.18&        0.17&        0.18&        0.12\\
            &\footnotesize(0.39)&\footnotesize(0.39)&\footnotesize(0.38)&\footnotesize(0.37)&\footnotesize(0.38)&\footnotesize(0.39)&\footnotesize(0.38)&\footnotesize(0.38)&\footnotesize(0.33)\\
Education (years)&       11.75&       11.95&       11.44&       11.49&       11.43&       11.16&       11.46&       11.57&       10.64\\
            &\footnotesize(2.83)&\footnotesize(2.68)&\footnotesize(3.00)&\footnotesize(2.94)&\footnotesize(3.03)&\footnotesize(3.34)&\footnotesize(2.99)&\footnotesize(2.98)&\footnotesize(2.94)\\
Age (years) &       44.57&       44.79&       44.26&       44.61&       44.14&       44.55&       44.10&       44.57&       40.85\\
            &\footnotesize(18.74)&\footnotesize(19.37)&\footnotesize(17.79)&\footnotesize(18.39)&\footnotesize(17.59)&\footnotesize(20.71)&\footnotesize(17.21)&\footnotesize(17.31)&\footnotesize(16.13)\\
Employed    &        0.62&        0.63&        0.61&        0.61&        0.61&        0.43&        0.63&        0.64&        0.57\\
            &\footnotesize(0.48)&\footnotesize(0.48)&\footnotesize(0.49)&\footnotesize(0.49)&\footnotesize(0.49)&\footnotesize(0.49)&\footnotesize(0.48)&\footnotesize(0.48)&\footnotesize(0.50)\\
Gross monthly income (\euro)&     3229.37&     3133.36&     3377.11&     3281.72&     3408.50&     3462.38&     3404.47&     3428.64&     3214.59\\
            &\footnotesize(14725.16)&\footnotesize(18654.01)&\footnotesize(3874.09)&\footnotesize(3395.19)&\footnotesize(4018.87)&\footnotesize(5082.39)&\footnotesize(3928.07)&\footnotesize(4013.85)&\footnotesize(3169.58)\\
Year of first participation&        2006&        2002&        2013&        2012&        2013&        2013&        2013&        2013&        2015\\
            &\footnotesize(11.42)&\footnotesize(11.49)&\footnotesize(7.20)&\footnotesize(7.85)&\footnotesize(6.95)&\footnotesize(7.02)&\footnotesize(6.95)&\footnotesize(7.12)&\footnotesize(5.40)\\
Share in SOEP-Core&        0.90&        0.90&        0.91&        0.85&        0.93&        0.94&        0.93&        0.92&        0.96\\
            &\footnotesize(0.30)&\footnotesize(0.30)&\footnotesize(0.29)&\footnotesize(0.36)&\footnotesize(0.26)&\footnotesize(0.24)&\footnotesize(0.26)&\footnotesize(0.27)&\footnotesize(0.19)\\[0.3ex]
            \hline \\[-1.8ex]
            N &       57'650&       34'807&       22'843&        5'703&       17'140&        1'707&       15'433&       13'485&        1'948\\
            Share of total (in \%) & 100 & 60 & 40 & 10 & 30 & 3 & 27 & 23 & 3 \\[0.3ex]
            \bottomrule
            \bottomrule
        \end{tabular}  
        \begin{tablenotes}
            \item \footnotesize{Table \ref{tab:consent_SOEPtotal_men} provides the estimated mean and standard deviation (in brackets) of observable characteristics for different subsamples of SOEP based on inclusion of the question of consent in the respective survey wave (\textit{Asked}), the latest available consent decision (\textit{Consented}), and the match type (\textit{Matched}; \textit{Other match}, includes probabilistic and manual matches). All estimated statistics concerning time-varying variables are based on the last available observation for each male individual. $N$ denotes the total number of cross-sectional units in a given subsample. The shares are relative to the total sample size ($57'650$). The shares may not add up due to rounding errors. Data: SOEP, SOEP-CMI-ADIAB.}
        \end{tablenotes}
    \end{threeparttable}
\end{sidewaystable}


\end{document}